\pdfoutput=1
\documentclass[aps,prd,preprint,tightenlines,lineumbers,nofootinbib]{revtex4-1}
\usepackage{graphicx}
\usepackage{dcolumn}
\usepackage{bm}
\usepackage{lineno}
\usepackage{amsmath}

\usepackage[colorlinks=true,linkcolor=blue,filecolor=blue,urlcolor=blue,citecolor=blue,pdftex=true,
plainpages=false]{hyperref}

\newcommand{\etal}{{\it et al.}}

\begin{document}

\preprint{\tighten\vbox{\hbox{\hfil EFI 15-21}}}
\preprint{\tighten\vbox{\hbox{\hfil FERMILAB-PUB-15-384-T }}}

\title
{\LARGE Leptonic Decays of Charged Pseudoscalar Mesons $-$ 2015}

\author{Jonathan L. Rosner}
\affiliation{Enrico Fermi Institute, University of Chicago, Chicago, IL 60637}
\author{Sheldon Stone}
\affiliation{Department of Physics, Syracuse University, Syracuse, NY 13244 }
\author{Ruth S.~\surname{Van de Water}}
\affiliation{Fermi National Accelerator Laboratory, Batavia, IL, USA \\}

\date{\today \\ \bigskip}

\begin{abstract}
We review the physics of purely leptonic decays of $\pi^\pm$, $K^\pm$,
$D^{\pm}$, $D_s^\pm$, and $B^\pm$ pseudoscalar mesons.  The measured decay
rates are related to the product of the relevant weak-interaction-based CKM
matrix element of the constituent quarks and a strong interaction parameter
related to the overlap of the quark and antiquark wave-functions in the meson,
called the decay constant $f_P$.  
The leptonic decay constants for $\pi^\pm$, $K^\pm$, $D^{\pm}$, $D_s^\pm$, and $B^\pm$ mesons can be obtained with controlled theoretical uncertainties and high precision from {\it ab initio} lattice-QCD simulations.
The combination of experimental leptonic decay-rate measurements and theoretical decay-constant calculations enables the determination of several elements of the CKM matrix within the standard model.  These determinations are competitive with those obtained from semileptonic decays, and also complementary because they are sensitive to axial-vector (as opposed to vector) quark flavor-changing currents.  They can also be used to test the unitarity of the first and second rows of the CKM matrix.
Conversely, taking the CKM elements predicted by unitarity, one can infer ``experimental" values for $f_P$ that can be compared with theory.  These provide tests of lattice-QCD methods, provided new-physics contributions to leptonic decays are negligible at the current level of precision.
This mini-review was prepared for the Particle Data Group's 2016 edition, updating the versions in Refs.~\cite{Nakamura:2010zzi,Beringer:1900zz,Agashe:2014kda}.
\end{abstract}

\maketitle

\section{Introduction}

\begin{figure}[hbt]
\centering
\includegraphics[width=3.5in]{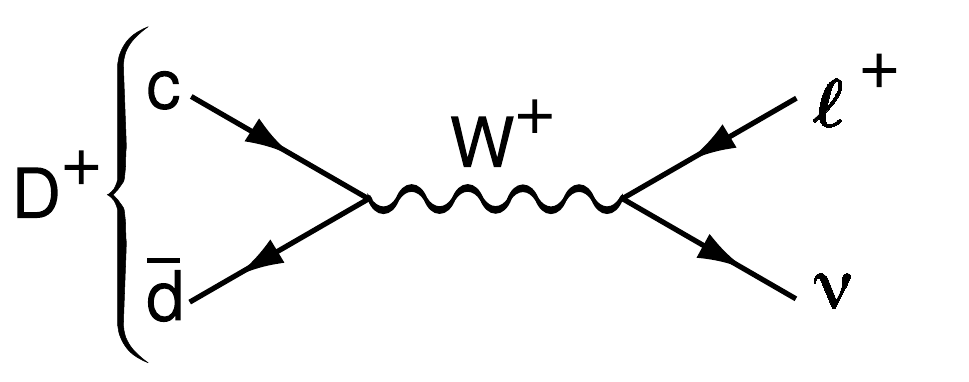}\vskip -0.02mm
\caption{The annihilation process for pure $D^+$ leptonic decays in the
standard model.
 } \label{Ptoellnu}
\end{figure}

Charged mesons formed from a quark and antiquark can decay to a
lepton-neutrino pair when these objects annihilate via a virtual
$W$ boson. Fig.~\ref{Ptoellnu} illustrates this process for the
purely leptonic decay of a $D^+$ meson.

Similar quark-antiquark annihilations via a virtual $W^+$ to
the $\ell^+ {\nu}$ final states occur for the $\pi^+$, $K^+$, $D_s^+$, and
$B^+$ mesons.  (Whenever psuedoscalar-meson charges are specified in this article, use of the charge-conjugate particles and corresponding decays are also implied.) Let $P$ be
any of these pseudoscalar mesons.  To lowest order, the decay width is
\begin{equation}
\Gamma(P\to \ell\nu) = {{G_F^2}\over 8\pi}f_{P}^2\ m_{\ell}^2M_{P}
\left(1-{m_{\ell}^2\over M_{P}^2}\right)^2 \left|V_{q_1
q_2}\right|^2~. \label{equ_rate}
\end{equation}

\noindent Here $M_{P}$ is the $P$ mass, $m_{\ell}$ is the $\ell$
mass, $V_{q_1 q_2}$ is the Cabibbo-Kobayashi-Maskawa (CKM) matrix
element between the constituent quarks $q_1 \bar q_2$ in $P$, and
$G_F$ is the Fermi coupling constant. The decay constant $f_P$ is proportional to the matrix element of the axial current
between the one-$P$-meson state and the vacuum:
\begin{equation}
	\langle 0 |  \bar{q}_1 \gamma_\mu \gamma_5 q_2 | P(p) \rangle = i p_\mu f_P  \,,
\end{equation}
and can be thought of as the ``wavefunction overlap" of the quark and antiquark.  In this article we use the convention in which $f_\pi \approx 130$~MeV.

The decay $P^\pm$ starts with a spin-0 meson, and ends up with a
left-handed neutrino or right-handed antineutrino.  By angular
momentum conservation, the $\ell^\pm$ must then also be left-handed
or right-handed, respectively. In the $m_\ell = 0$ limit, the decay
is forbidden, and can only occur as a result of the finite $\ell$
mass.  This helicity suppression is the origin of the $m_\ell^2$
dependence of the decay width. Radiative corrections are needed
when the final charged particle is an electron or muon; for the $\tau$ they are greatly suppressed due to the large lepton mass, and hence negligible.

Measurements of purely leptonic decay branching fractions and lifetimes allow an experimental determination of the product $\left|V_{q_1 q_2}\right| f_{P}$. If the decay constant $f_P$ is known to sufficient precision from theory, one can obtain the corresponding CKM element within the standard model.  If, on the other hand, one takes the value of $\left|V_{q_1 q_2}\right|$ assuming CKM unitarity, one can infer an ``experimental measurement" of the decay constant that can then be compared with theory.

The importance of measuring $\Gamma(P\to \ell\nu)$ depends on the particle being considered.  Leptonic decays of charged pseudoscalar mesons occur at tree level within the standard model.  Thus one does not expect large new-physics contributions to measurements of $\Gamma(P \to \ell\nu)$ for the lighter mesons $P = \pi^+, K^+$, and these processes in principle provide clean standard-model determinations of $V_{ud}$ and $V_{us}$.  The situation is different for leptonic decays of charm and bottom mesons.  The presence of new heavy particles such as charged Higgs bosons or leptoquarks could lead to observable effects in $\Gamma(P \to \ell\nu)$ for $P = D^+_{(s)}, B^+$~\cite{Hou:1992sy,Akeroyd:2002pi,Akeroyd:2003zr,Akeroyd:2003jb,Dobrescu:2008er}.  Thus the determination of $|V_{ub}|$ from $B^+ \to \tau \nu$ decay, in particular, should be considered a probe of new physics.  More generally, the ratio of leptonic decays to $\tau \nu$ over $\mu \nu$ final states probes lepton universality~\cite{Hou:1992sy,Hewett:1995aw}.

The determinations of CKM elements from leptonic decays of charged pseudoscalar mesons provide complementary information to those from other decay processes.  The decay $P \to \ell\nu$ proceeds in the standard model via the axial-vector current $\bar{q_1} \gamma_\mu \gamma_5 q_2$, whereas semileptonic pseudoscalar meson decays $P_1 \to P_2 \ell\nu$ proceed via the vector current $\bar{q_1} \gamma_\mu q_2$.  Thus the comparison of determinations of $\left|V_{q_1 q_2}\right|$ from leptonic and semileptonic decays tests the $V-A$ structure of the standard-model electroweak charged-current interaction.  More generally, a small right-handed admixture to the standard-model weak current would lead to discrepancies between $\left|V_{q_1 q_2}\right|$ obtained from leptonic pseudoscalar-meson decays, exclusive semileptonic pseudoscalar-meson decays, exclusive semileptonic baryon decays, and inclusive semileptonic decays~\cite{Crivellin:2009sd,Bernlochner:2014ova}.

Both measurements of the decay rates $\Gamma(P \to \ell\nu)$ and theoretical calculations of the decay constants 
$f_P$ for $P = \pi^+, K^+, D_{(s)}^+$ from numerical lattice-QCD simulations are now quite precise.  As a result, the elements of the first row of the CKM matrix $|V_{ud}|$ and $|V_{us}|$ can be obtained to sub-percent precision from $\pi^+ \to \ell \nu$ and $K^+ \to \ell \nu$, where the limiting error is from theory.  The elements of the second row of the CKM matrix $|V_{cd(s)}|$ can be obtained from leptonic decays of charged pseudoscalar mesons to few-percent precision, where here the limiting error is from experiment.  These enable stringent tests of the unitarity of the first and second rows of the CKM matrix. 

\bigskip

This review is organized as follows.  Because the experimental and theoretical issues associated with measurements of pions and kaons, charmed mesons, and bottom mesons differ, we discuss each one separately.  We begin with the pion and kaon system in Sec.~\ref{sec:PiKAll}.   First, in Sec.~\ref{sec:PiKExp} we review current measurements of the experimental decay rates.  We provide tables of branching-ratio measurements and determinations of the product $|V_{ud(s)}| f_{\pi^+(K^+)}$, as well as average values for these quantities including correlations and other effects needed to combine results.  Then, in Sec.~\ref{sec:PiKTheory} we summarize the status of theoretical calculations of the decay constants.  We provide tables of  recent lattice-QCD results for $f_{\pi^+}$, $f_{K^+}$, and their ratio from simulations including dynamical $u, d, s,$ and (in some cases $c$) quarks, and present averages for each of these quantities including correlations and strong SU(2)-isospin corrections as needed.  We note that, for the leptonic decay constants in Secs.~\ref{sec:PiKTheory},~\ref{sec:CharmTheory}, and~\ref{sec:BottomTheory}, when available we use preliminary averages from the Flavor Lattice Averaging Group~\cite{JuettnerLat15,FLAGInPrep} that update the determinations in Ref.~\cite{Aoki:2013ldr} to include results that have appeared since their most recent review, which dates from 2013.  We next discuss the charmed meson system in Sec.~\ref{sec:CharmAll}, again reviewing current experimental rate measurements in Sec.~\ref{sec:CharmExp} and theoretical decay-constant calculations in Sec.~\ref{sec:CharmTheory}.  Last, we discuss the bottom meson system in Sec.~\ref{sec:BottomAll}, following the same organization as the two previous sections.   

After having established the status of both experimental measurements and theoretical calculations of leptonic charged pseudoscalar-meson decays, we discuss some implications for phenomenology in Sec.~\ref{sec:Pheno}.  We combine the average ${\cal{B}}(P \to \ell\nu)$ with the average $f_P$ to obtain the relevant CKM elements from leptonic decays, and then compare them with determinations from other processes.  We also use the CKM elements obtained from leptonic decays to test the unitarity of the first and second rows of the CKM matrix.  Further, as in previous reviews, we combine the experimental ${\cal{B}}(P \to \ell\nu)$s with the associated CKM elements obtained from CKM unitarity to infer ``experimental" values for the decay constants; the comparison with theory
 provides a test of lattice and other QCD approaches assuming that new-physics
contributions to these processes are not significant.

\section{Pions and kaons} \label{sec:PiKAll}

\subsection{Experimental rate measurements} \label{sec:PiKExp}

The leading-order expression for $\Gamma(P\to \ell\nu)$ in Eq.~(\ref{equ_rate}) is modified by radiative corrections arising from diagrams involving photons, in some cases with additional quark loops.   These electroweak and ``hadronic" contributions can be combined
into an overall factor that multiplies the rate in the presence of only the
strong interaction ($\Gamma^{(0)}$) as follows (cf.\ Refs.~\cite{Marciano:2004uf,%
Cirigliano:2011ny}, and references therein): 
\begin{equation}
 \Gamma(P\to \ell\nu) = \Gamma^{(0)} \left[ 1 + \frac{\alpha}{\pi} C_P \right]
 \, ~~,\label{eq:RadRate}
\end{equation}
where $C_P$ differs for $P = \pi, K$.  
The inclusion of these corrections is
numerically important given the level of precision achieved on the experimental
measurements of the $\pi^{\pm} \to \mu^{\pm} \nu$ and $K^{\pm} \to \mu^{\pm}
\nu$ decay widths.  The explicit expression for the term in brackets above
including all known electroweak and hadronic contributions is given in
Eq.~(114) of Ref.~\cite{Cirigliano:2007ga}.  It includes the universal
short-distance electroweak correction obtained by Sirlin~\cite{Sirlin:1981ie},
the universal long-distance correction for a point-like meson from
Kinoshita~\cite{Kinoshita:1959ha}, and corrections that depend on the hadronic
structure~\cite{Knecht:1999ag}.  We evaluate $\delta_P \equiv
(\alpha/\pi ) C_P$ using the latest experimentally-measured meson and lepton
masses and coupling constants from the Particle Data
group~\cite{Agashe:2014kda}, and taking the low-energy constants (LECs) that
parameterize the hadronic contributions from Refs.~\cite{Ananthanarayan:2004qk,%
DescotesGenon:2005pw,Cirigliano:2007ga}.  
The finite non-logarithmic parts of
the LECs were estimated within the large-$N_C$ approximation assuming that
contributions from the lowest-lying resonances dominate.  We therefore
conservatively assign a 100\% uncertainty to the LECs, which leads to a $\pm
0.9$ error in $C_{\pi,K}$.\footnote{This uncertainty on $C_{\pi,K}$ is smaller
than the error estimated by Marciano and Sirlin in Ref.~\cite{Marciano:1993sh},
which predates the calculations of the hadronic-structure contributions in
Refs.~\cite{Knecht:1999ag,Ananthanarayan:2004qk,DescotesGenon:2005pw,%
Cirigliano:2007ga}.  The hadronic LECs incorporate the large short-distance
electroweak logarithm discussed in Ref.~\cite{Marciano:1993sh}, and their
dependence on the chiral renormalization scale cancels the scale-dependence
induced by chiral loops, thereby removing the dominant scale uncertainty of the
Marciano--Sirlin analysis~\cite{Marciano:1993sh}.}  We obtain the following
correction factors to the individual charged pion and kaon decay widths:
\begin{equation}
 \delta_{\pi} = 0.0176(21) \quad \rm{and} \quad  \delta_{K} =
 0.0107(21) \,. \label{eq:delta_EM_Gamma}
\end{equation}
The error on the ratio of kaon-to-pion leptonic decay widths is under better
theoretical control because the hadronic contributions from low-energy
constants estimated within the large-$N_c$ framework cancel at lowest order in
the chiral expansion.  For the ratio, we use the correction factor 
\begin{equation}
	 \delta_{K/\pi} = -0.0069(17) \,, \label{eq:delta_EM_Ratio}
\end{equation}
where we take the estimated error due to higher-order corrections in the chiral
expansion from Ref.~\cite{Cirigliano:2011tm}.

The sum of branching fractions for $\pi^- \to \mu^- \bar \nu$ and $\pi^- \to
\mu^- \bar \nu \gamma$ is 99.98770(4)\% \cite{Agashe:2014kda}.  The two modes
are difficult to separate experimentally, so we use this sum.  Together with
the lifetime 26.033(5) ns \cite{Agashe:2014kda} this implies $\Gamma(\pi^-
\to \mu^- \bar \nu [\gamma]) = 3.8408(7) \times 10^7$ s$^{-1}$.  The right-hand
side of Eq.\ (1) is modified by the factor $1.0176 \pm 0.0021$ mentioned above
to include photon emission and radiative corrections \cite{Marciano:1993sh,%
Cirigliano:2007xi}.  The decay rate together with the masses from the 2014 PDG review~\cite{Agashe:2014kda} gives
\begin{equation} \label{eqn:fpiv}
 f_{\pi^-}|V_{ud}| = (127.13 \pm 0.02 \pm 0.13)~{\rm MeV} \,,
\end{equation}
where the errors are from the experimental rate measurement and the radiative correction 
factor  $\delta_{\pi}$ in Eq.~(\ref{eq:delta_EM_Gamma}), respectively.
\noindent The uncertainty is dominated by that from theoretical estimate of the
hadronic structure-dependent radiative corrections, which include next-to-leading order contributions of ${\cal O}(e^2 p_{\pi, K}^2)$ in chiral perturbation theory~\cite{Cirigliano:2007ga}.

The data on $K_{\mu 2}$ decays have been updated recently through a global fit
to branching ratios and lifetime measurements \cite{Moulson:2014cra}: ${\cal B}
(K^-\to\mu^-\bar\nu [\gamma]) = 63.58(11)\%$ and $\tau_{K^\pm}=12.384(15)$ ns. 
The improvement in the branching ratio is primarily due to a new measurement
of ${\cal B}(K^\pm \to \pi^\pm \pi^+ \pi^-)$ from KLOE-2~\cite{Babusci:2014hxa},
which is correlated with ${\cal B}(K^\pm_{\mu 2})$ through the constraint that
the sum of individual branching ratios must equal unity.  The sum of branching
fractions for $K^- \to \mu^- \bar \nu$ and $K^- \to \mu^- \bar \nu \gamma$ and
the lifetime imply $\Gamma(K^- \to \mu^- \bar \nu [\gamma]) = 5.134(11) \times
10^7$ s$^{-1}$.  Again taking the 2014 PDG masses~\cite{Agashe:2014kda}, this decay
rate implies
\begin{equation} \label{eqn:fkv}
 f_{K^+} |V_{us}|  = (35.09 \pm 0.04 \pm 0.04)~{\rm MeV} \,,
\end{equation}
where the errors are from the experimental rate measurement and the radiative correction
factor  $\delta_{K}$, respectively.

Short-distance radiative corrections cancel in the ratio of pion-to-kaon decay 
rates~\cite{Antonelli:2010yf}:
\begin{equation} \label{eqn:wratio}
\frac{\Gamma_{K_{\ell 2 [\gamma]}}}{\Gamma_{\pi_{\ell 2 [\gamma]}}}
 = \frac{|V_{us}^2| f^2_{K^-}}{|V_{ud}|^2 f^2_{\pi^-}}
 \frac{m_K(1 - m_\ell^2/m_K^2)^2}{m_\pi(1 - m_\ell^2/m_\pi^2)^2}
 (1 + \delta_{K/\pi}) \,,
\end{equation}
where $\delta_{K/\pi}$ is given in Eq.~(\ref{eq:delta_EM_Ratio}).
The left-hand side of Eq.\ (\ref{eqn:wratio}) is 1.3367(28), yielding
\begin{equation} \label{eqn:vfratio}
\frac{|V_{us}| f_{K^-}}{|V_{ud}| f_{\pi^-}} = 0.27599 \pm 0.00029 \pm 0.00024 \,,
\end{equation}
where the first uncertainty is due to the branching fractions and the second is
due to $\delta_{K/\pi}$.  
Here the estimated error on the hadronic structure-dependent radiative corrections is
commensurate with the experimental error.

In summary, the main experimental results pertaining to charged pion and kaon leptonic decays are
\begin{eqnarray}
 |V_{ud}| f_{\pi^-} &=& (127.13 \pm 0.02 \pm 0.13)~{\rm MeV} \,, \label{eq:fPi_Vud} \\
 |V_{us}| f_{K^+} &=& (35.09 \pm 0.04 \pm 0.04)~{\rm MeV} \,, \label{eq:fK_Vus}  \\
 \frac{|V_{us}| f_{K^+}}{|V_{ud}|f_{\pi^-}} &=& 0.27599 \pm 0.00029 \pm 0.00024 \,, \label{eq:fK_Vus_over_fPi_Vud}
\end{eqnarray}
where the errors are from the experimental uncertainties in the branching fractions and the theoretical uncertainties in the radiative correction factors $\delta_{P}$, respectively.

\subsection{Theoretical decay-constant calculations} \label{sec:PiKTheory}

Table~\ref{tab:fkfpi} presents recent lattice-QCD calculations of the charged pion and kaon decay constants and their ratio from simulations with three ($N_f = 2+1$) or four flavors ($N_f = 2+1+1$) of dynamical quarks.  The results have been obtained using several independent sets of gauge-field configurations, and a variety of lattice fermion actions that are sensitive to different systematic uncertainties.\footnote{See the PDG mini-review on ``Lattice Quantum Chromodynamics"~\cite{PDGLQCD2015} for a general review of numerical lattice-QCD simulations.  Details on the different methods used in modern lattice-QCD calculations are provided in Appendix~A of the FLAG ``Review of lattice results concerning low energy particle physics"~\cite{Aoki:2013ldr}.} ÊThe lattice-QCD uncertainties on both the individual decay constants and their ratio have now reached sub-percent precision.  The SU(3)-breaking ratio $f_{K^+}/f_{\pi^+}$ can be obtained with especially small errors because statistical errors associated with the Monte Carlo simulations are correlated between the numerator and denominator, as are some systematics.  The good agreement between these largely independent determinations indicates that the lattice-QCD uncertainties are controlled and that the associated error estimates are reliable.\footnote{The recent review~\cite{Kronfeld:2012uk} summarizes the large body of evidence validating the methods employed in modern lattice-QCD simulations.} 

\begin{table}[tb]
\caption{Recent lattice-QCD results for $f_{\pi^+}$, $f_{K^+}$, and their ratio. The upper and lower panels show $(2+1+1)$-flavor and $(2+1)$-flavor determinations, respectively.  When two errors are shown, they are statistical and systematic, respectively.  
Results for $f_\pi$ and $f_K$ in the isospin-symmetric limit $m_u = m_d$ are noted with an ``$^\ast$"; they are corrected for isospin breaking via Eqs.~(\ref{eq:IsospinConvertPi})--(\ref{eq:IsospinConvertPiKRatio}) before computing the averages.
Unpublished results noted with a ``$^{\dag}$" or ``$^{\ddag}$" are not included in the averages.
 \label{tab:fkfpi}}
\begin{center}
\begin{tabular}{l@{\quad}c@{\quad}c@{\quad}c@{\quad}c} \hline\hline
Reference & $N_f$ & $f_{\pi^+}$(MeV) & $f_{K^+}$(MeV) & $f_{K^+}/f_{\pi^+}$ \\ \hline 
ETM 14~\cite{Carrasco:2014poa}$^\S$	&	2+1+1 & -- & 154.4(1.5)(1.3) & 1.184(12)(11)	\\
Fermilab/MILC 14~\cite{Bazavov:2014wgs}$^\S$	&	2+1+1 & -- & 155.92(13)($^{+42}_{-34}$) & 1.1956(10)($^{+26}_{-18}$)	\\
HPQCD 13~\cite{Dowdall:2013rya}$^\S$ & 2+1+1 & -- &155.37(20)(28) & 1.1916(15)(16) \\ \hline
FLAG 15 average~\cite{JuettnerLat15,FLAGInPrep}$^{\mathparagraph}$ & 2+1+1 & -- & 155.6(0.4) & 1.193(3) \\ \hline
RBC/UKQCD 14~\cite{Blum:2014tka}$^{\ast,\dag}$ & 2+1 & 130.19(89) & 155.51(83) & 1.1945(45) \\
RBC/UKQCD 12~\cite{Arthur:2012yc}$^{\ast}$ & 2+1 & 127(3)(3) & 152(3)(2) & 1.199(12)(14) \\
Laiho \&  Van de Water 11~\cite{Laiho:2011np}$^\ddag$ & 2+1 & $130.53(87)(210)$ & $156.8(1.0)(1.7)$ &$1.202(11)(9)(2)(5)$\\
MILC 10~\cite{Bazavov:2010hj} & 2+1 &129.2(0.4)(1.4) & 156.1(4)($^{+6}_{-9}$) & 1.197(2)($^{+3}_{-7}$) \\
BMW 10~\cite{Durr:2010hr}$^\ast$ & 2+1 & -- & -- & 1.192(7)(6)	\\
HPQCD/UKQCD 07~\cite{Follana:2007uv}$^\ast$ & 2+1 & 132(2)  &   157(2) & 1.189(2)(7) \\ \hline
FLAG 15 average~\cite{JuettnerLat15,FLAGInPrep}$^{\mathparagraph}$ & 2+1 & 130.2(1.4) &  155.9(0.9)  &  1.192(5)  \\ \hline
Our average & Both & 130.2(1.7) &  155.6(0.4)  &  1.1928(26) \\
\hline\hline
\end{tabular}
\\\vspace{5mm}
\begin{minipage}{\linewidth}
\begin{itemize}
{\footnotesize 
\item[$^{\S}$] PDG 2014 value of $f_{\pi^+} = 130.41(21)$~MeV used to set absolute lattice scale. \\\vspace{-2mm}
\item[$^{\mathparagraph}$] Preliminary numbers shown here may change if further new lattice-QCD calculations are published before the deadline for inclusion in the final 2015 FLAG review. \\\vspace{-2mm}
\item[$^{\dag}$] Preprint submitted to Phys.\ Rev.\ D.  Published RBC/UKQCD~12 results included in $N_f = 2+1$ average. \\\vspace{-2mm}
\item[$^{\ddag}$] Lattice 2011 conference proceedings. \\
}
\end{itemize}
\end{minipage}
\end{center}
\end{table}

Table~\ref{tab:fkfpi} also shows the 2015 preliminary three- and four-flavor averages for the pion and kaon decay constants and their ratio from the Flavour Lattice Averaging Group (FLAG)~\cite{JuettnerLat15,FLAGInPrep} in the lines labeled ``FLAG 15 average."  These preliminary updates of the 2013 FLAG averages~\cite{Aoki:2013ldr} include only those results from Table~\ref{tab:fkfpi} that are published in refereed journals, or that are straightforward conference updates of published analyses.
In the (2+1+1)-flavor averages, the statistical errors of HPQCD and Fermilab/MILC were conservatively treated as 100\% correlated because the calculations employed some of the same gauge-field configurations.  The errors have also been increased by the $\sqrt{\chi^2/{\rm dof}}$ to reflect a slight tension between the results.    
  There are no four-flavor lattice-QCD results for the pion decay constant in Table~\ref{tab:fkfpi} because all of the calculations listed use the quantity $f_{\pi^+}$ to fix the absolute lattice scale needed to convert from lattice-spacing units to GeV~\cite{Dowdall:2013rya,Bazavov:2014wgs,Carrasco:2014poa}.

All of the results in Table~\ref{tab:fkfpi} were obtained using isospin-symmetric gauge-field configurations, {\it i.e.}, the dynamical up and down quarks have the same mass.  Most calculations of pion and kaon decay constants now include the dominant effect of nondegenerate up- and down-quark masses by evaluating the masses of the constituent light (valence) quarks in the pion at the physical up- and down-quark masses, respectively, and evaluating the mass of the valence light quark in the kaon at the physical $m_u$.  Those results obtained with degenerate up and down valence quarks are corrected for isospin breaking using chiral perturbation theory ($\chi$PT) before being averaged.  The isospin-breaking corrections at next-to-leading order in $\chi$PT can be parameterized as~\cite{Gasser:1984gg,Cirigliano:2011tm}
\begin{eqnarray}
	f_\pi &=& f_{\pi^+} \,\label{eq:IsospinConvertPi} \\
	f_K &=& f_{K^+} \left( 1 - \delta_{\rm SU(2)}/2 \right) \, \\
	\frac{f_K}{f_\pi} & = & \frac{1}{\sqrt{\delta_{\rm SU(2)}+1}} \frac{f_{K^+}}{f_{\pi^+}} \label{eq:IsospinConvertPiKRatio}
\end{eqnarray}
where the expression for $\delta_{\rm SU(2)}$ in terms of the quark masses, meson masses, and decay constants, is given in Eq.~(37) of Ref.~\cite{Aoki:2013ldr}.  Numerically, values of $\delta_{\rm SU(2)} \approx -0.004$ were employed by FLAG to obtain the (2+1)-flavor averages in Table~\ref{tab:fkfpi}, but some direct lattice-QCD calculations of $\delta_{\rm SU(2)}$ give larger values~\cite{deDivitiis:2011eh,Dowdall:2013rya,Carrasco:2014poa} and further studies are needed.

To obtain the best decay-constant values for comparison with experimental rate measurements and other phenomenological applications, we combine the available $(2+1)$- and $(2+1+1)$-flavor lattice-QCD results, first accounting for the omission of charm sea quarks in the three-flavor simulations.
The error introduced by omitting charm sea quarks can be roughly estimated by expanding the charm-quark determinant in powers of $1/m_c$~\cite{Nobes:2005yh}; the resulting leading contribution is of order $\alpha_s \left( \Lambda_{\rm QCD} / 2 m_c\right)^2$~\cite{Bazavov:2016nty}.  Taking the $\overline{\rm MS}$ values $\overline m_c(\overline m_c) = 1.275$~GeV, $\overline\Lambda_{\rm QCD} \sim 340$~MeV from FLAG~\cite{Aoki:2013ldr}, and $\overline\alpha(\overline m _c) \sim 0.4$, leads to an estimate of about 0.7\% for the contribution to the decay constants from charm sea quarks.  The charm sea-quark contribution to ratios of decay constants is expected to be further suppressed by the SU(3)-breaking factor $(m_s - m_d) / \Lambda_{\rm QCD}$, and hence about 0.2\%.

We can compare these power-counting estimates of charm sea-quark contributions to the observed differences between the (2+1)- and (2+1+1)-flavor lattice-QCD averages for kaon, $D_{(s)}$-meson, and $B_{(s)}$-decay constants and ratios in Tables~\ref{tab:fkfpi}, \ref{tab:LatfD}, and~\ref{tab:LatfB}.   Of these, the kaon decay constants have been calculated most precisely, and the and three- and four-flavor averages for $f_{K^+}$ and $f_{K^+}/f_{\pi^+}$ agree within sub-percent errors.
Within present uncertainties, however, effects of this size in pseudoscalar-meson decay constants cannot be ruled out.   Therefore, to be conservative, in this review we add in quadrature additional systematic errors of $0.7\%$ and 0.2\% to all (2+1)-flavor decay-constant and decay-constant-ratio averages, respectively, to account for the omission of charm sea quarks.  Numerically, this increases the errors by at most about 50\% for $f_{K^+}$ and less for all other decay constants and ratios, indicating that the published (2+1)-flavor lattice-QCD results and uncertainties are reliable.

Our final preferred theoretical values for the charged pion and kaon decay constants are
\begin{eqnarray}
	{\rm Our~averages}: \quad && f_{\pi^+} = 130.2(1.7)~{\rm MeV}  \,, \;\; f_{K^+} = 155.6(0.4)~{\rm MeV} \,, \;\; \frac{f_{K^+}}{f_{\pi^+}}  = 1.1928(26) \,, \qquad \label{eq:PDG_fPi_fK}
\end{eqnarray}
where $f_{\pi^+}$ is simply the (2+1)-flavor FLAG average with the error increased by the estimated 0.7\% charm sea-quark contribution.  
For $f_{K^+}$ and $f_{K^+}/f_{\pi^+}$, we take a simple weighted average of the (2+1)- and (2+1+1)-flavor FLAG values, because they are each obtained from a sufficient number of independent calculations that we do not expect there to be significant correlations.
In practice, the addition of the charm sea-quark error has a tiny impact on our final values in Eq.~(\ref{eq:PDG_fPi_fK}), increasing the uncertainty on $f_{\pi^+}$ by 0.3~MeV, and the central value for $f_{K^+}/f_{\pi^+}$ by one in the last digit.

\section{Charmed mesons} \label{sec:CharmAll}

\subsection{Experimental rate measurements} \label{sec:CharmExp}

Measurements
have been made for $D^+\to\mu^+\nu$,  $D_s^+\to\mu^+\nu$, and $D_s^+\to\tau^+\nu$. Only an upper limit has been determined for $D^+\to\tau^+
\nu$. Both CLEO-c and BES have made measurements of  $D^+$ decay using $e^+e^-$
collisions at the $\psi(3770)$ resonant energy where $D^-D^+$ pairs are
copiously produced. They fully reconstruct one of the $D$'s, say the $D^-$.
Counting the number of these events provides the normalization for the
branching fraction measurement. They then find a candidate $\mu^+$, and then
form the missing-mass squared, $MM^2=\left(E_{\rm CM}-E_{D^-}\right)^2-
\left(\overrightarrow{p}_{\!\rm CM}-\overrightarrow{p}_{\!D^-}-
\overrightarrow{p}_{\!\mu^+}\right)^2$, taking into account their knowledge of
the center-of-mass energy, $E_{\rm CM}$,  and momentum, $p_{\rm CM}$, that
equals zero in $e^+e^-$ collisions.  A peak at zero $MM^2$ inplies the
existence of a missing neutrino and hence the $\mu^+\nu$ decay of the $D^+$.
CLEO-c does not explicitly identify the muon, so their data consists of a
combination of $\mu^+\nu$ and $\tau^+\nu$, $\tau^+\to\pi^+\nu$ events.  This
permits them to do two fits: in one they fit for the individual components, and
in the other they fix the ratio of $\tau^+\nu/\mu^+\nu$ events to be that given
by the standard-model expectation. Thus, the latter measurement should be used for standard-model
comparisons and the other for new-physics searches.  Our average uses the fixed
ratio value.  The measurements are shown in Table~\ref{tab:fDp}.

\begin{table}[b]
\caption{Experimental results for ${\cal{B}}(D^+\to \mu^+\nu)$, ${\cal{B}}
(D^+\to \tau^+\nu)$, and {$|V_{cd}|f_{D^+}$}. 
Numbers for {$|V_{cd}| f_{D^+}$} have been
extracted using updated values for masses (see text).
Radiative corrections are included. Systematic uncertainties arising from the $D^+$
lifetime and mass are included. For the average $\mu^+\nu$ number we use the CLEO-c result for $\mu^+\nu^++\tau^+\nu$.
\label{tab:fDp}}
\begin{center}
\begin{tabular}{llccc}
 \hline\hline
& Experiment & Mode & ${\cal{B}}$ & {$|V_{cd}|f_{D^+}$ (MeV)} \\ \hline
& CLEO-c \cite{Artuso:2005ym,Eisenstein:2008aa}& $\mu^+\nu$& $(3.93\pm
0.35\pm 0.09)\times 10^{-4}$ & $47.07 \pm 2.10\pm 0.57$\\
& CLEO-c \cite{Artuso:2005ym,Eisenstein:2008aa}& $\mu^+\nu$+$\tau^+\nu$ & $(3.82\pm
0.32\pm 0.09)\times 10^{-4}$ & $46.41 \pm 1.94\pm 0.57$\\ 
& BES \cite{Ablikim:2013uvu}& $\mu^+\nu$& $(3.71\pm
0.19\pm 0.06)\times 10^{-4}$ & $45.73\pm 1.17\pm 0.38$\\
\hline
& Our average
& Lines 2+3 & $(3.74\pm 0.17)\times 10^{-4}$ & $45.91\pm 1.05$ \\
\hline
& CLEO-c \cite{Artuso:2007zg,Alexander:2009ux} & $\tau^+\nu$ & $<1.2\times 10^{-3}$ &
\\ \hline\hline
\end{tabular}
\end{center}
\end{table}

To extract the value of $|V_{cd}| f_{D^+}$ we use the well-measured $D^+$ lifetime of 
1.040(7) ps. The $\mu^+\nu$
results include a 1\% correction (lowering) of the rate due to the presence of
the radiative $\mu^+\nu\gamma$ final state based on the estimate by
Dobrescu and Kronfeld \cite{Dobrescu:2008er}. 

We now discuss the
$D_s^+$. Measurements of the leptonic decay rate have been made by several groups and are
listed in Table~\ref{tab:fDs} \cite{Artuso:2007zg,Alexander:2009ux,Zupanc:2013byn,Naik:2009tk,Ecklund:2007aa,Onyisi:2009th,%
delAmoSanchez:2010jg}.  We exclude older values obtained by normalizing to $D_s^+$ decay
modes that are not well defined. Many measurements, for example, used the
$\phi\pi^+$ mode. This decay is a subset of the $D_s^+\to K^+ K^- \pi^+$
channel which has interferences from other modes populating the $K^+K^-$ mass
region near the $\phi$, the most prominent of which is the $f_0(980)$. Thus
the extraction of the effective $\phi\pi^+$ rate is sensitive to the mass
resolution of the experiment and the cuts used to define the $\phi$ mass region
\cite{Alexander:2008aa}.\footnote{We have not included the BaBar result for ${\cal B}(D_s^+ \to \mu^+ \nu)$ reported in Ref.~\cite{Aubert:2006sd} because this measurement determined the ratio of the leptonic decay rate to the hadronic decay rate $\Gamma(D_s^+\to\ell^+\nu)/\Gamma(D_s^+\to\phi\pi^+)$.}

\begin{table}[t]
\caption{Experimental results for ${\cal{B}}(D_s^+\to \mu^+\nu)$, ${\cal{B}}
(D_s^+\to \tau^+\nu)$, and {$|V_{cs}| f_{D_s^+}$}. Numbers for {$|V_{cs}| f_{D_s^+}$} have been
extracted using updated values for masses (see text). The systematic
uncertainty for correlated error on the $D_s^+$ lifetime is included. The mass uncertainties are also common, but negligible. Common systematic errors in each
experiment have been taken into account in the averages. 
\label{tab:fDs}}
\begin{center}
\begin{tabular}{lllcc}
 \hline\hline
& Experiment & Mode & ${\cal{B}}$(\%) &{$|V_{cs}| f_{D_s^+}$} (MeV) \\ \hline
& CLEO-c \cite{Artuso:2007zg,Alexander:2009ux}& $\mu^+\nu$& $(0.565\pm
0.045\pm 0.017)$ & $250.8\pm 10.0\pm 4.2$\\
& BaBar\footnote{We do not use a previous unpublished BaBar result from a subsample of data that uses a different technique for obtaining the branching fraction normalization \cite{Lees:2010qj}.} \cite{delAmoSanchez:2010jg}& $\mu^+\nu$& $(0.602\pm
0.038\pm 0.034)$ & $258.9\pm ~8.2\pm 7.5$\\
&Belle \cite{Zupanc:2013byn}
& $\mu^+\nu$ & $(0.531\pm 0.028\pm 0.020) $ & $243.1
\pm ~6.4 \pm 4.9$ \\
\hline
& Our average & $\mu^+\nu$ & $(0.556\pm 0.024)$ & $248.8\pm 5.8$ \\
\hline
& CLEO-c \cite{Artuso:2007zg,Alexander:2009ux} & $\tau^+\nu~(\pi^+\overline{\nu})$ & $(6.42\pm 0.81\pm
0.18) $ & $270.8\pm 17.1 \pm 4.2 $ \\
& CLEO-c \cite{Naik:2009tk} & $\tau^+\nu~(\rho^+\overline{\nu})$ & $(5.52\pm 0.57
\pm 0.21)$ & $251.1\pm 13.0 \pm 5.1 $ \\
& CLEO-c \cite{Ecklund:2007aa,Onyisi:2009th} & $\tau^+\nu~(e^+\nu\overline{\nu})$ &
$(5.30\pm 0.47\pm 0.22)$ & $246.1\pm 10.9 \pm 5.4 $ \\
& BaBar \cite{delAmoSanchez:2010jg} & $\tau^+\nu~(e^+(\mu^+)\nu\overline{\nu})$ &
$(5.00\pm 0.35\pm 0.49)$ & $239.0\pm 8.4 \pm 11.9 $ \\
&Belle \cite{Zupanc:2013byn}
& $\tau^+\nu~(\pi^+\overline{\nu})$  & $(6.04\pm 0.43^{+0.46}_{-0.40})
$ & $262.7\pm 9.3^{+10.2}_{-8.9} $ \\
&Belle \cite{Zupanc:2013byn}
& $\tau^+\nu~(e^+\nu\overline{\nu})$  & $(5.37\pm 0.33^{+0.35}_{-0.31})
 $ & $247.7\pm 7.6^{+8.3}_{-7.4}~ $ \\
&Belle \cite{Zupanc:2013byn}
& $\tau^+\nu~(\mu^+\nu\overline{\nu})$  & $(5.86\pm 0.37^{+0.34}_{-0.59})
$ & $258.7\pm 8.2^{+7.7}_{-13.2}  $ \\ \hline
& Our average & $\tau^+\nu$ & $(5.56\pm 0.22)$ & $252.1\pm 5.2$\\\hline
& Our average & $\mu^+\nu + \tau^+\nu$&  & $250.9\pm 4.0$\\ \hline\hline
\end{tabular}
\end{center}
\end{table}

To find decays in the $\mu^+ \nu$  signal channels, CLEO, BaBar and Belle rely
on fully reconstructing all the final state particles except for neutrinos and
using a missing-mass technique to infer the existence of the neutrino. CLEO
uses $e^+e^-\to D_sD_s^*$ collisions at 4170 MeV, while Babar and Belle use
$e^+e^-\to D Kn\pi D_s^*$ collisions at energies near the $\Upsilon(4S)$.
CLEO does a similar analysis as was done for the $D^+$ above.  Babar and Belle
do a similar $MM^2$ calculation by using the reconstructed hadrons, the photon
from the $D_s^{*+}$ decay and a detected $\mu^+$. To get the normalization they
do a $MM^2$ fit without the $\mu^+$ and use the signal at the $D_s^+$ mass
squared to determine the total $D_s^+$ yield. 

When selecting the $\tau^+\to\pi^+\bar{\nu}$ and $\tau^+\to\rho^+\bar{\nu}$
decay modes, CLEO uses both the calculation of the missing-mass and the fact that
there should be no extra energy in the event beyond that deposited by the
measured tagged $D_s^-$ and the $\tau^+$ decay products. The $\tau^+\to
e^+\nu\bar{\nu}$ mode, however, uses only extra energy.  Babar and Belle also
use the extra energy to discriminate signal from background in their
$\tau^+\nu$ measurements.

We extract the decay constant times the CKM factor from the measured branching ratios using the
$D_s^+$ mass of 1.96830(11) GeV,  the $\tau^+$ mass of 1.77682(16) GeV, and a
lifetime of 0.500(7) ps \cite{Agashe:2014kda}.  CLEO has
included the radiative correction of 1\% in the $\mu^+\nu$ rate listed in the
Table \cite{Dobrescu:2008er}~(the $\tau^+\nu$ rates need not be corrected). Other
theoretical calculations show that the $\gamma\mu^+\nu$ rate is a factor of
40--100 below the $\mu^+\nu$ rate for charm \cite{Burdman:1994ip,Atwood:1994za,Colangelo:1995sm,Khodjamirian:1995uc,Eilam:1995zv,Geng:1997ws,Geng:2000if,Korchemsky:1999qb,Hwang:2005uk,Lu:2002mn}. As this is a
small effect we do not attempt to correct the other measurements. The values for $f_{D_s^+}|V_{cs}|$ are in good agreement for the two decay modes. Our average value including both the $\mu^+\nu$ and $\tau^+\nu$ final states is $250.9\pm 4.0$~MeV.

\subsection{Theoretical decay-constant calculations} \label{sec:CharmTheory}
Table~\ref{tab:LatfD} presents recent theoretical calculations of the charged $D^+$- and $D_s$-meson decay constants and their ratio.  The upper two panels show results from lattice-QCD simulations with three ($N_f = 2+1$) or four flavors ($N_f = 2+1+1$) of dynamical quarks.  Although there are fewer available results than for the pion and kaon sector, both $f_{D^+}$ and $f_{D_s}$ have been obtained using multiple sets of gauge-field configurations with different lattice fermion actions, providing independent confirmation.  For comparison, the bottom panel of Table~\ref{tab:LatfD} shows non-lattice determinations from QCD sum rules and the light-front quark model; only results which include uncertainty estimates are shown.  The lattice and non-lattice results agree, but the uncertainties on $D^+_{(s)}$-meson decay constants from lattice QCD have now reached significantly greater precision than those from other approaches.

\begin{table}[tb]
\caption{Recent theoretical determinations of $f_{D^+}$, $f_{D_s}$, and their ratio. The upper panels show results from lattice-QCD simulations with $(2+1+1)$ and $(2+1)$ dynamical quark flavors, respectively. Statistical and systematic errors are quoted separately.  Lattice-QCD results for $f_D$ and $f_{D_s}/f_D$ in the isospin-symmetric limit $m_u = m_d$ are noted with an ``$^\ast$".  
The bottom panel shows estimates from QCD sum rules (QCD SR) and the light-front quark model (LFQM).  These are not used to obtain our preferred decay-constant values. \label{tab:LatfD}}
\begin{center}
\begin{tabular}{l@{\quad}c@{\quad}c@{\quad}c@{\quad}c@{\quad}c} \hline\hline
Reference & Method & $N_f$ & $f_{D^+}$(MeV) & $f_{D_s}$(MeV) & $f_{D_s}/f_{D^+}$ \\ \hline 
ETM 14~\cite{Carrasco:2014poa}$^\ast$ & LQCD & 2+1+1 & 207.4(3.7)(0.9) & 247.2(3.9)(1.4) & 1.192(19)(11)	\\
Fermilab/MILC 14~\cite{Bazavov:2014wgs} & LQCD 	& 2+1+1 & 212.6(0.4)($^{+1.0}_{-1.2}$) & 249.0(0.3)($^{+1.1}_{-1.5}$) & 1.1712(10)($^{+29}_{-32}$)	\\ \hline
Average & LQCD 	& 2+1+1 &  212.2(1.5) &  248.8(1.3) &  1.172(3)	\\ \hline
$\chi$QCD 14~\cite{Yang:2014sea}$^{\ast}$ & LQCD & 2+1 & -- & 254(2)(4) & -- \\
HPQCD 12~\cite{Na:2012iu}$^\ast$ & LQCD & 2+1 & 208.3(1.0)(3.3) & -- & 1.187(4)(12) \\
Fermilab/MILC 11~\cite{Bazavov:2011aa} & LQCD & 2+1 & 218.9(9.2)(6.6) & 260.1(8.9)(6.1) &   1.188(14)(21) \\
HPQCD 10~\cite{Davies:2010ip}$^\ast$ & LQCD & 2+1 & -- & 248.0(1.4)(2.1) & -- \\ \hline
Average & LQCD & 2+1 & 209.2(3.3) & 249.8(2.3) & 1.187(12) \\ \hline
Our average & LQCD & Both & 211.9(1.1) & 249.0(1.2) & 1.173(3) \\ \hline\hline
Wang 15~\cite{Wang:2015mxa}$^{\S}$ & QCD SR & & 208(10) & 240(10) & 1.15(6) \\ 
Gelhausen 13~\cite{Gelhausen:2013wia} & QCD SR & & 201$\left(^{+12}_{-13}\right)$ & 238$\left(^{+13}_{-23}\right)$  &   1.15$\left(^{+0.04}_{-0.05}\right)$ \\
Narison 12~\cite{Narison:2012xy} &QCD SR & & 204(6)  & 246(6) & 1.21(4) \\
Lucha 11~\cite{Lucha:2011zp} & QCD SR & & 206.2(8.9) & 245.3(16.3) & 1.193(26) \\ 
Hwang 09~\cite{Hwang:2009qz} & LFQM & & -- & 264.5(17.5)$^{\mathparagraph}$ & 1.29(7) \\
\hline\hline
\end{tabular}
\\\vspace{5mm}
\begin{minipage}{\linewidth}
\begin{itemize}
{\footnotesize 
\item[$^{\S}$] Obtained using $m_c^{\overline{{\rm MS}}}$; results using $m_c^{\rm pole}$ are also given in the paper. \\\vspace{-2mm}
\item[$^{\mathparagraph}$] Obtained by combining PDG value $f_D = 205.8(8.9)$~MeV~\cite{Amsler:Update} with $f_{D_s}/f_D$ from this work. \\
}
\end{itemize}
\end{minipage}
\end{center}
\end{table}

The lattice-QCD results in Table~\ref{tab:LatfD} were all obtained using isospin-symmetric gauge-field configurations.  The two calculations by the Fermilab Lattice and MILC Collaborations~\cite{Bazavov:2011aa,Bazavov:2014wgs}, however, include the dominant strong isospin-breaking contribution by evaluating the mass of the valence light quark in the $D^+$-meson decay constant at the physical down-quark mass.  Reference~\cite{Bazavov:2014wgs} provides a determination of the size of this correction, 
\begin{equation}
	f_{D^+} - f_D = 0.47(1)(^{+25}_{-6})~{\rm MeV} \,, \label{eq:SU2_fD}
\end{equation}
where $f_D$ is the value of the $D$-meson decay constant evaluated at the average up-down quark mass.  Equation~(\ref{eq:SU2_fD}) implies that the correction to the SU(3)$_f$-breaking ratio is
 \begin{equation}
 	\frac{f_{D_s}}{f_{D^+}} - \frac{f_{D_s}}{f_D} = -0.0026 \,, \label{eq:SU2_fDsfD}
 \end{equation}
taking the central values for $f_{D^+}$ and $f_{D_s}$ from the same work. Because the errors on the calculations listed in Table~\ref{tab:LatfD} that neglect isospin breaking are still about 5--8 $\times$ larger than the sizes of the shifts in Eqs.~(\ref{eq:SU2_fD})--(\ref{eq:SU2_fDsfD}), we do not correct any results {\it a posteriori} for this effect in the current review.  Nevertheless, we strongly encourage future lattice-QCD publications to present results for both the ${D^+}$- and ${D^0}$-meson decay constants.  
Including the effect of isospin breaking will be essential once lattice-QCD calculations of $f_D$ and $f_{D_s}/f_D$ reach the level of precision in Eqs.~(\ref{eq:SU2_fD})--(\ref{eq:SU2_fDsfD}).

We average the lattice-QCD results in Table~\ref{tab:LatfD} accounting for possible correlations between them following the approach established by Laiho \etal\ \cite{Laiho:2009eu}.
Whenever we have reason to believe that a source of uncertainty is correlated between two results, we conservatively take the correlation to be 100\% when calculating the average.   We then construct the correlation matrix for the set of lattice-QCD results using the prescription of Schmelling~\cite{Schmelling:1994pz}.

We first separately average the three- and four-flavor results for the charged $D_{(s)}^+$-meson decay constants and their ratio.  There have been no new three-flavor lattice-QCD calculations of $f_{D^+}$ or $f_{D_{s}^+}/f_{D^+}$ since 2013, so we take the (2+1)-flavor averages from FLAG~\cite{Aoki:2013ldr}.  In this average, the statistical errors were treated as 100\% correlated between the results of Fermilab/MILC~\cite{Bazavov:2011aa} and HPQCD~\cite{Na:2012iu} because the calculations employed some of the same ensembles of gauge-field configurations.   
For $f_{D_s}$, we average the (2+1)-flavor results given in Table~\ref{tab:LatfD}, again treating the Fermilab/MILC~\cite{Bazavov:2011aa} and HPQCD~\cite{Davies:2010ip} statistical errors as correlated, and taking the $\chi$QCD result~\cite{Yang:2014sea} to be independent.
For the $(2+1+1)$-flavor $D_{(s)}$-meson decay constants, we take a simple weighted average of the ETM~\cite{Carrasco:2014poa} and Fermilab/MILC 14 results~\cite{Bazavov:2014wgs} in Table~\ref{tab:LatfD}.
We expect them to be independent because the calculations use different light-quark and gluon actions and different treatments of the chiral-continuum extrapolation.  Our separate three- and four-flavor averages are listed in the lines labeled ``Average"  in Table~\ref{tab:LatfD}, where the errors on the (2+1)-flavor $f_{D_s}$ and (2+1+1)-flavor $f_D$ averages have been rescaled by the factors $\sqrt{(\chi^2/{\rm dof})}= 1.1$ and $\sqrt{(\chi^2/{\rm dof})}= 1.3$, respectively.\footnote{After this article was submitted for review, preliminary (2+1)- and (2+1+1)-flavor FLAG averages for $f_{D}$, $f_{D_s}$, and $f_{D_{s}}/f_{D}$ were presented in Ref.~\cite{Vladikas:2015bra} that are identical to our separate averages in Table~\ref{tab:LatfD}.}   

To obtain the single-best values of the $D_{(s)}^+$-meson decay constants for phenomenology applications, we combine the available $(2+1)$- and $(2+1+1)$-flavor lattice-QCD results, which are compatible within the current level of precision.  We account for the omission of charm sea-quark contributions in the three-flavor calculations by adding to the errors on the (2+1)-flavor averages in Table~\ref{tab:LatfD} our power-counting estimates of charm sea-quark errors from Sec.~\ref{sec:PiKTheory}.
Because the estimated charm sea-quark errors of 0.7\% for decay constants and 0.2\% for decay-constant ratios are less than those on the (2+1)-flavor averages, adding them in quadrature has a small impact on the total uncertainties.  The error increase is at most about 25\% for $f_{D_s}$, and below 10\% for both $f_{D^+}$ and $f_{D_s}/f_{D^+}$.
Our final preferred theoretical values for the charged $D^+_{(s)}$-meson decay constants are given by the weighted average of the entries in the two lines labeled ``Average" in Table~\ref{tab:LatfD}, after including the additional charm sea-quark errors in the (2+1)-flavor entries:
\begin{eqnarray}
	{\rm Our~averages}: \quad && f_{D^+} = 211.9(1.1)~{\rm MeV} \,, \;\; f_{D_s} = 249.0(1.2)~{\rm MeV} \,, \;\; \frac{f_{D_s}}{f_{D^+}} = 1.173(3) \,. \qquad \label{eq:fD_PDG}
\end{eqnarray}
In practice, the errors on the (2+1+1)-flavor averages are so much smaller than on the (2+1)-flavor averages that the combination in Eq.~(\ref{eq:fD_PDG}) is almost identical to the (2+1+1)-flavor average in Table~\ref{tab:LatfD}.  The most precise result from Fermilab/MILC, in particular, has a large weight in the average.

\section{Bottom mesons} \label{sec:BottomAll}

\subsection{Experimental rate measurements} \label{sec:BottomExp}

The Belle and BaBar collaborations have found evidence for $B^-\to\tau^-
\overline{\nu}$ decay in $e^+e^-\to B^-B^+$ collisions at the $\Upsilon(4S)$
energy.  The analysis relies on reconstructing a hadronic or semi-leptonic $B$
decay tag, finding a $\tau$ candidate in the remaining track and photon
candidates, and examining the extra energy in the event which should be close
to zero for a real $\tau^-$ decay to $e^- \nu \bar \nu$ or $\mu^- \nu \bar \nu$
opposite a $B^+$ tag. While the BaBar results have remained unchanged, Belle
reanalyzed both samples of their data.  The
branching fraction using hadronic tags changed from $1.79\,^{+0.56\,+0.46}_{-0.49\,-0.51}\times
10^{-4}$ \cite{Ikado:2006un} to $0.72^{+0.27}_{-0.25}\pm0.11\times10^{-4}$
\cite{Adachi:2012mm}, while the corresponding change using semileponic tags was from $1.54^{+0.38+0.29}_{-0.37-0.31}$  to $1.25\pm0.28\pm0.27$. These changes demonstrate the difficulty of the analysis. The results are
listed in Table~\ref{tab:Btotaunu}.

\begin{table}[htb]
\caption{Experimental results for ${\cal{B}}(B^-\to \tau^-\overline{\nu})$ {and $|V_{ub}|f_{B^+}$}. }
\label{tab:Btotaunu}
\begin{center}
\begin{tabular}{lllcc} \hline\hline
&Experiment & Tag &${\cal{B}}$ (units of $10^{-4}$) & {$|V_{ub}|f_{B^+}$ (MeV)} \hfill\\
\hline\\[-2.5ex] 
&Belle~\cite{Adachi:2012mm}&Hadronic&$0.72^{+0.27}_{-0.25}\pm0.11$\\
&Belle~\cite{Kronenbitter:2015kls}&Semileptonic&$1.25\pm0.28\pm0.27$\\
&Belle~\cite{Kronenbitter:2015kls}&Average&$0.91 \pm 0.22$ &$0.72\pm 0.09$\\\hline\\[-2.5ex] 
&BaBar~\cite{Lees:2012ju} & Hadronic & $1.83\,^{+0.53}_{-0.49}\pm0.24$\\
&BaBar~\cite{Aubert:2009wt} & Semileptonic & $1.7\pm 0.8\pm 0.2$\\
&BaBar~\cite{Lees:2012ju}  & Average & $1.79 \pm 0.48$& $1.01\pm 0.14$\\\hline
& Our average  & &$1.06\pm0.20$&$0.77\pm0.07$\\
\hline\hline
\end{tabular}
\end{center}
\end{table}

There are large backgrounds under the signals in all cases. The systematic
errors are also quite large. Thus, the significances are
not that large.  Belle quotes 4.6$\sigma$ for their combined hadronic
and semileptonic tags, while BaBar quotes 3.3$\sigma$ and 2.3
$\sigma$,  for hadronic and semileptonic tags.  Greater precision is necessary to determine if any effects beyond the Standard Model are present.

To extract the value of $|V_{ub}| f_{B^+}$ we use the PDG 2014 value of the $B^+$ lifetime of 1.638$\pm$0.004~ps, and the $\tau^+$ and $B^+$ masses of 1.77684 and 5.27926~GeV, respectively.

\subsection{Theoretical decay-constant calculations} \label{sec:BottomTheory}
Tables~\ref{tab:LatfB} and~\ref{tab:LatfB0} present theoretical calculations of the $B^+$-, $B^0$-, and $B_s$-meson decay constants and their ratios.  (The decay constants of the neutral $B^0$ and $B_s$ mesons enter the rates for the rare leptonic decays $B_{d,s} \to \mu^+\mu^-$.)   The upper two panels show results from lattice-QCD simulations with three ($N_f = 2+1$) or four flavors ($N_f = 2+1+1$) of dynamical quarks.  For all decay constants, calculations using different gauge-field configurations, light-quark actions, and $b$-quark actions provide independent confirmation.  For comparison, the bottom panel of Table~\ref{tab:LatfB} shows non-lattice determinations of the $B_{(s)}$-meson decay constants which include error estimates.  These are consistent with the lattice values, but with much larger uncertainties.

\begin{table}[tb]
\caption{Recent theoretical determinations of $f_{B^+}$, $f_{B_s}$, and their ratio. The upper panels show results from lattice-QCD simulations with $(2+1+1)$ and $(2+1)$ dynamical quark flavors, respectively. For some of the lattice-QCD results, statistical and systematic errors are quoted separately.  
Lattice-QCD results for $f_B$ and $f_{B_s}/f_B$ in the isospin-symmetric limit $m_u = m_d$ are noted with an ``$^\ast$"; they are corrected by the factors in Eqs.~(\ref{eq:SU2_fB}) and~(\ref{eq:SU2_fBsfB}), respectively, before computing the averages.
Preliminary conference results noted with a ``$^\dag\!$" are not included in the averages. The bottom panel shows estimates from QCD sum rules and the light-front quark model, which are not used to obtain our preferred decay-constant values. \label{tab:LatfB}}
\begin{center}
\begin{tabular}{l@{\quad}c@{\quad}c@{\quad}c@{\quad}c@{\quad}c} \hline\hline
Reference & Method & $N_f$ & $f_{B^+}$(MeV) & $f_{B_s}$(MeV) & $f_{B_s}/f_{B^+}$ \\ \hline 
ETM 13~\cite{Carrasco:2013naa}$^{\ast,\dag}$	 & LQCD & 2+1+1	& 196(9) & 235(9) & 1.201(25)	\\ 
HPQCD 13~\cite{Dowdall:2013tga} & LQCD &	2+1+1 & 184(4) & 224(5) & 1.217(8)	\\ \hline
Average & LQCD &	2+1+1 & 184(4)  & 224(5) & 1.217(8) \\ \hline
Aoki 14~\cite{Aoki:2014nga}$^{\ast,\ddag}$	& LQCD  & 2+1 & 218.8(6.5)(30.8) & 263.5(4.8)(36.7) & 1.193(20)(44) \\
RBC/UKQCD 14~\cite{Christ:2014uea} & LQCD & 2+1 & 195.6(6.4)(13.3) & 235.4(5.2)(11.1) & 1.223(14)(70) \\
HPQCD 12~\cite{Na:2012kp}$^\ast$ & LQCD & 2+1 & 191(1)(8) & 228(3)(10) & 1.188(12)(13) \\
HPQCD 12~\cite{Na:2012kp}$^\ast$ & LQCD & 2+1 & 189(3)(3)$^\star$ & -- & -- \\
HPQCD 11~\cite{McNeile:2011ng} & LQCD & 2+1 & -- & 225(3)(3) & -- \\
Fermilab/MILC 11~\cite{Bazavov:2011aa} & LQCD & 2+1 & 196.9(5.5)(7.0) & 242.0(5.1)(8.0) & 1.229(13)(23) \\\hline
Average & LQCD & 2+1 & 189.9(4.2) & 228.6(3.8) & 1.210(15) \\\hline
Our average & LQCD & Both & 187.1(4.2) & 227.2(3.4)  &  1.215(7) \\\hline\hline
Wang 15~\cite{Wang:2015mxa}$^{\S}$ & QCD SR & & 194(15) & 231(16) & 1.19(10) \\ 
Baker 13~\cite{Baker:2013mwa} & QCD SR & & 186(14) & 222 (12) & 1.19(4)  \\ 
Lucha 13~\cite{Lucha:2013gta} & QCD SR & & 192.0(14.6) & 228.0(19.8) & 1.184(24) \\ 
Gelhausen 13~\cite{Gelhausen:2013wia} & QCD SR & & 207$\left(^{+17}_{-9}\right)$ & 242$\left(^{+17}_{-12}\right)$ &  1.17$\left(^{+3}_{-4}\right)$ \\  
Narison 12~\cite{Narison:2012xy} &QCD SR & & 206(7) & 234(5)  & 1.14(3)  \\
Hwang 09~\cite{Hwang:2009qz} & LFQM & & -- & 270.0(42.8)$^{\mathparagraph}$ & 1.32(8) \\  
\hline\hline
\end{tabular}
\\\vspace{5mm}
\begin{minipage}{\linewidth}
\begin{itemize}
{\footnotesize 
\item[$^{\dag}$] Lattice 2013 conference proceedings. \\\vspace{-2mm}
\item[$^{\ddag}$] Obtained with static $b$ quarks ({\it i.e.} $m_b \to \infty$). \\\vspace{-2mm}
\item[$^{\star}$] Obtained by combining $f_{B_s}$ from HPQCD 11 with $f_{B_s}/f_B$ from this work.  Approximate statistical (systematic) error obtained from quadrature sum of individual statistical (systematic) errors. \\\vspace{-2mm}
\item[$^{\S}$] Obtained using $m_b^{\overline{{\rm MS}}}$; results using $m_b^{\rm pole}$ are also given in the paper.\\\vspace{-2mm}
\item[$^{\mathparagraph}$] Obtained by combining PDG value $f_B = 204(31)$~MeV~\cite{Amsler:Update} with $f_{B_s}/f_B$ from this work.\\
}
\end{itemize}
\end{minipage}
\end{center}
\end{table}

\begin{table}[tb]
\caption{Recent lattice-QCD determinations of $f_{B^0}$ and $f_{B_s}/f_{B^0}$. 
Results obtained in the isospin-symmetric limit $m_u = m_d$ are noted with an ``$^\ast$", while those for the $B^+$-meson are noted with an ``$^\S$".
Although the quoted results are identical those in Table~\ref{tab:LatfB}, they are corrected by different factors in Eqs.~(\ref{eq:SU2_fB})--(\ref{eq:SU2_fBsfB0}) before computing the averages.  Other labels and descriptions are the same as in Table~\ref{tab:LatfB}. \label{tab:LatfB0}}
\begin{center}
\begin{tabular}{l@{\quad}c@{\quad}c@{\quad}c@{\quad}c@{\quad}c} \hline\hline
Reference & Method & $N_f$ & $f_{B^0}$(MeV) & $f_{B_s}/f_{B^0}$ \\ \hline 
ETM 13~\cite{Carrasco:2013naa}$^{\ast,\dag}$	 & LQCD & 2+1+1	& 196(9) & 1.201(25)	\\ 
HPQCD 13~\cite{Dowdall:2013tga} & LQCD &	2+1+1 & 188(4) & 1.194(7)	\\ \hline
Average & LQCD &	2+1+1 & 188(4) & 1.194(7) \\ \hline
Aoki 14~\cite{Aoki:2014nga}$^{\ast,\ddag}$	& LQCD  & 2+1 & 218.8(6.5)(30.8) & 1.193(20)(44) \\
RBC/UKQCD 14~\cite{Christ:2014uea} & LQCD & 2+1 & 199.5(6.2)(12.6) & 1.197(13)(49) \\
HPQCD 12~\cite{Na:2012kp}$^\ast$ & LQCD & 2+1 & 191(1)(8) & 1.188(12)(13) \\
HPQCD 12~\cite{Na:2012kp}$^\ast$ & LQCD & 2+1 & 189(3)(3)$^\star$ & -- \\
Fermilab/MILC 11$^\S$~\cite{Bazavov:2011aa} & LQCD & 2+1 & 196.9(5.5)(7.0)  & 1.229(13)(23) \\\hline
Average & LQCD & 2+1 & 193.6(4.2) & 1.187(15)  \\\hline
Our average & LQCD & Both & 190.9(4.1) & 1.192(6) \\
\hline\hline
\end{tabular}
\\\vspace{5mm}
\begin{minipage}{\linewidth}
\begin{itemize}
{\footnotesize 
\item[$^{\dag}$] Lattice 2013 conference proceedings. \\\vspace{-2mm}
\item[$^{\ddag}$] Obtained with static $b$ quarks ({\it i.e.} $m_b \to \infty$). \\\vspace{-2mm}
\item[$^{\star}$] Obtained by combining $f_{B_s}$ from HPQCD 11 with $f_{B_s}/f_B$ from this work.  Approximate statistical (systematic) error obtained from quadrature sum of individual statistical (systematic) errors. \\
}
\end{itemize}
\end{minipage}
\end{center}
\end{table}

The lattice-QCD results in Tables~\ref{tab:LatfB} and~\ref{tab:LatfB0} were all obtained using isospin-symmetric gauge-field configurations.  The most recent calculations of $f_{B^+}$ by the HPQCD, Fermilab/MILC, and RBC/UKQCD Collaborations~\cite{Bazavov:2011aa,Dowdall:2013tga,Christ:2014uea}, however, include the dominant effect of nondegenerate up- and down-quark masses by evaluating the decay constant with the valence light-quark mass fixed to the physical up-quark mass.
HPQCD and RBC/UKQCD also calculate $f_{B^0}$ by fixing the valence light-quark mass equal to the physical down-quark mass~\cite{Dowdall:2013tga,Christ:2014uea}; they find differences between the $B^+$- and $B^0$-meson decay constants of $f_{B^0} - f_{B^+} \approx 4$~MeV and $f_{B_s}/f_{B^+} - f_{B_s}/f_{B^+} \approx 0.025$.   
Inspection of Tables~\ref{tab:LatfB} and~\ref{tab:LatfB0} shows that these differences are comparable to the error on the HPQCD~12 result for $f_B$~\cite{Na:2012kp}, and to the errors on the Fermilab/MILC, HPQCD~12, and ETM results for $f_{B_s}/f_B$~\cite{Na:2012kp,Carrasco:2013naa}, none of which account for isospin-breaking.
Therefore, to enable comparison with experimental measurements, in this review we correct those lattice-QCD results for $B$-meson decay constants obtained with degenerate up and down valence quarks {\it a posteriori} for isospin breaking before computing our averages.   For the correction factors, we use the differences obtained empirically by HPQCD in Ref.~\cite{Dowdall:2013tga}\footnote{The correlated uncertainties were provided by HPQCD via private communication.}
\begin{eqnarray}
	f_{B^+} - f_B &=& -1.9(5)~{\rm MeV}\,, \label{eq:SU2_fB} \\
	\frac{f_{B_s}}{f_{B^+}} - \frac{f_{B_s}}{f_B} &=& 0.012(4) \,, \label{eq:SU2_fBsfB} \\
	f_{B^0} - f_B &= & 1.7(5)~{\rm MeV}\,, \label{eq:SU2_fB0} \\
	\frac{f_{B_s}}{f_{B^0}} - \frac{f_{B_s}}{f_B} &=& -0.011(4) \,. \label{eq:SU2_fBsfB0}
\end{eqnarray}
%
The isospin-breaking correction factors  in Eqs.~(\ref{eq:SU2_fB})--(\ref{eq:SU2_fBsfB0}) are well determined because of cancellations between correlated errors in the differences.

We first average the published (2+1)-flavor lattice-QCD results for the charged and neutral $B_{(s)}$-meson decay constants and their ratios in Tables~\ref{tab:LatfB} and~\ref{tab:LatfB0}, accounting for possibly correlated uncertainties.   We treat the statistical errors as correlated between the calculations of Aoki \etal\ and RBC/UKQCD because they employ the same gauge-field configurations~\cite{Christ:2014uea,Aoki:2014nga}.\footnote{There may be mild correlations between some sub-dominant systematic errors of Aoki \etal\ and RBC/UKQCD, who use the same determinations of the absolute lattice scale and the physical light- and strange-quark masses from Ref.~\cite{Aoki:2010dy}, and who use the same power-counting estimates for the light-quark and gluon discretization errors.  The effects of any correlations between these systematics, however, would be too small to impact the numerical values of the averages.}  
We also treat the statistical errors as correlated between the HPQCD and Fermilab/MILC calculations because they analyze an overlapping set of gauge-field configurations~\cite{McNeile:2011ng,Bazavov:2011aa,Na:2012kp}.  For $f_{B_s}$, we include HPQCD's results from both 2011~\cite{McNeile:2011ng} and 2012~\cite{Na:2012kp}, which were obtained using different $b$-quark actions, but on some of the same gauge-field configurations.  HPQCD 11 and 12 also use the same determination of the absolute lattice scale, which is the second-largest source of systematic uncertainty in both calculations.  We therefore treat the statistical and scale errors as correlated between HPQCD's (2+1)-flavor $f_{B_s}$ results.  HPQCD also presents two results for $f_B$ in Ref.~\cite{Na:2012kp}.  The more precise value is obtained by combining the ratio $f_{B_s}/f_B$ from this work with $f_{B_s}$ from Ref.~\cite{McNeile:2011ng}, but an associated error budget is not provided.  Because this would be needed to estimate correlations between the two $f_B$ determinations, we include only HPQCD's more precise (2+1)-flavor result for $f_B$ in our average.
Our separate three- and four-flavor averages for the $B^+$-, $B^0$-, and $B_s$-meson decay constants and ratios are listed in the lines labeled ``Average" in Tables~\ref{tab:LatfB} and~\ref{tab:LatfB0},  where the error on the (2+1)-flavor $f_{B_s}$ average has been rescaled by the factor $\sqrt{(\chi^2/{\rm dof})} = 1.2$ to account for the tension among results.
Our (2+1+1)-flavor ``averages" are identical to the ``HPQCD 13" entries in Tables~\ref{tab:LatfB} and~\ref{tab:LatfB0}, whcih are the only published four-flavor results available.  

To obtain the single-best values of the $B_{(s)}$-meson decay constants for phenomenology applications, we combine the available $(2+1)$- and $(2+1+1)$-flavor lattice-QCD results, which are compatible within the current level of precision. Because the four-flavor ``average" is obtained from only a single result, we do not simply combine the two lines labeled ``Average" in Tables~\ref{tab:LatfB} and~\ref{tab:LatfB0}, which would weight the four-flavor result too heavily.   Instead, we form a single average including the published (2+1)-flavor results and the (2+1+1)-flavor result from HPQCD 13.  We account for the omission of charm sea-quark contributions in the three-flavor calculations by adding to the errors on the (2+1)-flavor averages in Tables~\ref{tab:LatfB} and~\ref{tab:LatfB0} our power-counting estimates of charm sea-quark errors from Sec.~\ref{sec:PiKTheory}, taking charm sea-quark error to be 100\% correlated between the three-flavor results.  Because the estimated charm sea-quark errors of 0.7\% for decay constants and 0.2\% for decay-constant ratios are much less than those on the (2+1)-flavor averages, adding them in quadrature has a tiny impact on the total uncertainties.  The largest observed change is an 0.3~MeV increase on the error $f_{B_s}$ from HPQCD 11, and most are negligible.  In the combined three- and four-flavor average we also consider correlations between the results of HPQCD 12 and HPQCD 13 because, although they employ different gauge-field configurations, they both use NRQCD for the $b$-quark action and the bottom-light axial-vector current.\footnote{HPQCD 13 uses a 1-loop radiatively improved $b$-quark action, whereas HPQCD 12 uses tree-level action coefficients.  Both include the same contributions to the currents at one loop, but renormalisation details differ.}  We take both the operator-matching and relativistic errors, which are the dominant uncertainties in the decay constants, to be correlated between the two calculations.  
Our final preferred theoretical values for the charged $B^+$ and neutral $B^0_{(s)}$-meson decay constants and their ratio are
\begin{eqnarray}
	{\rm Our~averages}: \quad && f_{B^+} = 187.1(4.2)~{\rm MeV} \,, \;\;  f_{B_s} = 227.2(3.4)~{\rm MeV} \,, \;\;  \frac{f_{B_s}}{f_{B^+}} = 1.215(7) \,, \qquad \label{eq:fB+_PDG} \\
	&& f_{B^0} = 190.9(4.1)~{\rm MeV} \,, \;\;  \frac{f_{B_s}}{f_{B^0}} = 1.192(6) \,. \qquad \label{eq:fB0_PDG}
\end{eqnarray}
The errors on $f_B^+$, $f_B^0$, and $f_{B_s}$ after combining the three- and four-flavor results are only slightly smaller than those of the separate averages due to the correlations assumed.

\section{Phenomenological implications} \label{sec:Pheno}

\subsection{$|V_{ud}|$, $|V_{us}|$, and status of first-row unitarity} \label{sec:FirstRow}

Using the average values for $f_{\pi^+} |V_{ud}|$, $f_{K^+} |V_{us}|$, and their ratio from Eqs.~(\ref{eq:fPi_Vud})--(\ref{eq:fK_Vus_over_fPi_Vud}) and for $f_{\pi^+}$, $f_{K^+}$, and their ratio from Eq.~(\ref{eq:PDG_fPi_fK}), we obtain the following determinations of the CKM matrix elements $|V_{ud}|$, $|V_{us}|$, and their ratio from leptonic decays within the standard model:
\begin{equation}
	|V_{ud}| = 0.9764(2)(127)(10) \,, \quad |V_{us}| = 0.2255(3)(6)(3), \quad \frac{|V_{us}|}{|V_{ud}|}  = 0.2314(2)(5)(2)  \,, \label{eq:VCKM_FirstRow}
\end{equation}
where the errors are from the experimental branching fraction(s), the pseudoscalar decay constant(s), and radiative corrections, respectively.  These results enable a precise test of the unitarity of the first row of the CKM matrix from leptonic decays alone (the contribution from $|V_{ub}|$ is negligible).  Using the values of $|V_{ud}|$ and $|V_{us}|$ from Eq.~(\ref{eq:VCKM_FirstRow}), we find 
\begin{equation}
	|V_{ud}|^2 + |V_{us}|^2  + |V_{ub}|^2 -1 = 0.004(25) \,,
\end{equation}
which is consistent with three-generation unitarity at the sub-percent level.

The determinations of $|V_{ud}|$ and $|V_{us}|$ from leptonic decays in Eq.~(\ref{eq:VCKM_FirstRow}) can be compared to those obtained from other processes.  The result above for $|V_{ud}|$ agrees with the determination from superallowed $\beta$-decay, $|V_{ud}| = 0.97417(21)$~\cite{Hardy:2014qxa}, but has an error more than fifty times larger that is primarily due to the uncertainty in the theoretical determination of $f_{\pi^+}$.  The CKM element $|V_{us}|$ can be determined from semileptonic $K^+ \to \pi^0 \ell^+ \nu$ decay.  Here experimental measurements provide a value for the product $f_+^{K\pi}(0) |V_{us}|$, where $f_+^{K\pi}(0)$ is the form-factor at zero four-momentum transfer between the initial state kaon and  the final state pion. Taking the most recent experimental determination of $|V_{us}| f_+^{K\pi}(0) = 0.2165(4)$ from Moulson~\cite{Moulson:2014cra}\footnote{This is an update of the 2010 Flavianet review~\cite{Antonelli:2010yf} that includes new measurements of the $K_s$ lifetime~\cite{Antonelli:2010aa,Abouzaid:2010ny}, Re($\epsilon'/\epsilon$)~\cite{Abouzaid:2010ny}, and ${\cal B}(K^\pm \to \pi^\pm \pi^+ \pi^-)$~\cite{Babusci:2014hxa}.  The latter measurement is the primary source of the reduced error on ${\cal B}(K_{\ell 3}$), via the constraint that the sum of all branching ratios must equal unity.} and the preliminary 2015 (2+1+1)-flavor FLAG average for $f_+(0)^{K\pi} = 0.9704(24)(22)$~\cite{JuettnerLat15,FLAGInPrep}\footnote{This result comes from the calculation of FNAL/MILC in Ref.~\cite{Bazavov:2013maa}.  For comparison, the 2015 preliminary (2+1)-flavor FLAG average based on the calculations of FNAL/MILC~\cite{Bazavov:2012cd} and RBC/UKQCD~\cite{Boyle:2015hfa} is $f_+(0)^{K\pi} = 0.9677(37)$ .} gives $|V_{us}| = 0.22310(74)_{\rm thy} (41)_{\rm exp}$ from $K_{\ell 3}$ decay. The determinations of $|V_{us}|$ from leptonic and semileptonic kaon decays are both quite precise (with the error from leptonic decay being about 20\% smaller), but the central values differ by 2.2$\sigma$.   Finally, the combination of the ratio $|V_{us}|/|V_{ud}|$ from leptonic decays [Eq.~(\ref{eq:VCKM_FirstRow})] with $|V_{ud}|$ from $\beta$ decay implies an alternative determination of $|V_{us}| = 0.2254(6)$ which agrees with the value from leptonic kaon decay, but disagrees with the $K_{\ell 3}$-decay result at the 2.2$\sigma$ level.  Collectively, these results indicate that that there is some tension between theoretical calculations and/or measurements of leptonic pion and kaon decays, semileptonic kaon decays, and superallowed $\beta$-decay.  Although this may be due to the presence of new physics, it is also important to revisit the quoted uncertainties on both the theoretical and experimental inputs.

Finally, we combine the experimental measurements of $f_{\pi^+} |V_{ud}|$, $f_{K^+} |V_{us}|$ from leptonic pseudoscalar-meson decays in Eqs.~(\ref{eq:fPi_Vud}) and~(\ref{eq:fK_Vus}) with determinations of the CKM elements from other decays or unitarity to infer ``experimental" values for the decay constants.  Assuming that there are no significant new-physics contributions to any of the input processes, the comparison of these results with theoretical calculations of the decay constants enables a test of lattice-QCD methods.  Taking $|V_{ud}|$ from superallowed $\beta$-decay~\cite{Towner:2014uta} leads to
\begin{equation}
f_{\pi^-}^{\rm ``exp"} = 130.50 (1)(3)(13)~{\rm MeV} \,,
\end{equation}
where the uncertainties are from the errors on $\Gamma$, $|V_{ud}|$, and higher-order corrections, respectively. This agrees with the theoretical value $f_{\pi^+} = 130.2(1.7)~{\rm MeV}$ in Eq.~(\ref{eq:PDG_fPi_fK}) obtained from an average of recent (2+1)-flavor lattice-QCD results~\cite{Follana:2007uv,Bazavov:2010hj,Arthur:2012yc}.  We take the value $|V_{us}| = 0.22534(65)$ from the most recent global unitarity-triangle fit of the UTfit Collaboration~\cite{Bona:2006ah} because there is tension between the values of $|V_{us}|$ obtained from leptonic and semileptonic kaon decays.  This implies
\begin{equation}
	f_{K^-}^{\rm ``exp"} = 155.72(17)(45)(16)~{\rm MeV} \,
\end{equation}
where the uncertainties are from the errors on $\Gamma$, $|V_{us}|$, and higher-order corrections, respectively.  This agrees with the theoretical value $f_{K^+} = 155.6(0.4)~{\rm MeV}$ in Eq.~(\ref{eq:PDG_fPi_fK}) obtained from an average of recent three and four-flavor lattice-QCD results~\cite{Follana:2007uv,Bazavov:2010hj,Arthur:2012yc,Dowdall:2013rya,Bazavov:2014wgs,Carrasco:2014poa}.

\subsection{$|V_{cd}|$, $|V_{cs}|$, and status of second-row unitarity} \label{sec:SecondRow}

Using the average values for $|V_{cd}|f_{D^+}$ and $|V_{cs}| f_{D_s^+}$ from Tables~\ref{tab:fDp} and~\ref{tab:fDs}, and for $f_{D^+}$ and $f_{D_s^+}$ from Eq.~(\ref{eq:fD_PDG}), we obtain the following determinations of the CKM matrix elements $|V_{cd}|$ and $|V_{cs}|$, and from leptonic decays within the standard model:
\begin{equation}
	|V_{cd}| = 0.217(5)(1) \quad {\rm and} \quad |V_{cs}| = 1.007(16)(5) \,, \label{eq:VCKM_SecondRow}
\end{equation}
where the errors are from experiment and theory, respectively, and are currently limited by the measured uncertainties on the decay rates.  The central value of $|V_{cs}|$ is greater than one, but is compatible with unity within the error.  The above results for $|V_{cd}|$ and $|V_{cs}|$ do not include higher-order electroweak and hadronic corrections to the rate, in analogy to Eq.~(\ref{eq:RadRate}).  These corrections have not been computed for $D_{(s)}^+$-meson leptonic decays, but are estimated to be about to be about 1--2\% for charged pion and kaon decays (see Sec.~\ref{sec:PiKExp}).  Now that the uncertainties on  $|V_{cd}|$ and $|V_{cs}|$ from leptonic decays are at this level, we hope that the needed theoretical calculations will be undertaken.

The CKM elements $|V_{cd}|$ and $|V_{cs}|$ can also be obtained from semileptonic $D^+ \to \pi^0 \ell^+ \nu$ and $D_{s}^+ \to K^0 \ell^+ \nu$ decays, respectively.  Here experimental measurements determine the product of the form factor times the CKM element, and theory provides the value for the form factor at zero four-momentum transfer between the initial $D_{(s)}$ meson and the final pion or kaon.  We combine the latest experimental averages for $f_+^{D\pi}(0)|V_{cd}| = 0.1425(19)$ and $f_+^{D_sK}(0)|V_{cs}| = 0.728(5)$ from the Heavy Flavor Averaging Group (HFAG)~\cite{Amhis:2014hma} with the zero-momentum-transfer form factors $f_+^{D\pi}(0) = 0.666(29)$ and $f_+^{D_sK}(0) = 0.747(19)$ calculated in (2+1)-flavor lattice QCD by the HPQCD Collaboration~\cite{Na:2010uf,Na:2011mc} to obtain $|V_{cd}| = 0.2140(97)$ and $|V_{cs}| = 0.9746(257)$ from semileptonic $D_{(s)}$-meson decays.  The values of $|V_{cd}|$ from leptonic and semileptonic decays agree, while those for $|V_{cs}|$ are compatible at the 1.1$\sigma$ level.  The determinations of $|V_{cd}|$ and $|V_{cs}|$ from leptonic decays in Eq.~(\ref{eq:VCKM_SecondRow}), however, are 2.0$\times$ and 1.6$\times$ more precise than those from semileptonic decays, respectively.

The results for $|V_{cd}|$ and $|V_{cs}|$ from Eq.~(\ref{eq:VCKM_SecondRow}) enable a test of the unitarity of the second row of the CKM matrix.  We obtain
\begin{equation}
	|V_{cd}|^2 + |V_{cs}|^2  + |V_{cb}|^2 -1 = 0.064(36) \,, \label{eq:SecondRowUnitarity}
\end{equation}
which is in slight tension with three-generation unitarity at the 2$\sigma$ level.  Because the contribution to Eq.~(\ref{eq:SecondRowUnitarity}) from $|V_{cb}|$ is so small, we obtain the same result taking $|V_{cb}|^{\rm incl.} \times 10^3 = 42.21(78)$ from inclusive $B\to X_c \ell\nu$ decay~\cite{Alberti:2014yda} or $|V_{cb}|^{\rm excl.} \times 10^3 = 39.04(75)$ from exclusive $B\to D^* \ell\nu$ decay at zero recoil~\cite{Bailey:2014tva}.

We can also combine the experimental measurements of $f_{D^+} |V_{cd}| = 45.91(1.05)$~MeV and $f_{D_s^+} |V_{cs}| = 250.9(4.0)$~MeV from leptonic pseudoscalar-meson decays from Tables~\ref{tab:fDp} and~\ref{tab:fDs} with determinations of $|V_{cd}|$ and $|V_{cs}|$ from CKM unitarity to infer ``experimental" values for the decay constants within the standard model.  For this purpose, we obtain the values of $|V_{cd}|$ and $|V_{cs}|$ by relating them to other CKM elements using the Wolfenstein parameterization~\cite{Wolfenstein:1983yz}.  We take $|V_{cd}|$ to equal the value of $|V_{us}|$ minus the leading correction~\cite{Charles:2004jd}:
\begin{eqnarray}
	|V_{cd}| &=& |V_{us}| \left| -1 + \frac{|V_{cb}|^2}{2} (1 - 2(\rho + i\eta )) \right| \\
		& = & |V_{us}| \left( \left[ -1 + (1 - 2 \rho) \frac{|V_{cb}|^2}{2} \right]^2  + \eta^2 |V_{cb}|^4  \right)^{1/2} \,\, .
\end{eqnarray}
Using $|V_{us}| =0.2255(3)(6)(3)$ from leptonic kaon decay [Eq.~(\ref{eq:VCKM_FirstRow})], inclusive $|V_{cb}|$ as above, and $\left(\rho, \eta \right) = \left( 0.136(24), 0.361(14) \right)$ from CKM unitarity~\cite{Bona:2006ah} $|V_{cd}|= $0.2254(7).  We take $|V_{cs}| = |V_{ud}| - |V_{cb}|^2/2$ \cite{Charles:2004jd}, using $|V_{ud}| = 0.97417(21)$ from $\beta$ decay~\cite{Hardy:2014qxa}, giving $|V_{cs}| = 0.9733(2)$.  Given these choices, we find
\begin{equation}
	f_{D^+}^{\rm \, ``exp"} = 203.7(4.7)(0.6)~{\rm MeV} \quad {\rm and} \quad f_{D_s^+}^{\rm \, ``exp"} = 257.8(4.1)(0.1) ~{\rm MeV} \,,
\end{equation}
where the uncertainties are from the errors on $\Gamma$ and $|V_{us}|$ (or $|V_{ud}|$), respectively.  These disagree with the theoretical values $f_{D^+} = 211.9(1.1)~{\rm MeV}$ and $f_{D_s^+} = 249.0(1.2)~{\rm MeV}$ in Eq.~(\ref{eq:fD_PDG}) obtained from averaging recently published three and four-flavor lattice-QCD results at the 1.7$\sigma$ and 2.0$\sigma$ levels, respectively.  The significances of the tensions are sensitive, however, to the choices made for $|V_{us}|$ and $|V_{ud}|$.  Thus resolving the inconsistencies between determinations of elements of the first row of the CKM matrix discussed previously in Sec.~\ref{sec:FirstRow} may also reduce the mild tensions observed here.

\subsection{$|V_{ub}|$ and other applications} \label{sec:Vub}

Using the average value for $|V_{ub}|f_{B^+}$ from Table~\ref{tab:Btotaunu}, and for $f_{B^+}$ from Eq.~(\ref{eq:fB+_PDG}), we obtain the following determination of the CKM matrix element $|V_{ub}|$ from leptonic decays within the standard model:
\begin{equation}
	|V_{ub}| = 4.12(37)(9) \times 10^{-3}  \,, \label{eq:VubLeptonic}
\end{equation}
where the errors are from experiment and theory, respectively.   We note, however, that decays involving the third generation of quarks and leptons may be particularly sensitive to new physics associated with electroweak symmetry breaking due to their larger masses~\cite{Hou:1992sy,Akeroyd:2003zr}, so Eq.~(\ref{eq:VubLeptonic}) is more likely to be influenced by new physics than the determinations of the elements of the first and second rows of the CKM matrix in the previous sections.

The CKM element $|V_{ub}|$ can also be obtained from semileptonic $B$-meson decays.  Over the past several years there has remained a persistent 2-3$\sigma$ tension between the determinations of $|V_{ub}|$ from exclusive $B\to\pi\ell\nu$ decay and from inclusive $B \to X_u \ell \nu$ decay, where $X_u$ denotes all hadrons which contain a constituent up quark~\cite{Antonelli:2009ws,Butler:2013kdw,Amhis:2014hma,Agashe:2014kda,Bevan:2014iga}.  The currently most precise determination of $|V_{ub}|^{\rm excl} = 3.72(16)\times 10^{-3} $ is obtained from a joint $z$-fit of the vector and scalar form factors $f_+^{B\pi}(q^2)$ and $f_0^{B\pi}(q^2)$ calculated in (2+1)-flavor lattice QCD by the FNAL/MILC Collaboration~\cite{Lattice:2015tia} and experimental measurements of the differential decay rate from BaBar~\cite{delAmoSanchez:2010af,Lees:2012vv} and~Belle~\cite{Ha:2010rf,Sibidanov:2013rkk}.  On the other hand, the most recent PDG average of inclusive determinations obtained using the theoretical frameworks in Refs.~\cite{Bosch:2004bt,Andersen:2005mj,Gambino:2007rp} is $|V_{ub}|^{\rm incl} = 4.49(16)\left(^{+16}_{-18}\right) \times 10^{-3}$~\cite{PDGVubVcb2015}.
The result for $|V_{ub}|$ from leptonic $B\to\tau\nu$ decay in Eq.~(\ref{eq:VubLeptonic}) is compatible with determinations from both exclusive and inclusive semileptonic $B$-meson decays.

The CKM element $|V_{ub}|$ can now also be obtained from semileptonic $\Lambda_b$ decays.  Specifically, the recent LHCb measurement of the ratio of decay rates for $\Lambda_b \to p \ell\nu$ over $\Lambda_b \to \Lambda_c \ell\nu$~\cite{Aaij:2015bfa}, when combined with the ratio of form factors from (2+1)-flavor lattice QCD~\cite{Detmold:2015aaa}, enables the first determination of the ratio of CKM elements $|V_{ub}|/|V_{cb}| = 0.083(4)(4)$ from baryonic decay.  Taking $|V_{cb}|^{\rm incl} = 42.21(78) \times 10^{-3} $~\cite{Alberti:2014yda} for the denominator,\footnote{This differs from the choice for $|V_{cb}|$ made by LHCb~\cite{Aaij:2015bfa}, who use the determination from exclusive $B\to D^{(*)} \ell\nu$ decays at zero recoil~\cite{PDGVubVcb}.  The Belle Experiment recently obtained a new measurement of the $B\to D\ell\nu$ differential decay rate~\cite{Glattauer:2015teq} and determination of $|V_{cb}| = 40.83(1.13) \times 10^{-3}$.  They find that the inclusion of experimental and theoretical nonzero-recoil information increases the value for $|V_{cb}|$ compared to when only zero-recoil information is used, and leads to agreement with the inclusive result.} we obtain $|V_{ub}| = 3.50(17)(17)(6) \times 10^{-3}$ from exclusive $\Lambda_b$ semileptonic decays, where the errors are from experiment, the form factors, and $|V_{cb}|$, respectively.  The result for $|V_{ub}|$ from leptonic $B\to\tau\nu$ decay in Eq.~(\ref{eq:VubLeptonic}) is 1.4$\sigma$ higher than the determination from $b$-baryon decays.

Given these results, the ``$V_{ub}$" puzzle still stands, and the determination from leptonic $B^+$-meson decay is not yet sufficiently precise to weigh in on the discrepancy.  New and improved experimental measurements and theoretical calculations of other $b\to u$ flavor-changing processes, however, are providing additional information and sharpening the picture of the various tensions.  Further, the error on $|V_{ub}|$ from $B\to\tau\nu$ decay will shrink once improved rate measurements from the Belle~II experiment are available.

Finally, we can combine the experimental measurement of $|V_{ub}|f_{B^+}$ from leptonic $B^+$-meson decays in Table~\ref{tab:Btotaunu} with a determination of the CKM element $|V_{ub}|$ from elsewhere to infer an ``experimental" values for $f_{B^+}$ within the standard model.  This, of course, assumes that there are no significant new-physics contributions to $B^+ \to \tau \nu$, which may turn out not to be the case.  Further, one does not know {\it a priori} what value to take for $|V_{ub}|$ given the inconsistencies between the various determinations discussed above.  We therefore take the PDG weighted average of the determinations from inclusive and exclusive semileptonic $B$-meson decays $|V_{ub}|^{\rm excl+incl} = 4.09(39) \times 10^{-3}$~\cite{PDGVubVcb2015}, where the error has been rescaled by the $\sqrt{\chi^2/{\rm dof}} = 2.6$ to account for the disagreement.  Using this result we obtain
\begin{equation}
	f_{B^+}^{\rm ``exp"} = 188(17)(18)~{\rm MeV} \,,
\end{equation}	
where the uncertainties are from the errors on $\Gamma$ and $|V_{ub}|$, respectively.  This agrees within large uncertainties with the theoretical value $f_{B^+} = 187.1(4.2)~{\rm MeV}$ in Eq.~(\ref{eq:fB+_PDG}) obtained from an average of recent three and four-flavor lattice-QCD results~\cite{Na:2012kp,Bazavov:2011aa,Dowdall:2013tga,Christ:2014uea}.

\section*{Acknowledgements}
We thank V.~Cirigliano, C.~Davies, A. El~Khadra, A. Khodjamirian, J. Laiho,
W. Marciano, M. Moulson, S. Narison, S. Sharpe, and Z.-G. Wang for useful discussions
and references.  We thank P.~Boyle, M. Della~Morte, D.~Lin and S. Simula for
providing information on the preliminary FLAG-3 lattice averages.
We gratefully acknowledge support of the U. S. National Science Foundation
and the U. S. Department of Energy through Grant No.\ DE-FG02-13ER41598.  The
work of J. L. R. was performed in part at the Aspen Center for Physics, which
is supported by National Science Foundation grant PHY-1066293.  Fermilab is
operated by Fermi Research Alliance, LLC, under Contract No.~DE-AC02-07CH11359
with the U.S.\ Department of Energy.

Create the reference section using BibTeX:
\bibliography{PDGRefs}

\begin{thebibliography}{124}%
\makeatletter
\providecommand \@ifxundefined [1]{%
 \@ifx{#1\undefined}
}%
\providecommand \@ifnum [1]{%
 \ifnum #1\expandafter \@firstoftwo
 \else \expandafter \@secondoftwo
 \fi
}%
\providecommand \@ifx [1]{%
 \ifx #1\expandafter \@firstoftwo
 \else \expandafter \@secondoftwo
 \fi
}%
\providecommand \natexlab [1]{#1}%
\providecommand \enquote  [1]{``#1''}%
\providecommand \bibnamefont  [1]{#1}%
\providecommand \bibfnamefont [1]{#1}%
\providecommand \citenamefont [1]{#1}%
\providecommand \href@noop [0]{\@secondoftwo}%
\providecommand \href [0]{\begingroup \@sanitize@url \@href}%
\providecommand \@href[1]{\@@startlink{#1}\@@href}%
\providecommand \@@href[1]{\endgroup#1\@@endlink}%
\providecommand \@sanitize@url [0]{\catcode `\\12\catcode `\$12\catcode
  `\&12\catcode `\#12\catcode `\^12\catcode `\_12\catcode `\%12\relax}%
\providecommand \@@startlink[1]{}%
\providecommand \@@endlink[0]{}%
\providecommand \url  [0]{\begingroup\@sanitize@url \@url }%
\providecommand \@url [1]{\endgroup\@href {#1}{\urlprefix }}%
\providecommand \urlprefix  [0]{URL }%
\providecommand \Eprint [0]{\href }%
\providecommand \doibase [0]{http://dx.doi.org/}%
\providecommand \selectlanguage [0]{\@gobble}%
\providecommand \bibinfo  [0]{\@secondoftwo}%
\providecommand \bibfield  [0]{\@secondoftwo}%
\providecommand \translation [1]{[#1]}%
\providecommand \BibitemOpen [0]{}%
\providecommand \bibitemStop [0]{}%
\providecommand \bibitemNoStop [0]{.\EOS\space}%
\providecommand \EOS [0]{\spacefactor3000\relax}%
\providecommand \BibitemShut  [1]{\csname bibitem#1\endcsname}%
\let\auto@bib@innerbib\@empty
\bibitem [{\citenamefont {Nakamura}\ \emph {et~al.}(2010)\citenamefont
  {Nakamura} \emph {et~al.}}]{Nakamura:2010zzi}%
  \BibitemOpen
  \bibfield  {author} {\bibinfo {author} {\bibfnamefont {K.}~\bibnamefont
  {Nakamura}} \emph {et~al.} (\bibinfo {collaboration} {Particle Data Group}),\
  }\href {\doibase 10.1088/0954-3899/37/7A/075021} {\bibfield  {journal}
  {\bibinfo  {journal} {J. Phys.}\ }\textbf {\bibinfo {volume} {G37}},\
  \bibinfo {pages} {075021} (\bibinfo {year} {2010})}\BibitemShut {NoStop}%
\bibitem [{\citenamefont {Beringer}\ \emph {et~al.}(2012)\citenamefont
  {Beringer} \emph {et~al.}}]{Beringer:1900zz}%
  \BibitemOpen
  \bibfield  {author} {\bibinfo {author} {\bibfnamefont {J.}~\bibnamefont
  {Beringer}} \emph {et~al.} (\bibinfo {collaboration} {Particle Data Group}),\
  }\href {\doibase 10.1103/PhysRevD.86.010001} {\bibfield  {journal} {\bibinfo
  {journal} {Phys. Rev.}\ }\textbf {\bibinfo {volume} {D86}},\ \bibinfo {pages}
  {010001} (\bibinfo {year} {2012})}\BibitemShut {NoStop}%
\bibitem [{\citenamefont {Olive}\ \emph {et~al.}(2014)\citenamefont {Olive}
  \emph {et~al.}}]{Agashe:2014kda}%
  \BibitemOpen
  \bibfield  {author} {\bibinfo {author} {\bibfnamefont {K.~A.}\ \bibnamefont
  {Olive}} \emph {et~al.} (\bibinfo {collaboration} {Particle Data Group}),\
  }\href {\doibase 10.1088/1674-1137/38/9/090001} {\bibfield  {journal}
  {\bibinfo  {journal} {Chin. Phys.}\ }\textbf {\bibinfo {volume} {C38}},\
  \bibinfo {pages} {090001} (\bibinfo {year} {2014})}\BibitemShut {NoStop}%
\bibitem [{\citenamefont {Hou}(1993)}]{Hou:1992sy}%
  \BibitemOpen
  \bibfield  {author} {\bibinfo {author} {\bibfnamefont {W.-S.}\ \bibnamefont
  {Hou}},\ }\href {\doibase 10.1103/PhysRevD.48.2342} {\bibfield  {journal}
  {\bibinfo  {journal} {Phys. Rev.}\ }\textbf {\bibinfo {volume} {D48}},\
  \bibinfo {pages} {2342} (\bibinfo {year} {1993})}\BibitemShut {NoStop}%
\bibitem [{\citenamefont {Akeroyd}\ and\ \citenamefont
  {Recksiegel}(2003{\natexlab{a}})}]{Akeroyd:2002pi}%
  \BibitemOpen
  \bibfield  {author} {\bibinfo {author} {\bibfnamefont {A.~G.}\ \bibnamefont
  {Akeroyd}}\ and\ \bibinfo {author} {\bibfnamefont {S.}~\bibnamefont
  {Recksiegel}},\ }\href {\doibase 10.1016/S0370-2693(02)03293-8} {\bibfield
  {journal} {\bibinfo  {journal} {Phys. Lett.}\ }\textbf {\bibinfo {volume}
  {B554}},\ \bibinfo {pages} {38} (\bibinfo {year} {2003}{\natexlab{a}})},\
  \Eprint {http://arxiv.org/abs/hep-ph/0210376} {arXiv:hep-ph/0210376 [hep-ph]}
  \BibitemShut {NoStop}%
\bibitem [{\citenamefont {Akeroyd}\ and\ \citenamefont
  {Recksiegel}(2003{\natexlab{b}})}]{Akeroyd:2003zr}%
  \BibitemOpen
  \bibfield  {author} {\bibinfo {author} {\bibfnamefont {A.~G.}\ \bibnamefont
  {Akeroyd}}\ and\ \bibinfo {author} {\bibfnamefont {S.}~\bibnamefont
  {Recksiegel}},\ }\href {\doibase 10.1088/0954-3899/29/10/301} {\bibfield
  {journal} {\bibinfo  {journal} {J. Phys.}\ }\textbf {\bibinfo {volume}
  {G29}},\ \bibinfo {pages} {2311} (\bibinfo {year} {2003}{\natexlab{b}})},\
  \Eprint {http://arxiv.org/abs/hep-ph/0306037} {arXiv:hep-ph/0306037 [hep-ph]}
  \BibitemShut {NoStop}%
\bibitem [{\citenamefont {Akeroyd}(2004)}]{Akeroyd:2003jb}%
  \BibitemOpen
  \bibfield  {author} {\bibinfo {author} {\bibfnamefont {A.~G.}\ \bibnamefont
  {Akeroyd}},\ }\href {\doibase 10.1143/PTP.111.295} {\bibfield  {journal}
  {\bibinfo  {journal} {Prog. Theor. Phys.}\ }\textbf {\bibinfo {volume}
  {111}},\ \bibinfo {pages} {295} (\bibinfo {year} {2004})},\ \Eprint
  {http://arxiv.org/abs/hep-ph/0308260} {arXiv:hep-ph/0308260 [hep-ph]}
  \BibitemShut {NoStop}%
\bibitem [{\citenamefont {Dobrescu}\ and\ \citenamefont
  {Kronfeld}(2008)}]{Dobrescu:2008er}%
  \BibitemOpen
  \bibfield  {author} {\bibinfo {author} {\bibfnamefont {B.~A.}\ \bibnamefont
  {Dobrescu}}\ and\ \bibinfo {author} {\bibfnamefont {A.~S.}\ \bibnamefont
  {Kronfeld}},\ }\href {\doibase 10.1103/PhysRevLett.100.241802} {\bibfield
  {journal} {\bibinfo  {journal} {Phys. Rev. Lett.}\ }\textbf {\bibinfo
  {volume} {100}},\ \bibinfo {pages} {241802} (\bibinfo {year} {2008})},\
  \Eprint {http://arxiv.org/abs/0803.0512} {arXiv:0803.0512 [hep-ph]}
  \BibitemShut {NoStop}%
\bibitem [{\citenamefont {Hewett}(1995)}]{Hewett:1995aw}%
  \BibitemOpen
  \bibfield  {author} {\bibinfo {author} {\bibfnamefont {J.~L.}\ \bibnamefont
  {Hewett}},\ }in\ \href
  {http://www-public.slac.stanford.edu/sciDoc/docMeta.aspx?slacPubNumber=SLAC-PUB-6821}
  {\emph {\bibinfo {booktitle} {{Heavy flavor physics. Proceedings, LISHEP 95,
  LAFEX International School on High-Energy Physics, session C, cbt Workshop,
  Rio de Janeiro, Brazil, February 21-23, 1995}}}}\ (\bibinfo {year} {1995})\
  \Eprint {http://arxiv.org/abs/hep-ph/9505246} {arXiv:hep-ph/9505246 [hep-ph]}
  \BibitemShut {NoStop}%
\bibitem [{\citenamefont {Crivellin}(2010)}]{Crivellin:2009sd}%
  \BibitemOpen
  \bibfield  {author} {\bibinfo {author} {\bibfnamefont {A.}~\bibnamefont
  {Crivellin}},\ }\href {\doibase 10.1103/PhysRevD.81.031301} {\bibfield
  {journal} {\bibinfo  {journal} {Phys. Rev.}\ }\textbf {\bibinfo {volume}
  {D81}},\ \bibinfo {pages} {031301} (\bibinfo {year} {2010})},\ \Eprint
  {http://arxiv.org/abs/0907.2461} {arXiv:0907.2461 [hep-ph]} \BibitemShut
  {NoStop}%
\bibitem [{\citenamefont {Bernlochner}\ \emph {et~al.}(2014)\citenamefont
  {Bernlochner}, \citenamefont {Ligeti},\ and\ \citenamefont
  {Turczyk}}]{Bernlochner:2014ova}%
  \BibitemOpen
  \bibfield  {author} {\bibinfo {author} {\bibfnamefont {F.~U.}\ \bibnamefont
  {Bernlochner}}, \bibinfo {author} {\bibfnamefont {Z.}~\bibnamefont {Ligeti}},
  \ and\ \bibinfo {author} {\bibfnamefont {S.}~\bibnamefont {Turczyk}},\ }\href
  {\doibase 10.1103/PhysRevD.90.094003} {\bibfield  {journal} {\bibinfo
  {journal} {Phys. Rev.}\ }\textbf {\bibinfo {volume} {D90}},\ \bibinfo {pages}
  {094003} (\bibinfo {year} {2014})},\ \Eprint {http://arxiv.org/abs/1408.2516}
  {arXiv:1408.2516 [hep-ph]} \BibitemShut {NoStop}%
\bibitem [{\citenamefont {J{\"u}ttner}(2015)}]{JuettnerLat15}%
  \BibitemOpen
  \bibfield  {author} {\bibinfo {author} {\bibfnamefont {A.}~\bibnamefont
  {J{\"u}ttner}},\ }\href@noop {} {\enquote {\bibinfo {title} {{Light Flavour
  Physics}},}\ }\bibinfo {howpublished}
  {\url{https://indico2.riken.jp/indico/contributionDisplay.py?sessionId=0&contribId=347&confId=1805}}
  (\bibinfo {year} {2015}),\ \bibinfo {note} {{plenary talk presented at
  Lattice 2015}}\BibitemShut {NoStop}%
\bibitem [{\citenamefont {Boyle}\ \emph
  {et~al.}(2015{\natexlab{a}})\citenamefont {Boyle}, \citenamefont {Kaneko},\
  and\ \citenamefont {Simula}}]{FLAGInPrep}%
  \BibitemOpen
  \bibfield  {author} {\bibinfo {author} {\bibfnamefont {P.~A.}\ \bibnamefont
  {Boyle}}, \bibinfo {author} {\bibfnamefont {T.}~\bibnamefont {Kaneko}}, \
  and\ \bibinfo {author} {\bibfnamefont {S.}~\bibnamefont {Simula}} (\bibinfo
  {collaboration} {Flavour Lattice Averaging Group}),\ }\href@noop {}
  {}\bibinfo {howpublished} {{private communication}} (\bibinfo {year}
  {2015}{\natexlab{a}})\BibitemShut {NoStop}%
\bibitem [{\citenamefont {Aoki}\ \emph {et~al.}(2014)\citenamefont {Aoki} \emph
  {et~al.}}]{Aoki:2013ldr}%
  \BibitemOpen
  \bibfield  {author} {\bibinfo {author} {\bibfnamefont {S.}~\bibnamefont
  {Aoki}} \emph {et~al.} (\bibinfo {collaboration} {Flavour Lattice Averaging
  Group}),\ }\href {\doibase 10.1140/epjc/s10052-014-2890-7} {\bibfield
  {journal} {\bibinfo  {journal} {Eur. Phys. J.}\ }\textbf {\bibinfo {volume}
  {C74}},\ \bibinfo {pages} {2890} (\bibinfo {year} {2014})},\ \Eprint
  {http://arxiv.org/abs/1310.8555} {arXiv:1310.8555 [hep-lat]} \BibitemShut
  {NoStop}%
\bibitem [{\citenamefont {Marciano}(2004)}]{Marciano:2004uf}%
  \BibitemOpen
  \bibfield  {author} {\bibinfo {author} {\bibfnamefont {W.~J.}\ \bibnamefont
  {Marciano}},\ }\href {\doibase 10.1103/PhysRevLett.93.231803} {\bibfield
  {journal} {\bibinfo  {journal} {Phys. Rev. Lett.}\ }\textbf {\bibinfo
  {volume} {93}},\ \bibinfo {pages} {231803} (\bibinfo {year} {2004})},\
  \Eprint {http://arxiv.org/abs/hep-ph/0402299} {arXiv:hep-ph/0402299 [hep-ph]}
  \BibitemShut {NoStop}%
\bibitem [{\citenamefont {Cirigliano}\ \emph {et~al.}(2012)\citenamefont
  {Cirigliano}, \citenamefont {Ecker}, \citenamefont {Neufeld}, \citenamefont
  {Pich},\ and\ \citenamefont {Portoles}}]{Cirigliano:2011ny}%
  \BibitemOpen
  \bibfield  {author} {\bibinfo {author} {\bibfnamefont {V.}~\bibnamefont
  {Cirigliano}}, \bibinfo {author} {\bibfnamefont {G.}~\bibnamefont {Ecker}},
  \bibinfo {author} {\bibfnamefont {H.}~\bibnamefont {Neufeld}}, \bibinfo
  {author} {\bibfnamefont {A.}~\bibnamefont {Pich}}, \ and\ \bibinfo {author}
  {\bibfnamefont {J.}~\bibnamefont {Portoles}},\ }\href {\doibase
  10.1103/RevModPhys.84.399} {\bibfield  {journal} {\bibinfo  {journal} {Rev.
  Mod. Phys.}\ }\textbf {\bibinfo {volume} {84}},\ \bibinfo {pages} {399}
  (\bibinfo {year} {2012})},\ \Eprint {http://arxiv.org/abs/1107.6001}
  {arXiv:1107.6001 [hep-ph]} \BibitemShut {NoStop}%
\bibitem [{\citenamefont {Cirigliano}\ and\ \citenamefont
  {Rosell}(2007{\natexlab{a}})}]{Cirigliano:2007ga}%
  \BibitemOpen
  \bibfield  {author} {\bibinfo {author} {\bibfnamefont {V.}~\bibnamefont
  {Cirigliano}}\ and\ \bibinfo {author} {\bibfnamefont {I.}~\bibnamefont
  {Rosell}},\ }\href {\doibase 10.1088/1126-6708/2007/10/005} {\bibfield
  {journal} {\bibinfo  {journal} {JHEP}\ }\textbf {\bibinfo {volume} {10}},\
  \bibinfo {pages} {005} (\bibinfo {year} {2007}{\natexlab{a}})},\ \Eprint
  {http://arxiv.org/abs/0707.4464} {arXiv:0707.4464 [hep-ph]} \BibitemShut
  {NoStop}%
\bibitem [{\citenamefont {Sirlin}(1982)}]{Sirlin:1981ie}%
  \BibitemOpen
  \bibfield  {author} {\bibinfo {author} {\bibfnamefont {A.}~\bibnamefont
  {Sirlin}},\ }\href {\doibase 10.1016/0550-3213(82)90303-0} {\bibfield
  {journal} {\bibinfo  {journal} {Nucl. Phys.}\ }\textbf {\bibinfo {volume}
  {B196}},\ \bibinfo {pages} {83} (\bibinfo {year} {1982})}\BibitemShut
  {NoStop}%
\bibitem [{\citenamefont {Kinoshita}(1959)}]{Kinoshita:1959ha}%
  \BibitemOpen
  \bibfield  {author} {\bibinfo {author} {\bibfnamefont {T.}~\bibnamefont
  {Kinoshita}},\ }\href {\doibase 10.1103/PhysRevLett.2.477} {\bibfield
  {journal} {\bibinfo  {journal} {Phys. Rev. Lett.}\ }\textbf {\bibinfo
  {volume} {2}},\ \bibinfo {pages} {477} (\bibinfo {year} {1959})}\BibitemShut
  {NoStop}%
\bibitem [{\citenamefont {Knecht}\ \emph {et~al.}(2000)\citenamefont {Knecht},
  \citenamefont {Neufeld}, \citenamefont {Rupertsberger},\ and\ \citenamefont
  {Talavera}}]{Knecht:1999ag}%
  \BibitemOpen
  \bibfield  {author} {\bibinfo {author} {\bibfnamefont {M.}~\bibnamefont
  {Knecht}}, \bibinfo {author} {\bibfnamefont {H.}~\bibnamefont {Neufeld}},
  \bibinfo {author} {\bibfnamefont {H.}~\bibnamefont {Rupertsberger}}, \ and\
  \bibinfo {author} {\bibfnamefont {P.}~\bibnamefont {Talavera}},\ }\href
  {\doibase 10.1007/s100529900265} {\bibfield  {journal} {\bibinfo  {journal}
  {Eur. Phys. J.}\ }\textbf {\bibinfo {volume} {C12}},\ \bibinfo {pages} {469}
  (\bibinfo {year} {2000})},\ \Eprint {http://arxiv.org/abs/hep-ph/9909284}
  {arXiv:hep-ph/9909284 [hep-ph]} \BibitemShut {NoStop}%
\bibitem [{\citenamefont {Ananthanarayan}\ and\ \citenamefont
  {Moussallam}(2004)}]{Ananthanarayan:2004qk}%
  \BibitemOpen
  \bibfield  {author} {\bibinfo {author} {\bibfnamefont {B.}~\bibnamefont
  {Ananthanarayan}}\ and\ \bibinfo {author} {\bibfnamefont {B.}~\bibnamefont
  {Moussallam}},\ }\href {\doibase 10.1088/1126-6708/2004/06/047} {\bibfield
  {journal} {\bibinfo  {journal} {JHEP}\ }\textbf {\bibinfo {volume} {06}},\
  \bibinfo {pages} {047} (\bibinfo {year} {2004})},\ \Eprint
  {http://arxiv.org/abs/hep-ph/0405206} {arXiv:hep-ph/0405206 [hep-ph]}
  \BibitemShut {NoStop}%
\bibitem [{\citenamefont {Descotes-Genon}\ and\ \citenamefont
  {Moussallam}(2005)}]{DescotesGenon:2005pw}%
  \BibitemOpen
  \bibfield  {author} {\bibinfo {author} {\bibfnamefont {S.}~\bibnamefont
  {Descotes-Genon}}\ and\ \bibinfo {author} {\bibfnamefont {B.}~\bibnamefont
  {Moussallam}},\ }\href {\doibase 10.1140/epjc/s2005-02316-8} {\bibfield
  {journal} {\bibinfo  {journal} {Eur. Phys. J.}\ }\textbf {\bibinfo {volume}
  {C42}},\ \bibinfo {pages} {403} (\bibinfo {year} {2005})},\ \Eprint
  {http://arxiv.org/abs/hep-ph/0505077} {arXiv:hep-ph/0505077 [hep-ph]}
  \BibitemShut {NoStop}%
\bibitem [{\citenamefont {Marciano}\ and\ \citenamefont
  {Sirlin}(1993)}]{Marciano:1993sh}%
  \BibitemOpen
  \bibfield  {author} {\bibinfo {author} {\bibfnamefont {W.~J.}\ \bibnamefont
  {Marciano}}\ and\ \bibinfo {author} {\bibfnamefont {A.}~\bibnamefont
  {Sirlin}},\ }\href {\doibase 10.1103/PhysRevLett.71.3629} {\bibfield
  {journal} {\bibinfo  {journal} {Phys. Rev. Lett.}\ }\textbf {\bibinfo
  {volume} {71}},\ \bibinfo {pages} {3629} (\bibinfo {year}
  {1993})}\BibitemShut {NoStop}%
\bibitem [{\citenamefont {Cirigliano}\ and\ \citenamefont
  {Neufeld}(2011)}]{Cirigliano:2011tm}%
  \BibitemOpen
  \bibfield  {author} {\bibinfo {author} {\bibfnamefont {V.}~\bibnamefont
  {Cirigliano}}\ and\ \bibinfo {author} {\bibfnamefont {H.}~\bibnamefont
  {Neufeld}},\ }\href {\doibase 10.1016/j.physletb.2011.04.038} {\bibfield
  {journal} {\bibinfo  {journal} {Phys. Lett.}\ }\textbf {\bibinfo {volume}
  {B700}},\ \bibinfo {pages} {7} (\bibinfo {year} {2011})},\ \Eprint
  {http://arxiv.org/abs/1102.0563} {arXiv:1102.0563 [hep-ph]} \BibitemShut
  {NoStop}%
\bibitem [{\citenamefont {Cirigliano}\ and\ \citenamefont
  {Rosell}(2007{\natexlab{b}})}]{Cirigliano:2007xi}%
  \BibitemOpen
  \bibfield  {author} {\bibinfo {author} {\bibfnamefont {V.}~\bibnamefont
  {Cirigliano}}\ and\ \bibinfo {author} {\bibfnamefont {I.}~\bibnamefont
  {Rosell}},\ }\href {\doibase 10.1103/PhysRevLett.99.231801} {\bibfield
  {journal} {\bibinfo  {journal} {Phys. Rev. Lett.}\ }\textbf {\bibinfo
  {volume} {99}},\ \bibinfo {pages} {231801} (\bibinfo {year}
  {2007}{\natexlab{b}})},\ \Eprint {http://arxiv.org/abs/0707.3439}
  {arXiv:0707.3439 [hep-ph]} \BibitemShut {NoStop}%
\bibitem [{\citenamefont {Moulson}(2014)}]{Moulson:2014cra}%
  \BibitemOpen
  \bibfield  {author} {\bibinfo {author} {\bibfnamefont {M.}~\bibnamefont
  {Moulson}},\ }in\ \href
  {http://inspirehep.net/record/1328784/files/arXiv:1411.5252.pdf} {\emph
  {\bibinfo {booktitle} {{8th International Workshop on the CKM Unitarity
  Triangle (CKM2014) Vienna, Austria, September 8-12, 2014}}}}\ (\bibinfo
  {year} {2014})\ \Eprint {http://arxiv.org/abs/1411.5252} {arXiv:1411.5252
  [hep-ex]} \BibitemShut {NoStop}%
\bibitem [{\citenamefont {Babusci}\ \emph {et~al.}(2014)\citenamefont {Babusci}
  \emph {et~al.}}]{Babusci:2014hxa}%
  \BibitemOpen
  \bibfield  {author} {\bibinfo {author} {\bibfnamefont {D.}~\bibnamefont
  {Babusci}} \emph {et~al.} (\bibinfo {collaboration} {KLOE KLOE-2}),\ }\href
  {\doibase 10.1016/j.physletb.2014.09.033} {\bibfield  {journal} {\bibinfo
  {journal} {Phys. Lett.}\ }\textbf {\bibinfo {volume} {B738}},\ \bibinfo
  {pages} {128} (\bibinfo {year} {2014})},\ \Eprint
  {http://arxiv.org/abs/1407.2028} {arXiv:1407.2028 [hep-ex]} \BibitemShut
  {NoStop}%
\bibitem [{\citenamefont {Antonelli}\ \emph
  {et~al.}(2010{\natexlab{a}})\citenamefont {Antonelli} \emph
  {et~al.}}]{Antonelli:2010yf}%
  \BibitemOpen
  \bibfield  {author} {\bibinfo {author} {\bibfnamefont {M.}~\bibnamefont
  {Antonelli}} \emph {et~al.} (\bibinfo {collaboration} {FlaviaNet Working
  Group on Kaon Decays}),\ }\href {\doibase 10.1140/epjc/s10052-010-1406-3}
  {\bibfield  {journal} {\bibinfo  {journal} {Eur. Phys. J.}\ }\textbf
  {\bibinfo {volume} {C69}},\ \bibinfo {pages} {399} (\bibinfo {year}
  {2010}{\natexlab{a}})},\ \Eprint {http://arxiv.org/abs/1005.2323}
  {arXiv:1005.2323 [hep-ph]} \BibitemShut {NoStop}%
\bibitem [{\citenamefont {Hashimoto}\ \emph {et~al.}(2015)\citenamefont
  {Hashimoto}, \citenamefont {Laiho},\ and\ \citenamefont
  {Sharpe}}]{PDGLQCD2015}%
  \BibitemOpen
  \bibfield  {author} {\bibinfo {author} {\bibfnamefont {S.}~\bibnamefont
  {Hashimoto}}, \bibinfo {author} {\bibfnamefont {J.}~\bibnamefont {Laiho}}, \
  and\ \bibinfo {author} {\bibfnamefont {S.~R.}\ \bibnamefont {Sharpe}},\
  }\href@noop {} {\enquote {\bibinfo {title} {{Lattice Quantum
  Chromodynamics}},}\ }\bibinfo {howpublished}
  {\url{http://pdg.lbl.gov/2015/reviews/rpp2015-rev-lattice-qcd.pdf}} (\bibinfo
  {year} {2015}),\ \bibinfo {note} {{review prepared for PDG 2015
  edition}}\BibitemShut {NoStop}%
\bibitem [{\citenamefont {Kronfeld}(2012)}]{Kronfeld:2012uk}%
  \BibitemOpen
  \bibfield  {author} {\bibinfo {author} {\bibfnamefont {A.~S.}\ \bibnamefont
  {Kronfeld}},\ }\href {\doibase 10.1146/annurev-nucl-102711-094942} {\bibfield
   {journal} {\bibinfo  {journal} {Ann. Rev. Nucl. Part. Sci.}\ }\textbf
  {\bibinfo {volume} {62}},\ \bibinfo {pages} {265} (\bibinfo {year} {2012})},\
  \Eprint {http://arxiv.org/abs/1203.1204} {arXiv:1203.1204 [hep-lat]}
  \BibitemShut {NoStop}%
\bibitem [{\citenamefont {Carrasco}\ \emph {et~al.}(2015)\citenamefont
  {Carrasco} \emph {et~al.}}]{Carrasco:2014poa}%
  \BibitemOpen
  \bibfield  {author} {\bibinfo {author} {\bibfnamefont {N.}~\bibnamefont
  {Carrasco}} \emph {et~al.} (\bibinfo {collaboration} {ETM}),\ }\href
  {\doibase 10.1103/PhysRevD.91.054507} {\bibfield  {journal} {\bibinfo
  {journal} {Phys. Rev.}\ }\textbf {\bibinfo {volume} {D91}},\ \bibinfo {pages}
  {054507} (\bibinfo {year} {2015})},\ \Eprint {http://arxiv.org/abs/1411.7908}
  {arXiv:1411.7908 [hep-lat]} \BibitemShut {NoStop}%
\bibitem [{\citenamefont {Bazavov}\ \emph
  {et~al.}(2014{\natexlab{a}})\citenamefont {Bazavov} \emph
  {et~al.}}]{Bazavov:2014wgs}%
  \BibitemOpen
  \bibfield  {author} {\bibinfo {author} {\bibfnamefont {A.}~\bibnamefont
  {Bazavov}} \emph {et~al.} (\bibinfo {collaboration} {Fermilab Lattice and
  MILC}),\ }\href {\doibase 10.1103/PhysRevD.90.074509} {\bibfield  {journal}
  {\bibinfo  {journal} {Phys. Rev.}\ }\textbf {\bibinfo {volume} {D90}},\
  \bibinfo {pages} {074509} (\bibinfo {year} {2014}{\natexlab{a}})},\ \Eprint
  {http://arxiv.org/abs/1407.3772} {arXiv:1407.3772 [hep-lat]} \BibitemShut
  {NoStop}%
\bibitem [{\citenamefont {Dowdall}\ \emph
  {et~al.}(2013{\natexlab{a}})\citenamefont {Dowdall}, \citenamefont {Davies},
  \citenamefont {Lepage},\ and\ \citenamefont {McNeile}}]{Dowdall:2013rya}%
  \BibitemOpen
  \bibfield  {author} {\bibinfo {author} {\bibfnamefont {R.}~\bibnamefont
  {Dowdall}}, \bibinfo {author} {\bibfnamefont {C.}~\bibnamefont {Davies}},
  \bibinfo {author} {\bibfnamefont {G.}~\bibnamefont {Lepage}}, \ and\ \bibinfo
  {author} {\bibfnamefont {C.}~\bibnamefont {McNeile}} (\bibinfo
  {collaboration} {HPQCD}),\ }\href {\doibase 10.1103/PhysRevD.88.074504}
  {\bibfield  {journal} {\bibinfo  {journal} {Phys.Rev.}\ }\textbf {\bibinfo
  {volume} {D88}},\ \bibinfo {pages} {074504} (\bibinfo {year}
  {2013}{\natexlab{a}})},\ \Eprint {http://arxiv.org/abs/1303.1670}
  {arXiv:1303.1670 [hep-lat]} \BibitemShut {NoStop}%
\bibitem [{\citenamefont {Blum}\ \emph {et~al.}(2014)\citenamefont {Blum} \emph
  {et~al.}}]{Blum:2014tka}%
  \BibitemOpen
  \bibfield  {author} {\bibinfo {author} {\bibfnamefont {T.}~\bibnamefont
  {Blum}} \emph {et~al.} (\bibinfo {collaboration} {RBC/UKQCD}),\ }\href@noop
  {} {\  (\bibinfo {year} {2014})},\ \Eprint {http://arxiv.org/abs/1411.7017}
  {arXiv:1411.7017 [hep-lat]} \BibitemShut {NoStop}%
\bibitem [{\citenamefont {Arthur}\ \emph {et~al.}(2013)\citenamefont {Arthur}
  \emph {et~al.}}]{Arthur:2012yc}%
  \BibitemOpen
  \bibfield  {author} {\bibinfo {author} {\bibfnamefont {R.}~\bibnamefont
  {Arthur}} \emph {et~al.} (\bibinfo {collaboration} {RBC/UKQCD}),\ }\href
  {\doibase 10.1103/PhysRevD.87.094514} {\bibfield  {journal} {\bibinfo
  {journal} {Phys. Rev.}\ }\textbf {\bibinfo {volume} {D87}},\ \bibinfo {pages}
  {094514} (\bibinfo {year} {2013})},\ \Eprint {http://arxiv.org/abs/1208.4412}
  {arXiv:1208.4412 [hep-lat]} \BibitemShut {NoStop}%
\bibitem [{\citenamefont {Laiho}\ and\ \citenamefont {Van~de
  Water}(2011)}]{Laiho:2011np}%
  \BibitemOpen
  \bibfield  {author} {\bibinfo {author} {\bibfnamefont {J.}~\bibnamefont
  {Laiho}}\ and\ \bibinfo {author} {\bibfnamefont {R.~S.}\ \bibnamefont {Van~de
  Water}},\ }\bibfield  {booktitle} {\emph {\bibinfo {booktitle} {{Proceedings,
  29th International Symposium on Lattice field theory (Lattice 2011)}}},\
  }\href@noop {} {\bibfield  {journal} {\bibinfo  {journal} {PoS}\ }\textbf
  {\bibinfo {volume} {LATTICE2011}},\ \bibinfo {pages} {293} (\bibinfo {year}
  {2011})},\ \Eprint {http://arxiv.org/abs/1112.4861} {arXiv:1112.4861
  [hep-lat]} \BibitemShut {NoStop}%
\bibitem [{\citenamefont {Bazavov}\ \emph {et~al.}(2010)\citenamefont {Bazavov}
  \emph {et~al.}}]{Bazavov:2010hj}%
  \BibitemOpen
  \bibfield  {author} {\bibinfo {author} {\bibfnamefont {A.}~\bibnamefont
  {Bazavov}} \emph {et~al.} (\bibinfo {collaboration} {MILC}),\ }\bibfield
  {booktitle} {\emph {\bibinfo {booktitle} {{Proceedings, 28th International
  Symposium on Lattice field theory (Lattice 2010)}}},\ }\href@noop {}
  {\bibfield  {journal} {\bibinfo  {journal} {PoS}\ }\textbf {\bibinfo {volume}
  {LATTICE2010}},\ \bibinfo {pages} {074} (\bibinfo {year} {2010})},\ \Eprint
  {http://arxiv.org/abs/1012.0868} {arXiv:1012.0868 [hep-lat]} \BibitemShut
  {NoStop}%
\bibitem [{\citenamefont {Durr}\ \emph {et~al.}(2010)\citenamefont {Durr},
  \citenamefont {Fodor}, \citenamefont {Hoelbling}, \citenamefont {Katz},
  \citenamefont {Krieg}, \citenamefont {Kurth}, \citenamefont {Lellouch},
  \citenamefont {Lippert}, \citenamefont {Ramos},\ and\ \citenamefont
  {Szabo}}]{Durr:2010hr}%
  \BibitemOpen
  \bibfield  {author} {\bibinfo {author} {\bibfnamefont {S.}~\bibnamefont
  {Durr}}, \bibinfo {author} {\bibfnamefont {Z.}~\bibnamefont {Fodor}},
  \bibinfo {author} {\bibfnamefont {C.}~\bibnamefont {Hoelbling}}, \bibinfo
  {author} {\bibfnamefont {S.~D.}\ \bibnamefont {Katz}}, \bibinfo {author}
  {\bibfnamefont {S.}~\bibnamefont {Krieg}}, \bibinfo {author} {\bibfnamefont
  {T.}~\bibnamefont {Kurth}}, \bibinfo {author} {\bibfnamefont
  {L.}~\bibnamefont {Lellouch}}, \bibinfo {author} {\bibfnamefont
  {T.}~\bibnamefont {Lippert}}, \bibinfo {author} {\bibfnamefont
  {A.}~\bibnamefont {Ramos}}, \ and\ \bibinfo {author} {\bibfnamefont {K.~K.}\
  \bibnamefont {Szabo}} (\bibinfo {collaboration} {BMW}),\ }\href {\doibase
  10.1103/PhysRevD.81.054507} {\bibfield  {journal} {\bibinfo  {journal} {Phys.
  Rev.}\ }\textbf {\bibinfo {volume} {D81}},\ \bibinfo {pages} {054507}
  (\bibinfo {year} {2010})},\ \Eprint {http://arxiv.org/abs/1001.4692}
  {arXiv:1001.4692 [hep-lat]} \BibitemShut {NoStop}%
\bibitem [{\citenamefont {Follana}\ \emph {et~al.}(2008)\citenamefont
  {Follana}, \citenamefont {Davies}, \citenamefont {Lepage},\ and\
  \citenamefont {Shigemitsu}}]{Follana:2007uv}%
  \BibitemOpen
  \bibfield  {author} {\bibinfo {author} {\bibfnamefont {E.}~\bibnamefont
  {Follana}}, \bibinfo {author} {\bibfnamefont {C.~T.~H.}\ \bibnamefont
  {Davies}}, \bibinfo {author} {\bibfnamefont {G.~P.}\ \bibnamefont {Lepage}},
  \ and\ \bibinfo {author} {\bibfnamefont {J.}~\bibnamefont {Shigemitsu}}
  (\bibinfo {collaboration} {HPQCD, UKQCD}),\ }\href {\doibase
  10.1103/PhysRevLett.100.062002} {\bibfield  {journal} {\bibinfo  {journal}
  {Phys. Rev. Lett.}\ }\textbf {\bibinfo {volume} {100}},\ \bibinfo {pages}
  {062002} (\bibinfo {year} {2008})},\ \Eprint {http://arxiv.org/abs/0706.1726}
  {arXiv:0706.1726 [hep-lat]} \BibitemShut {NoStop}%
\bibitem [{\citenamefont {Gasser}\ and\ \citenamefont
  {Leutwyler}(1985)}]{Gasser:1984gg}%
  \BibitemOpen
  \bibfield  {author} {\bibinfo {author} {\bibfnamefont {J.}~\bibnamefont
  {Gasser}}\ and\ \bibinfo {author} {\bibfnamefont {H.}~\bibnamefont
  {Leutwyler}},\ }\href {\doibase 10.1016/0550-3213(85)90492-4} {\bibfield
  {journal} {\bibinfo  {journal} {Nucl. Phys.}\ }\textbf {\bibinfo {volume}
  {B250}},\ \bibinfo {pages} {465} (\bibinfo {year} {1985})}\BibitemShut
  {NoStop}%
\bibitem [{\citenamefont {de~Divitiis}\ \emph {et~al.}(2012)\citenamefont
  {de~Divitiis} \emph {et~al.}}]{deDivitiis:2011eh}%
  \BibitemOpen
  \bibfield  {author} {\bibinfo {author} {\bibfnamefont {G.~M.}\ \bibnamefont
  {de~Divitiis}} \emph {et~al.},\ }\href {\doibase 10.1007/JHEP04(2012)124}
  {\bibfield  {journal} {\bibinfo  {journal} {JHEP}\ }\textbf {\bibinfo
  {volume} {04}},\ \bibinfo {pages} {124} (\bibinfo {year} {2012})},\ \Eprint
  {http://arxiv.org/abs/1110.6294} {arXiv:1110.6294 [hep-lat]} \BibitemShut
  {NoStop}%
\bibitem [{\citenamefont {Nobes}(2005)}]{Nobes:2005yh}%
  \BibitemOpen
  \bibfield  {author} {\bibinfo {author} {\bibfnamefont {M.}~\bibnamefont
  {Nobes}},\ }\href@noop {} {\  (\bibinfo {year} {2005})},\ \Eprint
  {http://arxiv.org/abs/hep-lat/0501009} {arXiv:hep-lat/0501009 [hep-lat]}
  \BibitemShut {NoStop}%
\bibitem [{\citenamefont {Bazavov}\ \emph {et~al.}(2016)\citenamefont {Bazavov}
  \emph {et~al.}}]{Bazavov:2016nty}%
  \BibitemOpen
  \bibfield  {author} {\bibinfo {author} {\bibfnamefont {A.}~\bibnamefont
  {Bazavov}} \emph {et~al.},\ }\href@noop {} {\  (\bibinfo {year} {2016})},\
  \Eprint {http://arxiv.org/abs/1602.03560} {arXiv:1602.03560 [hep-lat]}
  \BibitemShut {NoStop}%
\bibitem [{\citenamefont {Artuso}\ \emph {et~al.}(2005)\citenamefont {Artuso}
  \emph {et~al.}}]{Artuso:2005ym}%
  \BibitemOpen
  \bibfield  {author} {\bibinfo {author} {\bibfnamefont {M.}~\bibnamefont
  {Artuso}} \emph {et~al.} (\bibinfo {collaboration} {CLEO}),\ }\href {\doibase
  10.1103/PhysRevLett.95.251801} {\bibfield  {journal} {\bibinfo  {journal}
  {Phys. Rev. Lett.}\ }\textbf {\bibinfo {volume} {95}},\ \bibinfo {pages}
  {251801} (\bibinfo {year} {2005})},\ \Eprint
  {http://arxiv.org/abs/hep-ex/0508057} {arXiv:hep-ex/0508057 [hep-ex]}
  \BibitemShut {NoStop}%
\bibitem [{\citenamefont {Eisenstein}\ \emph {et~al.}(2008)\citenamefont
  {Eisenstein} \emph {et~al.}}]{Eisenstein:2008aa}%
  \BibitemOpen
  \bibfield  {author} {\bibinfo {author} {\bibfnamefont {B.~I.}\ \bibnamefont
  {Eisenstein}} \emph {et~al.} (\bibinfo {collaboration} {CLEO}),\ }\href
  {\doibase 10.1103/PhysRevD.78.052003} {\bibfield  {journal} {\bibinfo
  {journal} {Phys. Rev.}\ }\textbf {\bibinfo {volume} {D78}},\ \bibinfo {pages}
  {052003} (\bibinfo {year} {2008})},\ \Eprint {http://arxiv.org/abs/0806.2112}
  {arXiv:0806.2112 [hep-ex]} \BibitemShut {NoStop}%
\bibitem [{\citenamefont {Ablikim}\ \emph {et~al.}(2014)\citenamefont {Ablikim}
  \emph {et~al.}}]{Ablikim:2013uvu}%
  \BibitemOpen
  \bibfield  {author} {\bibinfo {author} {\bibfnamefont {M.}~\bibnamefont
  {Ablikim}} \emph {et~al.} (\bibinfo {collaboration} {BESIII}),\ }\href
  {\doibase 10.1103/PhysRevD.89.051104} {\bibfield  {journal} {\bibinfo
  {journal} {Phys. Rev.}\ }\textbf {\bibinfo {volume} {D89}},\ \bibinfo {pages}
  {051104} (\bibinfo {year} {2014})},\ \Eprint {http://arxiv.org/abs/1312.0374}
  {arXiv:1312.0374 [hep-ex]} \BibitemShut {NoStop}%
\bibitem [{\citenamefont {Artuso}\ \emph {et~al.}(2007)\citenamefont {Artuso}
  \emph {et~al.}}]{Artuso:2007zg}%
  \BibitemOpen
  \bibfield  {author} {\bibinfo {author} {\bibfnamefont {M.}~\bibnamefont
  {Artuso}} \emph {et~al.} (\bibinfo {collaboration} {CLEO}),\ }\href {\doibase
  10.1103/PhysRevLett.99.071802} {\bibfield  {journal} {\bibinfo  {journal}
  {Phys. Rev. Lett.}\ }\textbf {\bibinfo {volume} {99}},\ \bibinfo {pages}
  {071802} (\bibinfo {year} {2007})},\ \Eprint {http://arxiv.org/abs/0704.0629}
  {arXiv:0704.0629 [hep-ex]} \BibitemShut {NoStop}%
\bibitem [{\citenamefont {Alexander}\ \emph {et~al.}(2009)\citenamefont
  {Alexander} \emph {et~al.}}]{Alexander:2009ux}%
  \BibitemOpen
  \bibfield  {author} {\bibinfo {author} {\bibfnamefont {J.~P.}\ \bibnamefont
  {Alexander}} \emph {et~al.} (\bibinfo {collaboration} {CLEO}),\ }\href
  {\doibase 10.1103/PhysRevD.79.052001} {\bibfield  {journal} {\bibinfo
  {journal} {Phys. Rev.}\ }\textbf {\bibinfo {volume} {D79}},\ \bibinfo {pages}
  {052001} (\bibinfo {year} {2009})},\ \Eprint {http://arxiv.org/abs/0901.1216}
  {arXiv:0901.1216 [hep-ex]} \BibitemShut {NoStop}%
\bibitem [{\citenamefont {Zupanc}\ \emph {et~al.}(2013)\citenamefont {Zupanc}
  \emph {et~al.}}]{Zupanc:2013byn}%
  \BibitemOpen
  \bibfield  {author} {\bibinfo {author} {\bibfnamefont {A.}~\bibnamefont
  {Zupanc}} \emph {et~al.} (\bibinfo {collaboration} {Belle}),\ }\href
  {\doibase 10.1007/JHEP09(2013)139} {\bibfield  {journal} {\bibinfo  {journal}
  {JHEP}\ }\textbf {\bibinfo {volume} {1309}},\ \bibinfo {pages} {139}
  (\bibinfo {year} {2013})},\ \Eprint {http://arxiv.org/abs/1307.6240}
  {arXiv:1307.6240 [hep-ex]} \BibitemShut {NoStop}%
\bibitem [{\citenamefont {Naik}\ \emph {et~al.}(2009)\citenamefont {Naik} \emph
  {et~al.}}]{Naik:2009tk}%
  \BibitemOpen
  \bibfield  {author} {\bibinfo {author} {\bibfnamefont {P.}~\bibnamefont
  {Naik}} \emph {et~al.} (\bibinfo {collaboration} {CLEO}),\ }\href {\doibase
  10.1103/PhysRevD.80.112004} {\bibfield  {journal} {\bibinfo  {journal} {Phys.
  Rev.}\ }\textbf {\bibinfo {volume} {D80}},\ \bibinfo {pages} {112004}
  (\bibinfo {year} {2009})},\ \Eprint {http://arxiv.org/abs/0910.3602}
  {arXiv:0910.3602 [hep-ex]} \BibitemShut {NoStop}%
\bibitem [{\citenamefont {Ecklund}\ \emph {et~al.}(2008)\citenamefont {Ecklund}
  \emph {et~al.}}]{Ecklund:2007aa}%
  \BibitemOpen
  \bibfield  {author} {\bibinfo {author} {\bibfnamefont {K.~M.}\ \bibnamefont
  {Ecklund}} \emph {et~al.} (\bibinfo {collaboration} {CLEO}),\ }\href
  {\doibase 10.1103/PhysRevLett.100.161801} {\bibfield  {journal} {\bibinfo
  {journal} {Phys. Rev. Lett.}\ }\textbf {\bibinfo {volume} {100}},\ \bibinfo
  {pages} {161801} (\bibinfo {year} {2008})},\ \Eprint
  {http://arxiv.org/abs/0712.1175} {arXiv:0712.1175 [hep-ex]} \BibitemShut
  {NoStop}%
\bibitem [{\citenamefont {Onyisi}\ \emph {et~al.}(2009)\citenamefont {Onyisi}
  \emph {et~al.}}]{Onyisi:2009th}%
  \BibitemOpen
  \bibfield  {author} {\bibinfo {author} {\bibfnamefont {P.~U.~E.}\
  \bibnamefont {Onyisi}} \emph {et~al.} (\bibinfo {collaboration} {CLEO}),\
  }\href {\doibase 10.1103/PhysRevD.79.052002} {\bibfield  {journal} {\bibinfo
  {journal} {Phys. Rev.}\ }\textbf {\bibinfo {volume} {D79}},\ \bibinfo {pages}
  {052002} (\bibinfo {year} {2009})},\ \Eprint {http://arxiv.org/abs/0901.1147}
  {arXiv:0901.1147 [hep-ex]} \BibitemShut {NoStop}%
\bibitem [{\citenamefont {del Amo~Sanchez}\ \emph {et~al.}(2010)\citenamefont
  {del Amo~Sanchez} \emph {et~al.}}]{delAmoSanchez:2010jg}%
  \BibitemOpen
  \bibfield  {author} {\bibinfo {author} {\bibfnamefont {P.}~\bibnamefont {del
  Amo~Sanchez}} \emph {et~al.} (\bibinfo {collaboration} {BaBar}),\ }\href
  {\doibase 10.1103/PhysRevD.82.091103, 10.1103/PhysRevD.91.019901} {\bibfield
  {journal} {\bibinfo  {journal} {Phys. Rev.}\ }\textbf {\bibinfo {volume}
  {D82}},\ \bibinfo {pages} {091103} (\bibinfo {year} {2010})},\ \bibinfo
  {note} {[Erratum: Phys. Rev.D91,no.1,019901(2015)]},\ \Eprint
  {http://arxiv.org/abs/1008.4080} {arXiv:1008.4080 [hep-ex]} \BibitemShut
  {NoStop}%
\bibitem [{\citenamefont {Alexander}\ \emph {et~al.}(2008)\citenamefont
  {Alexander} \emph {et~al.}}]{Alexander:2008aa}%
  \BibitemOpen
  \bibfield  {author} {\bibinfo {author} {\bibfnamefont {J.~P.}\ \bibnamefont
  {Alexander}} \emph {et~al.} (\bibinfo {collaboration} {CLEO}),\ }\href
  {\doibase 10.1103/PhysRevLett.100.161804} {\bibfield  {journal} {\bibinfo
  {journal} {Phys. Rev. Lett.}\ }\textbf {\bibinfo {volume} {100}},\ \bibinfo
  {pages} {161804} (\bibinfo {year} {2008})},\ \Eprint
  {http://arxiv.org/abs/0801.0680} {arXiv:0801.0680 [hep-ex]} \BibitemShut
  {NoStop}%
\bibitem [{\citenamefont {Aubert}\ \emph {et~al.}(2007)\citenamefont {Aubert}
  \emph {et~al.}}]{Aubert:2006sd}%
  \BibitemOpen
  \bibfield  {author} {\bibinfo {author} {\bibfnamefont {B.}~\bibnamefont
  {Aubert}} \emph {et~al.} (\bibinfo {collaboration} {BaBar}),\ }\href
  {\doibase 10.1103/PhysRevLett.98.141801} {\bibfield  {journal} {\bibinfo
  {journal} {Phys. Rev. Lett.}\ }\textbf {\bibinfo {volume} {98}},\ \bibinfo
  {pages} {141801} (\bibinfo {year} {2007})},\ \Eprint
  {http://arxiv.org/abs/hep-ex/0607094} {arXiv:hep-ex/0607094 [hep-ex]}
  \BibitemShut {NoStop}%
\bibitem [{\citenamefont {Lees}\ \emph {et~al.}(2010)\citenamefont {Lees} \emph
  {et~al.}}]{Lees:2010qj}%
  \BibitemOpen
  \bibfield  {author} {\bibinfo {author} {\bibfnamefont {J.~P.}\ \bibnamefont
  {Lees}} \emph {et~al.} (\bibinfo {collaboration} {BaBar}),\ }\href@noop {} {\
   (\bibinfo {year} {2010})},\ \Eprint {http://arxiv.org/abs/1003.3063}
  {arXiv:1003.3063 [hep-ex]} \BibitemShut {NoStop}%
\bibitem [{\citenamefont {Burdman}\ \emph {et~al.}(1995)\citenamefont
  {Burdman}, \citenamefont {Goldman},\ and\ \citenamefont
  {Wyler}}]{Burdman:1994ip}%
  \BibitemOpen
  \bibfield  {author} {\bibinfo {author} {\bibfnamefont {G.}~\bibnamefont
  {Burdman}}, \bibinfo {author} {\bibfnamefont {J.~T.}\ \bibnamefont
  {Goldman}}, \ and\ \bibinfo {author} {\bibfnamefont {D.}~\bibnamefont
  {Wyler}},\ }\href {\doibase 10.1103/PhysRevD.51.111} {\bibfield  {journal}
  {\bibinfo  {journal} {Phys. Rev.}\ }\textbf {\bibinfo {volume} {D51}},\
  \bibinfo {pages} {111} (\bibinfo {year} {1995})},\ \Eprint
  {http://arxiv.org/abs/hep-ph/9405425} {arXiv:hep-ph/9405425 [hep-ph]}
  \BibitemShut {NoStop}%
\bibitem [{\citenamefont {Atwood}\ \emph {et~al.}(1996)\citenamefont {Atwood},
  \citenamefont {Eilam},\ and\ \citenamefont {Soni}}]{Atwood:1994za}%
  \BibitemOpen
  \bibfield  {author} {\bibinfo {author} {\bibfnamefont {D.}~\bibnamefont
  {Atwood}}, \bibinfo {author} {\bibfnamefont {G.}~\bibnamefont {Eilam}}, \
  and\ \bibinfo {author} {\bibfnamefont {A.}~\bibnamefont {Soni}},\ }\href
  {\doibase 10.1142/S0217732396001090} {\bibfield  {journal} {\bibinfo
  {journal} {Mod. Phys. Lett.}\ }\textbf {\bibinfo {volume} {A11}},\ \bibinfo
  {pages} {1061} (\bibinfo {year} {1996})},\ \Eprint
  {http://arxiv.org/abs/hep-ph/9411367} {arXiv:hep-ph/9411367 [hep-ph]}
  \BibitemShut {NoStop}%
\bibitem [{\citenamefont {Colangelo}\ \emph {et~al.}(1996)\citenamefont
  {Colangelo}, \citenamefont {De~Fazio},\ and\ \citenamefont
  {Nardulli}}]{Colangelo:1995sm}%
  \BibitemOpen
  \bibfield  {author} {\bibinfo {author} {\bibfnamefont {P.}~\bibnamefont
  {Colangelo}}, \bibinfo {author} {\bibfnamefont {F.}~\bibnamefont {De~Fazio}},
  \ and\ \bibinfo {author} {\bibfnamefont {G.}~\bibnamefont {Nardulli}},\
  }\href {\doibase 10.1016/0370-2693(96)00079-2} {\bibfield  {journal}
  {\bibinfo  {journal} {Phys. Lett.}\ }\textbf {\bibinfo {volume} {B372}},\
  \bibinfo {pages} {331} (\bibinfo {year} {1996})},\ \Eprint
  {http://arxiv.org/abs/hep-ph/9506332} {arXiv:hep-ph/9506332 [hep-ph]}
  \BibitemShut {NoStop}%
\bibitem [{\citenamefont {Khodjamirian}\ \emph {et~al.}(1995)\citenamefont
  {Khodjamirian}, \citenamefont {Stoll},\ and\ \citenamefont
  {Wyler}}]{Khodjamirian:1995uc}%
  \BibitemOpen
  \bibfield  {author} {\bibinfo {author} {\bibfnamefont {A.}~\bibnamefont
  {Khodjamirian}}, \bibinfo {author} {\bibfnamefont {G.}~\bibnamefont {Stoll}},
  \ and\ \bibinfo {author} {\bibfnamefont {D.}~\bibnamefont {Wyler}},\ }\href
  {\doibase 10.1016/0370-2693(95)00972-N} {\bibfield  {journal} {\bibinfo
  {journal} {Phys. Lett.}\ }\textbf {\bibinfo {volume} {B358}},\ \bibinfo
  {pages} {129} (\bibinfo {year} {1995})},\ \Eprint
  {http://arxiv.org/abs/hep-ph/9506242} {arXiv:hep-ph/9506242 [hep-ph]}
  \BibitemShut {NoStop}%
\bibitem [{\citenamefont {Eilam}\ \emph {et~al.}(1995)\citenamefont {Eilam},
  \citenamefont {Halperin},\ and\ \citenamefont {Mendel}}]{Eilam:1995zv}%
  \BibitemOpen
  \bibfield  {author} {\bibinfo {author} {\bibfnamefont {G.}~\bibnamefont
  {Eilam}}, \bibinfo {author} {\bibfnamefont {I.~E.}\ \bibnamefont {Halperin}},
  \ and\ \bibinfo {author} {\bibfnamefont {R.~R.}\ \bibnamefont {Mendel}},\
  }\href {\doibase 10.1016/0370-2693(95)01088-8} {\bibfield  {journal}
  {\bibinfo  {journal} {Phys. Lett.}\ }\textbf {\bibinfo {volume} {B361}},\
  \bibinfo {pages} {137} (\bibinfo {year} {1995})},\ \Eprint
  {http://arxiv.org/abs/hep-ph/9506264} {arXiv:hep-ph/9506264 [hep-ph]}
  \BibitemShut {NoStop}%
\bibitem [{\citenamefont {Geng}\ \emph {et~al.}(1998)\citenamefont {Geng},
  \citenamefont {Lih},\ and\ \citenamefont {Zhang}}]{Geng:1997ws}%
  \BibitemOpen
  \bibfield  {author} {\bibinfo {author} {\bibfnamefont {C.~Q.}\ \bibnamefont
  {Geng}}, \bibinfo {author} {\bibfnamefont {C.~C.}\ \bibnamefont {Lih}}, \
  and\ \bibinfo {author} {\bibfnamefont {W.-M.}\ \bibnamefont {Zhang}},\ }\href
  {\doibase 10.1103/PhysRevD.57.5697} {\bibfield  {journal} {\bibinfo
  {journal} {Phys. Rev.}\ }\textbf {\bibinfo {volume} {D57}},\ \bibinfo {pages}
  {5697} (\bibinfo {year} {1998})},\ \Eprint
  {http://arxiv.org/abs/hep-ph/9710323} {arXiv:hep-ph/9710323 [hep-ph]}
  \BibitemShut {NoStop}%
\bibitem [{\citenamefont {Geng}\ \emph {et~al.}(2000)\citenamefont {Geng},
  \citenamefont {Lih},\ and\ \citenamefont {Zhang}}]{Geng:2000if}%
  \BibitemOpen
  \bibfield  {author} {\bibinfo {author} {\bibfnamefont {C.~Q.}\ \bibnamefont
  {Geng}}, \bibinfo {author} {\bibfnamefont {C.~C.}\ \bibnamefont {Lih}}, \
  and\ \bibinfo {author} {\bibfnamefont {W.-M.}\ \bibnamefont {Zhang}},\ }\href
  {\doibase 10.1142/S021773230000267X} {\bibfield  {journal} {\bibinfo
  {journal} {Mod. Phys. Lett.}\ }\textbf {\bibinfo {volume} {A15}},\ \bibinfo
  {pages} {2087} (\bibinfo {year} {2000})},\ \Eprint
  {http://arxiv.org/abs/hep-ph/0012066} {arXiv:hep-ph/0012066 [hep-ph]}
  \BibitemShut {NoStop}%
\bibitem [{\citenamefont {Korchemsky}\ \emph {et~al.}(2000)\citenamefont
  {Korchemsky}, \citenamefont {Pirjol},\ and\ \citenamefont
  {Yan}}]{Korchemsky:1999qb}%
  \BibitemOpen
  \bibfield  {author} {\bibinfo {author} {\bibfnamefont {G.~P.}\ \bibnamefont
  {Korchemsky}}, \bibinfo {author} {\bibfnamefont {D.}~\bibnamefont {Pirjol}},
  \ and\ \bibinfo {author} {\bibfnamefont {T.-M.}\ \bibnamefont {Yan}},\ }\href
  {\doibase 10.1103/PhysRevD.61.114510} {\bibfield  {journal} {\bibinfo
  {journal} {Phys. Rev.}\ }\textbf {\bibinfo {volume} {D61}},\ \bibinfo {pages}
  {114510} (\bibinfo {year} {2000})},\ \Eprint
  {http://arxiv.org/abs/hep-ph/9911427} {arXiv:hep-ph/9911427 [hep-ph]}
  \BibitemShut {NoStop}%
\bibitem [{\citenamefont {Hwang}(2006)}]{Hwang:2005uk}%
  \BibitemOpen
  \bibfield  {author} {\bibinfo {author} {\bibfnamefont {C.-W.}\ \bibnamefont
  {Hwang}},\ }\href {\doibase 10.1140/epjc/s2006-02510-2} {\bibfield  {journal}
  {\bibinfo  {journal} {Eur. Phys. J.}\ }\textbf {\bibinfo {volume} {C46}},\
  \bibinfo {pages} {379} (\bibinfo {year} {2006})},\ \Eprint
  {http://arxiv.org/abs/hep-ph/0512006} {arXiv:hep-ph/0512006 [hep-ph]}
  \BibitemShut {NoStop}%
\bibitem [{\citenamefont {Lu}\ and\ \citenamefont {Song}(2003)}]{Lu:2002mn}%
  \BibitemOpen
  \bibfield  {author} {\bibinfo {author} {\bibfnamefont {C.-D.}\ \bibnamefont
  {Lu}}\ and\ \bibinfo {author} {\bibfnamefont {G.-L.}\ \bibnamefont {Song}},\
  }\href {\doibase 10.1016/S0370-2693(03)00549-5} {\bibfield  {journal}
  {\bibinfo  {journal} {Phys. Lett.}\ }\textbf {\bibinfo {volume} {B562}},\
  \bibinfo {pages} {75} (\bibinfo {year} {2003})},\ \Eprint
  {http://arxiv.org/abs/hep-ph/0212363} {arXiv:hep-ph/0212363 [hep-ph]}
  \BibitemShut {NoStop}%
\bibitem [{\citenamefont {Yang}\ \emph {et~al.}(2015)\citenamefont {Yang} \emph
  {et~al.}}]{Yang:2014sea}%
  \BibitemOpen
  \bibfield  {author} {\bibinfo {author} {\bibfnamefont {Y.-B.}\ \bibnamefont
  {Yang}} \emph {et~al.} (\bibinfo {collaboration} {$\chi$QCD}),\ }\href
  {\doibase 10.1103/PhysRevD.92.034517} {\bibfield  {journal} {\bibinfo
  {journal} {Phys. Rev.}\ }\textbf {\bibinfo {volume} {D92}},\ \bibinfo {pages}
  {034517} (\bibinfo {year} {2015})},\ \Eprint {http://arxiv.org/abs/1410.3343}
  {arXiv:1410.3343 [hep-lat]} \BibitemShut {NoStop}%
\bibitem [{\citenamefont {Na}\ \emph {et~al.}(2012{\natexlab{a}})\citenamefont
  {Na}, \citenamefont {Davies}, \citenamefont {Follana}, \citenamefont
  {Lepage},\ and\ \citenamefont {Shigemitsu}}]{Na:2012iu}%
  \BibitemOpen
  \bibfield  {author} {\bibinfo {author} {\bibfnamefont {H.}~\bibnamefont
  {Na}}, \bibinfo {author} {\bibfnamefont {C.~T.~H.}\ \bibnamefont {Davies}},
  \bibinfo {author} {\bibfnamefont {E.}~\bibnamefont {Follana}}, \bibinfo
  {author} {\bibfnamefont {G.~P.}\ \bibnamefont {Lepage}}, \ and\ \bibinfo
  {author} {\bibfnamefont {J.}~\bibnamefont {Shigemitsu}} (\bibinfo
  {collaboration} {HPQCD}),\ }\href {\doibase 10.1103/PhysRevD.86.054510}
  {\bibfield  {journal} {\bibinfo  {journal} {Phys. Rev.}\ }\textbf {\bibinfo
  {volume} {D86}},\ \bibinfo {pages} {054510} (\bibinfo {year}
  {2012}{\natexlab{a}})},\ \Eprint {http://arxiv.org/abs/1206.4936}
  {arXiv:1206.4936 [hep-lat]} \BibitemShut {NoStop}%
\bibitem [{\citenamefont {Bazavov}\ \emph {et~al.}(2012)\citenamefont {Bazavov}
  \emph {et~al.}}]{Bazavov:2011aa}%
  \BibitemOpen
  \bibfield  {author} {\bibinfo {author} {\bibfnamefont {A.}~\bibnamefont
  {Bazavov}} \emph {et~al.} (\bibinfo {collaboration} {Fermilab Lattice and
  MILC}),\ }\href {\doibase 10.1103/PhysRevD.85.114506} {\bibfield  {journal}
  {\bibinfo  {journal} {Phys. Rev.}\ }\textbf {\bibinfo {volume} {D85}},\
  \bibinfo {pages} {114506} (\bibinfo {year} {2012})},\ \Eprint
  {http://arxiv.org/abs/1112.3051} {arXiv:1112.3051 [hep-lat]} \BibitemShut
  {NoStop}%
\bibitem [{\citenamefont {Davies}\ \emph {et~al.}(2010)\citenamefont {Davies},
  \citenamefont {McNeile}, \citenamefont {Follana}, \citenamefont {Lepage},
  \citenamefont {Na},\ and\ \citenamefont {Shigemitsu}}]{Davies:2010ip}%
  \BibitemOpen
  \bibfield  {author} {\bibinfo {author} {\bibfnamefont {C.~T.~H.}\
  \bibnamefont {Davies}}, \bibinfo {author} {\bibfnamefont {C.}~\bibnamefont
  {McNeile}}, \bibinfo {author} {\bibfnamefont {E.}~\bibnamefont {Follana}},
  \bibinfo {author} {\bibfnamefont {G.~P.}\ \bibnamefont {Lepage}}, \bibinfo
  {author} {\bibfnamefont {H.}~\bibnamefont {Na}}, \ and\ \bibinfo {author}
  {\bibfnamefont {J.}~\bibnamefont {Shigemitsu}} (\bibinfo {collaboration}
  {HPQCD}),\ }\href {\doibase 10.1103/PhysRevD.82.114504} {\bibfield  {journal}
  {\bibinfo  {journal} {Phys. Rev.}\ }\textbf {\bibinfo {volume} {D82}},\
  \bibinfo {pages} {114504} (\bibinfo {year} {2010})},\ \Eprint
  {http://arxiv.org/abs/1008.4018} {arXiv:1008.4018 [hep-lat]} \BibitemShut
  {NoStop}%
\bibitem [{\citenamefont {Wang}(2015)}]{Wang:2015mxa}%
  \BibitemOpen
  \bibfield  {author} {\bibinfo {author} {\bibfnamefont {Z.-G.}\ \bibnamefont
  {Wang}},\ }\href {\doibase 10.1140/epjc/s10052-015-3653-9} {\bibfield
  {journal} {\bibinfo  {journal} {Eur. Phys. J.}\ }\textbf {\bibinfo {volume}
  {C75}},\ \bibinfo {pages} {427} (\bibinfo {year} {2015})},\ \Eprint
  {http://arxiv.org/abs/1506.01993} {arXiv:1506.01993 [hep-ph]} \BibitemShut
  {NoStop}%
\bibitem [{\citenamefont {Gelhausen}\ \emph {et~al.}(2013)\citenamefont
  {Gelhausen}, \citenamefont {Khodjamirian}, \citenamefont {Pivovarov},\ and\
  \citenamefont {Rosenthal}}]{Gelhausen:2013wia}%
  \BibitemOpen
  \bibfield  {author} {\bibinfo {author} {\bibfnamefont {P.}~\bibnamefont
  {Gelhausen}}, \bibinfo {author} {\bibfnamefont {A.}~\bibnamefont
  {Khodjamirian}}, \bibinfo {author} {\bibfnamefont {A.~A.}\ \bibnamefont
  {Pivovarov}}, \ and\ \bibinfo {author} {\bibfnamefont {D.}~\bibnamefont
  {Rosenthal}},\ }\href {\doibase 10.1103/PhysRevD.88.014015,
  10.1103/PhysRevD.91.099901, 10.1103/PhysRevD.89.099901} {\bibfield  {journal}
  {\bibinfo  {journal} {Phys.Rev.}\ }\textbf {\bibinfo {volume} {D88}},\
  \bibinfo {pages} {014015} (\bibinfo {year} {2013})},\ \Eprint
  {http://arxiv.org/abs/1305.5432} {arXiv:1305.5432 [hep-ph]} \BibitemShut
  {NoStop}%
\bibitem [{\citenamefont {Narison}(2013)}]{Narison:2012xy}%
  \BibitemOpen
  \bibfield  {author} {\bibinfo {author} {\bibfnamefont {S.}~\bibnamefont
  {Narison}},\ }\href {\doibase 10.1016/j.physletb.2012.10.057} {\bibfield
  {journal} {\bibinfo  {journal} {Phys. Lett.}\ }\textbf {\bibinfo {volume}
  {B718}},\ \bibinfo {pages} {1321} (\bibinfo {year} {2013})},\ \Eprint
  {http://arxiv.org/abs/1209.2023} {arXiv:1209.2023 [hep-ph]} \BibitemShut
  {NoStop}%
\bibitem [{\citenamefont {Lucha}\ \emph {et~al.}(2011)\citenamefont {Lucha},
  \citenamefont {Melikhov},\ and\ \citenamefont {Simula}}]{Lucha:2011zp}%
  \BibitemOpen
  \bibfield  {author} {\bibinfo {author} {\bibfnamefont {W.}~\bibnamefont
  {Lucha}}, \bibinfo {author} {\bibfnamefont {D.}~\bibnamefont {Melikhov}}, \
  and\ \bibinfo {author} {\bibfnamefont {S.}~\bibnamefont {Simula}},\ }\href
  {\doibase 10.1016/j.physletb.2011.05.031} {\bibfield  {journal} {\bibinfo
  {journal} {Phys. Lett.}\ }\textbf {\bibinfo {volume} {B701}},\ \bibinfo
  {pages} {82} (\bibinfo {year} {2011})},\ \Eprint
  {http://arxiv.org/abs/1101.5986} {arXiv:1101.5986 [hep-ph]} \BibitemShut
  {NoStop}%
\bibitem [{\citenamefont {Hwang}(2010)}]{Hwang:2009qz}%
  \BibitemOpen
  \bibfield  {author} {\bibinfo {author} {\bibfnamefont {C.-W.}\ \bibnamefont
  {Hwang}},\ }\href {\doibase 10.1103/PhysRevD.81.054022} {\bibfield  {journal}
  {\bibinfo  {journal} {Phys. Rev.}\ }\textbf {\bibinfo {volume} {D81}},\
  \bibinfo {pages} {054022} (\bibinfo {year} {2010})},\ \Eprint
  {http://arxiv.org/abs/0910.0145} {arXiv:0910.0145 [hep-ph]} \BibitemShut
  {NoStop}%
\bibitem [{\citenamefont {Amsler}\ \emph {et~al.}(2008)\citenamefont {Amsler}
  \emph {et~al.}}]{Amsler:Update}%
  \BibitemOpen
  \bibfield  {author} {\bibinfo {author} {\bibfnamefont {C.}~\bibnamefont
  {Amsler}} \emph {et~al.} (\bibinfo {collaboration} {Particle Data Group}),\
  }\href {\doibase 10.1016/j.physletb.2008.07.018} {\bibfield  {journal}
  {\bibinfo  {journal} {Phys. Lett.}\ }\textbf {\bibinfo {volume} {B667}},\
  \bibinfo {pages} {1} (\bibinfo {year} {2008})},\ \bibinfo {note} {{and 2009
  partial update for the 2010 edition}}\BibitemShut {NoStop}%
\bibitem [{\citenamefont {Laiho}\ \emph {et~al.}(2010)\citenamefont {Laiho},
  \citenamefont {Lunghi},\ and\ \citenamefont {Van~de Water}}]{Laiho:2009eu}%
  \BibitemOpen
  \bibfield  {author} {\bibinfo {author} {\bibfnamefont {J.}~\bibnamefont
  {Laiho}}, \bibinfo {author} {\bibfnamefont {E.}~\bibnamefont {Lunghi}}, \
  and\ \bibinfo {author} {\bibfnamefont {R.~S.}\ \bibnamefont {Van~de Water}},\
  }\href {\doibase 10.1103/PhysRevD.81.034503} {\bibfield  {journal} {\bibinfo
  {journal} {Phys. Rev.}\ }\textbf {\bibinfo {volume} {D81}},\ \bibinfo {pages}
  {034503} (\bibinfo {year} {2010})},\ \Eprint {http://arxiv.org/abs/0910.2928}
  {arXiv:0910.2928 [hep-ph]} \BibitemShut {NoStop}%
\bibitem [{\citenamefont {Schmelling}(1995)}]{Schmelling:1994pz}%
  \BibitemOpen
  \bibfield  {author} {\bibinfo {author} {\bibfnamefont {M.}~\bibnamefont
  {Schmelling}},\ }\href {\doibase 10.1088/0031-8949/51/6/002} {\bibfield
  {journal} {\bibinfo  {journal} {Phys. Scripta}\ }\textbf {\bibinfo {volume}
  {51}},\ \bibinfo {pages} {676} (\bibinfo {year} {1995})}\BibitemShut
  {NoStop}%
\bibitem [{\citenamefont {Vladikas}(2015)}]{Vladikas:2015bra}%
  \BibitemOpen
  \bibfield  {author} {\bibinfo {author} {\bibfnamefont {A.}~\bibnamefont
  {Vladikas}}\ }(\bibinfo {year} {2015})\ \Eprint
  {http://arxiv.org/abs/1509.01155} {arXiv:1509.01155 [hep-lat]} \BibitemShut
  {NoStop}%
\bibitem [{\citenamefont {Ikado}\ \emph {et~al.}(2006)\citenamefont {Ikado}
  \emph {et~al.}}]{Ikado:2006un}%
  \BibitemOpen
  \bibfield  {author} {\bibinfo {author} {\bibfnamefont {K.}~\bibnamefont
  {Ikado}} \emph {et~al.} (\bibinfo {collaboration} {Belle}),\ }\href {\doibase
  10.1103/PhysRevLett.97.251802} {\bibfield  {journal} {\bibinfo  {journal}
  {Phys. Rev. Lett.}\ }\textbf {\bibinfo {volume} {97}},\ \bibinfo {pages}
  {251802} (\bibinfo {year} {2006})},\ \Eprint
  {http://arxiv.org/abs/hep-ex/0604018} {arXiv:hep-ex/0604018 [hep-ex]}
  \BibitemShut {NoStop}%
\bibitem [{\citenamefont {Adachi}\ \emph {et~al.}(2013)\citenamefont {Adachi}
  \emph {et~al.}}]{Adachi:2012mm}%
  \BibitemOpen
  \bibfield  {author} {\bibinfo {author} {\bibfnamefont {I.}~\bibnamefont
  {Adachi}} \emph {et~al.} (\bibinfo {collaboration} {Belle}),\ }\href
  {\doibase 10.1103/PhysRevLett.110.131801} {\bibfield  {journal} {\bibinfo
  {journal} {Phys. Rev. Lett.}\ }\textbf {\bibinfo {volume} {110}},\ \bibinfo
  {pages} {131801} (\bibinfo {year} {2013})},\ \Eprint
  {http://arxiv.org/abs/1208.4678} {arXiv:1208.4678 [hep-ex]} \BibitemShut
  {NoStop}%
\bibitem [{\citenamefont {Kronenbitter}\ \emph {et~al.}(2015)\citenamefont
  {Kronenbitter} \emph {et~al.}}]{Kronenbitter:2015kls}%
  \BibitemOpen
  \bibfield  {author} {\bibinfo {author} {\bibfnamefont {B.}~\bibnamefont
  {Kronenbitter}} \emph {et~al.} (\bibinfo {collaboration} {Belle}),\ }\href
  {\doibase 10.1103/PhysRevD.92.051102} {\bibfield  {journal} {\bibinfo
  {journal} {Phys. Rev.}\ }\textbf {\bibinfo {volume} {D92}},\ \bibinfo {pages}
  {051102} (\bibinfo {year} {2015})},\ \Eprint
  {http://arxiv.org/abs/1503.05613} {arXiv:1503.05613 [hep-ex]} \BibitemShut
  {NoStop}%
\bibitem [{\citenamefont {Lees}\ \emph {et~al.}(2013)\citenamefont {Lees} \emph
  {et~al.}}]{Lees:2012ju}%
  \BibitemOpen
  \bibfield  {author} {\bibinfo {author} {\bibfnamefont {J.~P.}\ \bibnamefont
  {Lees}} \emph {et~al.} (\bibinfo {collaboration} {BaBar}),\ }\href {\doibase
  10.1103/PhysRevD.88.031102} {\bibfield  {journal} {\bibinfo  {journal} {Phys.
  Rev.}\ }\textbf {\bibinfo {volume} {D88}},\ \bibinfo {pages} {031102}
  (\bibinfo {year} {2013})},\ \Eprint {http://arxiv.org/abs/1207.0698}
  {arXiv:1207.0698 [hep-ex]} \BibitemShut {NoStop}%
\bibitem [{\citenamefont {Aubert}\ \emph {et~al.}(2010)\citenamefont {Aubert}
  \emph {et~al.}}]{Aubert:2009wt}%
  \BibitemOpen
  \bibfield  {author} {\bibinfo {author} {\bibfnamefont {B.}~\bibnamefont
  {Aubert}} \emph {et~al.} (\bibinfo {collaboration} {BaBar}),\ }\href
  {\doibase 10.1103/PhysRevD.81.051101} {\bibfield  {journal} {\bibinfo
  {journal} {Phys. Rev.}\ }\textbf {\bibinfo {volume} {D81}},\ \bibinfo {pages}
  {051101} (\bibinfo {year} {2010})},\ \Eprint {http://arxiv.org/abs/0912.2453}
  {arXiv:0912.2453 [hep-ex]} \BibitemShut {NoStop}%
\bibitem [{\citenamefont {Carrasco}\ \emph {et~al.}(2014)\citenamefont
  {Carrasco} \emph {et~al.}}]{Carrasco:2013naa}%
  \BibitemOpen
  \bibfield  {author} {\bibinfo {author} {\bibfnamefont {N.}~\bibnamefont
  {Carrasco}} \emph {et~al.} (\bibinfo {collaboration} {ETM}),\ }\bibfield
  {booktitle} {\emph {\bibinfo {booktitle} {{Proceedings, 31st International
  Symposium on Lattice Field Theory (Lattice 2013)}}},\ }\href@noop {}
  {\bibfield  {journal} {\bibinfo  {journal} {PoS}\ }\textbf {\bibinfo {volume}
  {LATTICE2013}},\ \bibinfo {pages} {313} (\bibinfo {year} {2014})},\ \Eprint
  {http://arxiv.org/abs/1311.2837} {arXiv:1311.2837 [hep-lat]} \BibitemShut
  {NoStop}%
\bibitem [{\citenamefont {Dowdall}\ \emph
  {et~al.}(2013{\natexlab{b}})\citenamefont {Dowdall}, \citenamefont {Davies},
  \citenamefont {Horgan}, \citenamefont {Monahan},\ and\ \citenamefont
  {Shigemitsu}}]{Dowdall:2013tga}%
  \BibitemOpen
  \bibfield  {author} {\bibinfo {author} {\bibfnamefont {R.}~\bibnamefont
  {Dowdall}}, \bibinfo {author} {\bibfnamefont {C.}~\bibnamefont {Davies}},
  \bibinfo {author} {\bibfnamefont {R.}~\bibnamefont {Horgan}}, \bibinfo
  {author} {\bibfnamefont {C.}~\bibnamefont {Monahan}}, \ and\ \bibinfo
  {author} {\bibfnamefont {J.}~\bibnamefont {Shigemitsu}} (\bibinfo
  {collaboration} {HPQCD}),\ }\href {\doibase 10.1103/PhysRevLett.110.222003}
  {\bibfield  {journal} {\bibinfo  {journal} {Phys.Rev.Lett.}\ }\textbf
  {\bibinfo {volume} {110}},\ \bibinfo {pages} {222003} (\bibinfo {year}
  {2013}{\natexlab{b}})},\ \Eprint {http://arxiv.org/abs/1302.2644}
  {arXiv:1302.2644 [hep-lat]} \BibitemShut {NoStop}%
\bibitem [{\citenamefont {Aoki}\ \emph {et~al.}(2015)\citenamefont {Aoki},
  \citenamefont {Ishikawa}, \citenamefont {Izubuchi}, \citenamefont {Lehner},\
  and\ \citenamefont {Soni}}]{Aoki:2014nga}%
  \BibitemOpen
  \bibfield  {author} {\bibinfo {author} {\bibfnamefont {Y.}~\bibnamefont
  {Aoki}}, \bibinfo {author} {\bibfnamefont {T.}~\bibnamefont {Ishikawa}},
  \bibinfo {author} {\bibfnamefont {T.}~\bibnamefont {Izubuchi}}, \bibinfo
  {author} {\bibfnamefont {C.}~\bibnamefont {Lehner}}, \ and\ \bibinfo {author}
  {\bibfnamefont {A.}~\bibnamefont {Soni}},\ }\href {\doibase
  10.1103/PhysRevD.91.114505} {\bibfield  {journal} {\bibinfo  {journal} {Phys.
  Rev.}\ }\textbf {\bibinfo {volume} {D91}},\ \bibinfo {pages} {114505}
  (\bibinfo {year} {2015})},\ \Eprint {http://arxiv.org/abs/1406.6192}
  {arXiv:1406.6192 [hep-lat]} \BibitemShut {NoStop}%
\bibitem [{\citenamefont {Christ}\ \emph {et~al.}(2015)\citenamefont {Christ},
  \citenamefont {Flynn}, \citenamefont {Izubuchi}, \citenamefont {Kawanai},
  \citenamefont {Lehner}, \citenamefont {Soni}, \citenamefont {Van~de Water},\
  and\ \citenamefont {Witzel}}]{Christ:2014uea}%
  \BibitemOpen
  \bibfield  {author} {\bibinfo {author} {\bibfnamefont {N.~H.}\ \bibnamefont
  {Christ}}, \bibinfo {author} {\bibfnamefont {J.~M.}\ \bibnamefont {Flynn}},
  \bibinfo {author} {\bibfnamefont {T.}~\bibnamefont {Izubuchi}}, \bibinfo
  {author} {\bibfnamefont {T.}~\bibnamefont {Kawanai}}, \bibinfo {author}
  {\bibfnamefont {C.}~\bibnamefont {Lehner}}, \bibinfo {author} {\bibfnamefont
  {A.}~\bibnamefont {Soni}}, \bibinfo {author} {\bibfnamefont {R.~S.}\
  \bibnamefont {Van~de Water}}, \ and\ \bibinfo {author} {\bibfnamefont
  {O.}~\bibnamefont {Witzel}} (\bibinfo {collaboration} {RBC/UKQCD}),\ }\href
  {\doibase 10.1103/PhysRevD.91.054502} {\bibfield  {journal} {\bibinfo
  {journal} {Phys. Rev.}\ }\textbf {\bibinfo {volume} {D91}},\ \bibinfo {pages}
  {054502} (\bibinfo {year} {2015})},\ \Eprint {http://arxiv.org/abs/1404.4670}
  {arXiv:1404.4670 [hep-lat]} \BibitemShut {NoStop}%
\bibitem [{\citenamefont {Na}\ \emph {et~al.}(2012{\natexlab{b}})\citenamefont
  {Na}, \citenamefont {Monahan}, \citenamefont {Davies}, \citenamefont
  {Horgan}, \citenamefont {Lepage},\ and\ \citenamefont
  {Shigemitsu}}]{Na:2012kp}%
  \BibitemOpen
  \bibfield  {author} {\bibinfo {author} {\bibfnamefont {H.}~\bibnamefont
  {Na}}, \bibinfo {author} {\bibfnamefont {C.~J.}\ \bibnamefont {Monahan}},
  \bibinfo {author} {\bibfnamefont {C.~T.~H.}\ \bibnamefont {Davies}}, \bibinfo
  {author} {\bibfnamefont {R.}~\bibnamefont {Horgan}}, \bibinfo {author}
  {\bibfnamefont {G.~P.}\ \bibnamefont {Lepage}}, \ and\ \bibinfo {author}
  {\bibfnamefont {J.}~\bibnamefont {Shigemitsu}} (\bibinfo {collaboration}
  {HPQCD}),\ }\href {\doibase 10.1103/PhysRevD.86.034506} {\bibfield  {journal}
  {\bibinfo  {journal} {Phys. Rev.}\ }\textbf {\bibinfo {volume} {D86}},\
  \bibinfo {pages} {034506} (\bibinfo {year} {2012}{\natexlab{b}})},\ \Eprint
  {http://arxiv.org/abs/1202.4914} {arXiv:1202.4914 [hep-lat]} \BibitemShut
  {NoStop}%
\bibitem [{\citenamefont {McNeile}\ \emph {et~al.}(2012)\citenamefont
  {McNeile}, \citenamefont {Davies}, \citenamefont {Follana}, \citenamefont
  {Hornbostel},\ and\ \citenamefont {Lepage}}]{McNeile:2011ng}%
  \BibitemOpen
  \bibfield  {author} {\bibinfo {author} {\bibfnamefont {C.}~\bibnamefont
  {McNeile}}, \bibinfo {author} {\bibfnamefont {C.~T.~H.}\ \bibnamefont
  {Davies}}, \bibinfo {author} {\bibfnamefont {E.}~\bibnamefont {Follana}},
  \bibinfo {author} {\bibfnamefont {K.}~\bibnamefont {Hornbostel}}, \ and\
  \bibinfo {author} {\bibfnamefont {G.~P.}\ \bibnamefont {Lepage}} (\bibinfo
  {collaboration} {HPQCD}),\ }\href {\doibase 10.1103/PhysRevD.85.031503}
  {\bibfield  {journal} {\bibinfo  {journal} {Phys. Rev.}\ }\textbf {\bibinfo
  {volume} {D85}},\ \bibinfo {pages} {031503} (\bibinfo {year} {2012})},\
  \Eprint {http://arxiv.org/abs/1110.4510} {arXiv:1110.4510 [hep-lat]}
  \BibitemShut {NoStop}%
\bibitem [{\citenamefont {Baker}\ \emph {et~al.}(2014)\citenamefont {Baker},
  \citenamefont {Bordes}, \citenamefont {Dominguez}, \citenamefont
  {Penarrocha},\ and\ \citenamefont {Schilcher}}]{Baker:2013mwa}%
  \BibitemOpen
  \bibfield  {author} {\bibinfo {author} {\bibfnamefont {M.~J.}\ \bibnamefont
  {Baker}}, \bibinfo {author} {\bibfnamefont {J.}~\bibnamefont {Bordes}},
  \bibinfo {author} {\bibfnamefont {C.~A.}\ \bibnamefont {Dominguez}}, \bibinfo
  {author} {\bibfnamefont {J.}~\bibnamefont {Penarrocha}}, \ and\ \bibinfo
  {author} {\bibfnamefont {K.}~\bibnamefont {Schilcher}},\ }\href {\doibase
  10.1007/JHEP07(2014)032} {\bibfield  {journal} {\bibinfo  {journal} {JHEP}\
  }\textbf {\bibinfo {volume} {07}},\ \bibinfo {pages} {032} (\bibinfo {year}
  {2014})},\ \Eprint {http://arxiv.org/abs/1310.0941} {arXiv:1310.0941
  [hep-ph]} \BibitemShut {NoStop}%
\bibitem [{\citenamefont {Lucha}\ \emph {et~al.}(2013)\citenamefont {Lucha},
  \citenamefont {Melikhov},\ and\ \citenamefont {Simula}}]{Lucha:2013gta}%
  \BibitemOpen
  \bibfield  {author} {\bibinfo {author} {\bibfnamefont {W.}~\bibnamefont
  {Lucha}}, \bibinfo {author} {\bibfnamefont {D.}~\bibnamefont {Melikhov}}, \
  and\ \bibinfo {author} {\bibfnamefont {S.}~\bibnamefont {Simula}},\ }\href
  {\doibase 10.1103/PhysRevD.88.056011} {\bibfield  {journal} {\bibinfo
  {journal} {Phys.Rev.}\ }\textbf {\bibinfo {volume} {D88}},\ \bibinfo {pages}
  {056011} (\bibinfo {year} {2013})},\ \Eprint {http://arxiv.org/abs/1305.7099}
  {arXiv:1305.7099 [hep-ph]} \BibitemShut {NoStop}%
\bibitem [{\citenamefont {Aoki}\ \emph {et~al.}(2011)\citenamefont {Aoki} \emph
  {et~al.}}]{Aoki:2010dy}%
  \BibitemOpen
  \bibfield  {author} {\bibinfo {author} {\bibfnamefont {Y.}~\bibnamefont
  {Aoki}} \emph {et~al.} (\bibinfo {collaboration} {RBC/UKQCD}),\ }\href
  {\doibase 10.1103/PhysRevD.83.074508} {\bibfield  {journal} {\bibinfo
  {journal} {Phys. Rev.}\ }\textbf {\bibinfo {volume} {D83}},\ \bibinfo {pages}
  {074508} (\bibinfo {year} {2011})},\ \Eprint {http://arxiv.org/abs/1011.0892}
  {arXiv:1011.0892 [hep-lat]} \BibitemShut {NoStop}%
\bibitem [{\citenamefont {Hardy}\ and\ \citenamefont
  {Towner}(2015)}]{Hardy:2014qxa}%
  \BibitemOpen
  \bibfield  {author} {\bibinfo {author} {\bibfnamefont {J.~C.}\ \bibnamefont
  {Hardy}}\ and\ \bibinfo {author} {\bibfnamefont {I.~S.}\ \bibnamefont
  {Towner}},\ }\href {\doibase 10.1103/PhysRevC.91.025501} {\bibfield
  {journal} {\bibinfo  {journal} {Phys. Rev.}\ }\textbf {\bibinfo {volume}
  {C91}},\ \bibinfo {pages} {025501} (\bibinfo {year} {2015})},\ \Eprint
  {http://arxiv.org/abs/1411.5987} {arXiv:1411.5987 [nucl-ex]} \BibitemShut
  {NoStop}%
\bibitem [{\citenamefont {Ambrosino}\ \emph {et~al.}(2011)\citenamefont
  {Ambrosino} \emph {et~al.}}]{Antonelli:2010aa}%
  \BibitemOpen
  \bibfield  {author} {\bibinfo {author} {\bibfnamefont {F.}~\bibnamefont
  {Ambrosino}} \emph {et~al.} (\bibinfo {collaboration} {KLOE}),\ }\href
  {\doibase 10.1140/epjc/s10052-011-1604-7} {\bibfield  {journal} {\bibinfo
  {journal} {Eur. Phys. J.}\ }\textbf {\bibinfo {volume} {C71}},\ \bibinfo
  {pages} {1604} (\bibinfo {year} {2011})},\ \Eprint
  {http://arxiv.org/abs/1011.2668} {arXiv:1011.2668 [hep-ex]} \BibitemShut
  {NoStop}%
\bibitem [{\citenamefont {Abouzaid}\ \emph {et~al.}(2011)\citenamefont
  {Abouzaid} \emph {et~al.}}]{Abouzaid:2010ny}%
  \BibitemOpen
  \bibfield  {author} {\bibinfo {author} {\bibfnamefont {E.}~\bibnamefont
  {Abouzaid}} \emph {et~al.} (\bibinfo {collaboration} {KTeV}),\ }\href
  {\doibase 10.1103/PhysRevD.83.092001} {\bibfield  {journal} {\bibinfo
  {journal} {Phys. Rev.}\ }\textbf {\bibinfo {volume} {D83}},\ \bibinfo {pages}
  {092001} (\bibinfo {year} {2011})},\ \Eprint {http://arxiv.org/abs/1011.0127}
  {arXiv:1011.0127 [hep-ex]} \BibitemShut {NoStop}%
\bibitem [{\citenamefont {Bazavov}\ \emph
  {et~al.}(2014{\natexlab{b}})\citenamefont {Bazavov} \emph
  {et~al.}}]{Bazavov:2013maa}%
  \BibitemOpen
  \bibfield  {author} {\bibinfo {author} {\bibfnamefont {A.}~\bibnamefont
  {Bazavov}} \emph {et~al.} (\bibinfo {collaboration} {Fermilab Lattice and
  MILC}),\ }\href {\doibase 10.1103/PhysRevLett.112.112001} {\bibfield
  {journal} {\bibinfo  {journal} {Phys. Rev. Lett.}\ }\textbf {\bibinfo
  {volume} {112}},\ \bibinfo {pages} {112001} (\bibinfo {year}
  {2014}{\natexlab{b}})},\ \Eprint {http://arxiv.org/abs/1312.1228}
  {arXiv:1312.1228 [hep-ph]} \BibitemShut {NoStop}%
\bibitem [{\citenamefont {Bazavov}\ \emph {et~al.}(2013)\citenamefont
  {Bazavov}, \citenamefont {Bernard}, \citenamefont {Bouchard}, \citenamefont
  {DeTar}, \citenamefont {Du} \emph {et~al.}}]{Bazavov:2012cd}%
  \BibitemOpen
  \bibfield  {author} {\bibinfo {author} {\bibfnamefont {A.}~\bibnamefont
  {Bazavov}}, \bibinfo {author} {\bibfnamefont {C.}~\bibnamefont {Bernard}},
  \bibinfo {author} {\bibfnamefont {C.}~\bibnamefont {Bouchard}}, \bibinfo
  {author} {\bibfnamefont {C.}~\bibnamefont {DeTar}}, \bibinfo {author}
  {\bibfnamefont {D.}~\bibnamefont {Du}},  \emph {et~al.} (\bibinfo
  {collaboration} {Fermilab Lattice and MILC}),\ }\href {\doibase
  10.1103/PhysRevD.87.073012} {\bibfield  {journal} {\bibinfo  {journal}
  {Phys.Rev.}\ }\textbf {\bibinfo {volume} {D87}},\ \bibinfo {pages} {073012}
  (\bibinfo {year} {2013})},\ \Eprint {http://arxiv.org/abs/1212.4993}
  {arXiv:1212.4993 [hep-lat]} \BibitemShut {NoStop}%
\bibitem [{\citenamefont {Boyle}\ \emph
  {et~al.}(2015{\natexlab{b}})\citenamefont {Boyle} \emph
  {et~al.}}]{Boyle:2015hfa}%
  \BibitemOpen
  \bibfield  {author} {\bibinfo {author} {\bibfnamefont {P.~A.}\ \bibnamefont
  {Boyle}} \emph {et~al.} (\bibinfo {collaboration} {RBC/UKQCD}),\ }\href
  {\doibase 10.1007/JHEP06(2015)164} {\bibfield  {journal} {\bibinfo  {journal}
  {JHEP}\ }\textbf {\bibinfo {volume} {06}},\ \bibinfo {pages} {164} (\bibinfo
  {year} {2015}{\natexlab{b}})},\ \Eprint {http://arxiv.org/abs/1504.01692}
  {arXiv:1504.01692 [hep-lat]} \BibitemShut {NoStop}%
\bibitem [{\citenamefont {Towner}\ and\ \citenamefont
  {Hardy}(2015)}]{Towner:2014uta}%
  \BibitemOpen
  \bibfield  {author} {\bibinfo {author} {\bibfnamefont {I.~S.}\ \bibnamefont
  {Towner}}\ and\ \bibinfo {author} {\bibfnamefont {J.~C.}\ \bibnamefont
  {Hardy}},\ }\href {\doibase 10.1103/PhysRevC.91.015501} {\bibfield  {journal}
  {\bibinfo  {journal} {Phys. Rev.}\ }\textbf {\bibinfo {volume} {C91}},\
  \bibinfo {pages} {015501} (\bibinfo {year} {2015})},\ \Eprint
  {http://arxiv.org/abs/1412.0727} {arXiv:1412.0727 [nucl-th]} \BibitemShut
  {NoStop}%
\bibitem [{\citenamefont {Bona}\ \emph {et~al.}(2006)\citenamefont {Bona} \emph
  {et~al.}}]{Bona:2006ah}%
  \BibitemOpen
  \bibfield  {author} {\bibinfo {author} {\bibfnamefont {M.}~\bibnamefont
  {Bona}} \emph {et~al.} (\bibinfo {collaboration} {UTfit}),\ }\href {\doibase
  10.1088/1126-6708/2006/10/081} {\bibfield  {journal} {\bibinfo  {journal}
  {JHEP}\ }\textbf {\bibinfo {volume} {10}},\ \bibinfo {pages} {081} (\bibinfo
  {year} {2006})},\ \bibinfo {note} {updated results and plots available at:
  \url{http://www.utfit.org}},\ \Eprint {http://arxiv.org/abs/hep-ph/0606167}
  {arXiv:hep-ph/0606167 [hep-ph]} \BibitemShut {NoStop}%
\bibitem [{\citenamefont {Amhis}\ \emph {et~al.}(2014)\citenamefont {Amhis}
  \emph {et~al.}}]{Amhis:2014hma}%
  \BibitemOpen
  \bibfield  {author} {\bibinfo {author} {\bibfnamefont {Y.}~\bibnamefont
  {Amhis}} \emph {et~al.} (\bibinfo {collaboration} {Heavy Flavor Averaging
  Group}),\ }\href@noop {} {\  (\bibinfo {year} {2014})},\ \Eprint
  {http://arxiv.org/abs/1412.7515} {arXiv:1412.7515 [hep-ex]} \BibitemShut
  {NoStop}%
\bibitem [{\citenamefont {Na}\ \emph {et~al.}(2010)\citenamefont {Na},
  \citenamefont {Davies}, \citenamefont {Follana}, \citenamefont {Lepage},\
  and\ \citenamefont {Shigemitsu}}]{Na:2010uf}%
  \BibitemOpen
  \bibfield  {author} {\bibinfo {author} {\bibfnamefont {H.}~\bibnamefont
  {Na}}, \bibinfo {author} {\bibfnamefont {C.~T.~H.}\ \bibnamefont {Davies}},
  \bibinfo {author} {\bibfnamefont {E.}~\bibnamefont {Follana}}, \bibinfo
  {author} {\bibfnamefont {G.~P.}\ \bibnamefont {Lepage}}, \ and\ \bibinfo
  {author} {\bibfnamefont {J.}~\bibnamefont {Shigemitsu}} (\bibinfo
  {collaboration} {HPQCD}),\ }\href {\doibase 10.1103/PhysRevD.82.114506}
  {\bibfield  {journal} {\bibinfo  {journal} {Phys. Rev.}\ }\textbf {\bibinfo
  {volume} {D82}},\ \bibinfo {pages} {114506} (\bibinfo {year} {2010})},\
  \Eprint {http://arxiv.org/abs/1008.4562} {arXiv:1008.4562 [hep-lat]}
  \BibitemShut {NoStop}%
\bibitem [{\citenamefont {Na}\ \emph {et~al.}(2011)\citenamefont {Na},
  \citenamefont {Davies}, \citenamefont {Follana}, \citenamefont {Koponen},
  \citenamefont {Lepage},\ and\ \citenamefont {Shigemitsu}}]{Na:2011mc}%
  \BibitemOpen
  \bibfield  {author} {\bibinfo {author} {\bibfnamefont {H.}~\bibnamefont
  {Na}}, \bibinfo {author} {\bibfnamefont {C.~T.~H.}\ \bibnamefont {Davies}},
  \bibinfo {author} {\bibfnamefont {E.}~\bibnamefont {Follana}}, \bibinfo
  {author} {\bibfnamefont {J.}~\bibnamefont {Koponen}}, \bibinfo {author}
  {\bibfnamefont {G.~P.}\ \bibnamefont {Lepage}}, \ and\ \bibinfo {author}
  {\bibfnamefont {J.}~\bibnamefont {Shigemitsu}} (\bibinfo {collaboration}
  {HPQCD}),\ }\href {\doibase 10.1103/PhysRevD.84.114505} {\bibfield  {journal}
  {\bibinfo  {journal} {Phys. Rev.}\ }\textbf {\bibinfo {volume} {D84}},\
  \bibinfo {pages} {114505} (\bibinfo {year} {2011})},\ \Eprint
  {http://arxiv.org/abs/1109.1501} {arXiv:1109.1501 [hep-lat]} \BibitemShut
  {NoStop}%
\bibitem [{\citenamefont {Alberti}\ \emph {et~al.}(2015)\citenamefont
  {Alberti}, \citenamefont {Gambino}, \citenamefont {Healey},\ and\
  \citenamefont {Nandi}}]{Alberti:2014yda}%
  \BibitemOpen
  \bibfield  {author} {\bibinfo {author} {\bibfnamefont {A.}~\bibnamefont
  {Alberti}}, \bibinfo {author} {\bibfnamefont {P.}~\bibnamefont {Gambino}},
  \bibinfo {author} {\bibfnamefont {K.~J.}\ \bibnamefont {Healey}}, \ and\
  \bibinfo {author} {\bibfnamefont {S.}~\bibnamefont {Nandi}},\ }\href
  {\doibase 10.1103/PhysRevLett.114.061802} {\bibfield  {journal} {\bibinfo
  {journal} {Phys. Rev. Lett.}\ }\textbf {\bibinfo {volume} {114}},\ \bibinfo
  {pages} {061802} (\bibinfo {year} {2015})},\ \Eprint
  {http://arxiv.org/abs/1411.6560} {arXiv:1411.6560 [hep-ph]} \BibitemShut
  {NoStop}%
\bibitem [{\citenamefont {Bailey}\ \emph {et~al.}(2014)\citenamefont {Bailey}
  \emph {et~al.}}]{Bailey:2014tva}%
  \BibitemOpen
  \bibfield  {author} {\bibinfo {author} {\bibfnamefont {J.~A.}\ \bibnamefont
  {Bailey}} \emph {et~al.} (\bibinfo {collaboration} {Fermilab Lattice,
  MILC}),\ }\href {\doibase 10.1103/PhysRevD.89.114504} {\bibfield  {journal}
  {\bibinfo  {journal} {Phys. Rev.}\ }\textbf {\bibinfo {volume} {D89}},\
  \bibinfo {pages} {114504} (\bibinfo {year} {2014})},\ \Eprint
  {http://arxiv.org/abs/1403.0635} {arXiv:1403.0635 [hep-lat]} \BibitemShut
  {NoStop}%
\bibitem [{\citenamefont {Wolfenstein}(1983)}]{Wolfenstein:1983yz}%
  \BibitemOpen
  \bibfield  {author} {\bibinfo {author} {\bibfnamefont {L.}~\bibnamefont
  {Wolfenstein}},\ }\href {\doibase 10.1103/PhysRevLett.51.1945} {\bibfield
  {journal} {\bibinfo  {journal} {Phys. Rev. Lett.}\ }\textbf {\bibinfo
  {volume} {51}},\ \bibinfo {pages} {1945} (\bibinfo {year}
  {1983})}\BibitemShut {NoStop}%
\bibitem [{\citenamefont {Charles}\ \emph {et~al.}(2005)\citenamefont
  {Charles}, \citenamefont {Hocker}, \citenamefont {Lacker}, \citenamefont
  {Laplace}, \citenamefont {Le~Diberder}, \citenamefont {Malcles},
  \citenamefont {Ocariz}, \citenamefont {Pivk},\ and\ \citenamefont
  {Roos}}]{Charles:2004jd}%
  \BibitemOpen
  \bibfield  {author} {\bibinfo {author} {\bibfnamefont {J.}~\bibnamefont
  {Charles}}, \bibinfo {author} {\bibfnamefont {A.}~\bibnamefont {Hocker}},
  \bibinfo {author} {\bibfnamefont {H.}~\bibnamefont {Lacker}}, \bibinfo
  {author} {\bibfnamefont {S.}~\bibnamefont {Laplace}}, \bibinfo {author}
  {\bibfnamefont {F.~R.}\ \bibnamefont {Le~Diberder}}, \bibinfo {author}
  {\bibfnamefont {J.}~\bibnamefont {Malcles}}, \bibinfo {author} {\bibfnamefont
  {J.}~\bibnamefont {Ocariz}}, \bibinfo {author} {\bibfnamefont
  {M.}~\bibnamefont {Pivk}}, \ and\ \bibinfo {author} {\bibfnamefont
  {L.}~\bibnamefont {Roos}} (\bibinfo {collaboration} {CKMfitter Group}),\
  }\href {\doibase 10.1140/epjc/s2005-02169-1} {\bibfield  {journal} {\bibinfo
  {journal} {Eur. Phys. J.}\ }\textbf {\bibinfo {volume} {C41}},\ \bibinfo
  {pages} {1} (\bibinfo {year} {2005})},\ \bibinfo {note} {updated results and
  plots available at: \url{http://ckmfitter.in2p3.fr}},\ \Eprint
  {http://arxiv.org/abs/hep-ph/0406184} {arXiv:hep-ph/0406184 [hep-ph]}
  \BibitemShut {NoStop}%
\bibitem [{\citenamefont {Antonelli}\ \emph
  {et~al.}(2010{\natexlab{b}})\citenamefont {Antonelli} \emph
  {et~al.}}]{Antonelli:2009ws}%
  \BibitemOpen
  \bibfield  {author} {\bibinfo {author} {\bibfnamefont {M.}~\bibnamefont
  {Antonelli}} \emph {et~al.},\ }\bibfield  {booktitle} {\emph {\bibinfo
  {booktitle} {{5th International Workshop on the CKM Unitarity Triangle (CKM
  2008) Rome, Italy, September 9-13, 2008}}},\ }\href {\doibase
  10.1016/j.physrep.2010.05.003} {\bibfield  {journal} {\bibinfo  {journal}
  {Phys. Rept.}\ }\textbf {\bibinfo {volume} {494}},\ \bibinfo {pages} {197}
  (\bibinfo {year} {2010}{\natexlab{b}})},\ \Eprint
  {http://arxiv.org/abs/0907.5386} {arXiv:0907.5386 [hep-ph]} \BibitemShut
  {NoStop}%
\bibitem [{\citenamefont {Butler}\ \emph {et~al.}(2013)\citenamefont {Butler}
  \emph {et~al.}}]{Butler:2013kdw}%
  \BibitemOpen
  \bibfield  {author} {\bibinfo {author} {\bibfnamefont {J.~N.}\ \bibnamefont
  {Butler}} \emph {et~al.} (\bibinfo {collaboration} {Quark Flavor Physics
  Working Group}),\ }in\ \href
  {http://inspirehep.net/record/1263372/files/arXiv:1311.1076.pdf} {\emph
  {\bibinfo {booktitle} {{Community Summer Study 2013: Snowmass on the
  Mississippi (CSS2013) Minneapolis, MN, USA, July 29-August 6, 2013}}}}\
  (\bibinfo {year} {2013})\ \Eprint {http://arxiv.org/abs/1311.1076}
  {arXiv:1311.1076 [hep-ex]} \BibitemShut {NoStop}%
\bibitem [{\citenamefont {Bevan}\ \emph {et~al.}(2014)\citenamefont {Bevan}
  \emph {et~al.}}]{Bevan:2014iga}%
  \BibitemOpen
  \bibfield  {author} {\bibinfo {author} {\bibfnamefont {A.~J.}\ \bibnamefont
  {Bevan}} \emph {et~al.} (\bibinfo {collaboration} {Belle, BaBar}),\ }\href
  {\doibase 10.1140/epjc/s10052-014-3026-9} {\bibfield  {journal} {\bibinfo
  {journal} {Eur. Phys. J.}\ }\textbf {\bibinfo {volume} {C74}},\ \bibinfo
  {pages} {3026} (\bibinfo {year} {2014})},\ \Eprint
  {http://arxiv.org/abs/1406.6311} {arXiv:1406.6311 [hep-ex]} \BibitemShut
  {NoStop}%
\bibitem [{\citenamefont {Bailey}\ \emph {et~al.}(2015)\citenamefont {Bailey}
  \emph {et~al.}}]{Lattice:2015tia}%
  \BibitemOpen
  \bibfield  {author} {\bibinfo {author} {\bibfnamefont {J.~A.}\ \bibnamefont
  {Bailey}} \emph {et~al.} (\bibinfo {collaboration} {Fermilab Lattice,
  MILC}),\ }\href {\doibase 10.1103/PhysRevD.92.014024} {\bibfield  {journal}
  {\bibinfo  {journal} {Phys. Rev.}\ }\textbf {\bibinfo {volume} {D92}},\
  \bibinfo {pages} {014024} (\bibinfo {year} {2015})},\ \Eprint
  {http://arxiv.org/abs/1503.07839} {arXiv:1503.07839 [hep-lat]} \BibitemShut
  {NoStop}%
\bibitem [{\citenamefont {del Amo~Sanchez}\ \emph {et~al.}(2011)\citenamefont
  {del Amo~Sanchez} \emph {et~al.}}]{delAmoSanchez:2010af}%
  \BibitemOpen
  \bibfield  {author} {\bibinfo {author} {\bibfnamefont {P.}~\bibnamefont {del
  Amo~Sanchez}} \emph {et~al.} (\bibinfo {collaboration} {BaBar}),\ }\href
  {\doibase 10.1103/PhysRevD.83.032007} {\bibfield  {journal} {\bibinfo
  {journal} {Phys. Rev.}\ }\textbf {\bibinfo {volume} {D83}},\ \bibinfo {pages}
  {032007} (\bibinfo {year} {2011})},\ \Eprint {http://arxiv.org/abs/1005.3288}
  {arXiv:1005.3288 [hep-ex]} \BibitemShut {NoStop}%
\bibitem [{\citenamefont {Lees}\ \emph {et~al.}(2012)\citenamefont {Lees} \emph
  {et~al.}}]{Lees:2012vv}%
  \BibitemOpen
  \bibfield  {author} {\bibinfo {author} {\bibfnamefont {J.~P.}\ \bibnamefont
  {Lees}} \emph {et~al.} (\bibinfo {collaboration} {BaBar}),\ }\href {\doibase
  10.1103/PhysRevD.86.092004} {\bibfield  {journal} {\bibinfo  {journal} {Phys.
  Rev.}\ }\textbf {\bibinfo {volume} {D86}},\ \bibinfo {pages} {092004}
  (\bibinfo {year} {2012})},\ \Eprint {http://arxiv.org/abs/1208.1253}
  {arXiv:1208.1253 [hep-ex]} \BibitemShut {NoStop}%
\bibitem [{\citenamefont {Ha}\ \emph {et~al.}(2011)\citenamefont {Ha} \emph
  {et~al.}}]{Ha:2010rf}%
  \BibitemOpen
  \bibfield  {author} {\bibinfo {author} {\bibfnamefont {H.}~\bibnamefont {Ha}}
  \emph {et~al.} (\bibinfo {collaboration} {Belle}),\ }\href {\doibase
  10.1103/PhysRevD.83.071101} {\bibfield  {journal} {\bibinfo  {journal} {Phys.
  Rev.}\ }\textbf {\bibinfo {volume} {D83}},\ \bibinfo {pages} {071101}
  (\bibinfo {year} {2011})},\ \Eprint {http://arxiv.org/abs/1012.0090}
  {arXiv:1012.0090 [hep-ex]} \BibitemShut {NoStop}%
\bibitem [{\citenamefont {Sibidanov}\ \emph {et~al.}(2013)\citenamefont
  {Sibidanov} \emph {et~al.}}]{Sibidanov:2013rkk}%
  \BibitemOpen
  \bibfield  {author} {\bibinfo {author} {\bibfnamefont {A.}~\bibnamefont
  {Sibidanov}} \emph {et~al.} (\bibinfo {collaboration} {Belle}),\ }\href
  {\doibase 10.1103/PhysRevD.88.032005} {\bibfield  {journal} {\bibinfo
  {journal} {Phys.Rev.}\ }\textbf {\bibinfo {volume} {D88}},\ \bibinfo {pages}
  {032005} (\bibinfo {year} {2013})},\ \Eprint {http://arxiv.org/abs/1306.2781}
  {arXiv:1306.2781 [hep-ex]} \BibitemShut {NoStop}%
\bibitem [{\citenamefont {Bosch}\ \emph {et~al.}(2004)\citenamefont {Bosch},
  \citenamefont {Lange}, \citenamefont {Neubert},\ and\ \citenamefont
  {Paz}}]{Bosch:2004bt}%
  \BibitemOpen
  \bibfield  {author} {\bibinfo {author} {\bibfnamefont {S.~W.}\ \bibnamefont
  {Bosch}}, \bibinfo {author} {\bibfnamefont {B.~O.}\ \bibnamefont {Lange}},
  \bibinfo {author} {\bibfnamefont {M.}~\bibnamefont {Neubert}}, \ and\
  \bibinfo {author} {\bibfnamefont {G.}~\bibnamefont {Paz}},\ }\href {\doibase
  10.1103/PhysRevLett.93.221801} {\bibfield  {journal} {\bibinfo  {journal}
  {Phys. Rev. Lett.}\ }\textbf {\bibinfo {volume} {93}},\ \bibinfo {pages}
  {221801} (\bibinfo {year} {2004})},\ \Eprint
  {http://arxiv.org/abs/hep-ph/0403223} {arXiv:hep-ph/0403223 [hep-ph]}
  \BibitemShut {NoStop}%
\bibitem [{\citenamefont {Andersen}\ and\ \citenamefont
  {Gardi}(2006)}]{Andersen:2005mj}%
  \BibitemOpen
  \bibfield  {author} {\bibinfo {author} {\bibfnamefont {J.~R.}\ \bibnamefont
  {Andersen}}\ and\ \bibinfo {author} {\bibfnamefont {E.}~\bibnamefont
  {Gardi}},\ }\href {\doibase 10.1088/1126-6708/2006/01/097} {\bibfield
  {journal} {\bibinfo  {journal} {JHEP}\ }\textbf {\bibinfo {volume} {01}},\
  \bibinfo {pages} {097} (\bibinfo {year} {2006})},\ \Eprint
  {http://arxiv.org/abs/hep-ph/0509360} {arXiv:hep-ph/0509360 [hep-ph]}
  \BibitemShut {NoStop}%
\bibitem [{\citenamefont {Gambino}\ \emph {et~al.}(2007)\citenamefont
  {Gambino}, \citenamefont {Giordano}, \citenamefont {Ossola},\ and\
  \citenamefont {Uraltsev}}]{Gambino:2007rp}%
  \BibitemOpen
  \bibfield  {author} {\bibinfo {author} {\bibfnamefont {P.}~\bibnamefont
  {Gambino}}, \bibinfo {author} {\bibfnamefont {P.}~\bibnamefont {Giordano}},
  \bibinfo {author} {\bibfnamefont {G.}~\bibnamefont {Ossola}}, \ and\ \bibinfo
  {author} {\bibfnamefont {N.}~\bibnamefont {Uraltsev}},\ }\href {\doibase
  10.1088/1126-6708/2007/10/058} {\bibfield  {journal} {\bibinfo  {journal}
  {JHEP}\ }\textbf {\bibinfo {volume} {10}},\ \bibinfo {pages} {058} (\bibinfo
  {year} {2007})},\ \Eprint {http://arxiv.org/abs/0707.2493} {arXiv:0707.2493
  [hep-ph]} \BibitemShut {NoStop}%
\bibitem [{\citenamefont {Kowalewski}\ and\ \citenamefont
  {Mannel}(2015)}]{PDGVubVcb2015}%
  \BibitemOpen
  \bibfield  {author} {\bibinfo {author} {\bibfnamefont {R.}~\bibnamefont
  {Kowalewski}}\ and\ \bibinfo {author} {\bibfnamefont {T.}~\bibnamefont
  {Mannel}},\ }\href@noop {} {\enquote {\bibinfo {title} {{Semileptonic
  $B$-meson decays and the determination of $V_{cb}$ and $V_{ub}$}},}\
  }\bibinfo {howpublished}
  {\url{http://pdg.lbl.gov/2015/reviews/rpp2015-rev-vcb-vub.pdf}} (\bibinfo
  {year} {2015}),\ \bibinfo {note} {{review prepared for PDG 2015
  edition}}\BibitemShut {NoStop}%
\bibitem [{\citenamefont {Aaij}\ \emph {et~al.}(2015)\citenamefont {Aaij} \emph
  {et~al.}}]{Aaij:2015bfa}%
  \BibitemOpen
  \bibfield  {author} {\bibinfo {author} {\bibfnamefont {R.}~\bibnamefont
  {Aaij}} \emph {et~al.} (\bibinfo {collaboration} {LHCb}),\ }\href {\doibase
  10.1038/nphys3415} {\bibfield  {journal} {\bibinfo  {journal} {Nature Phys.}\
  }\textbf {\bibinfo {volume} {11}} (\bibinfo {year} {2015}),\
  10.1038/nphys3415},\ \Eprint {http://arxiv.org/abs/1504.01568}
  {arXiv:1504.01568 [hep-ex]} \BibitemShut {NoStop}%
\bibitem [{\citenamefont {Detmold}\ \emph {et~al.}(2015)\citenamefont
  {Detmold}, \citenamefont {Lehner},\ and\ \citenamefont
  {Meinel}}]{Detmold:2015aaa}%
  \BibitemOpen
  \bibfield  {author} {\bibinfo {author} {\bibfnamefont {W.}~\bibnamefont
  {Detmold}}, \bibinfo {author} {\bibfnamefont {C.}~\bibnamefont {Lehner}}, \
  and\ \bibinfo {author} {\bibfnamefont {S.}~\bibnamefont {Meinel}},\ }\href
  {\doibase 10.1103/PhysRevD.92.034503} {\bibfield  {journal} {\bibinfo
  {journal} {Phys. Rev.}\ }\textbf {\bibinfo {volume} {D92}},\ \bibinfo {pages}
  {034503} (\bibinfo {year} {2015})},\ \Eprint
  {http://arxiv.org/abs/1503.01421} {arXiv:1503.01421 [hep-lat]} \BibitemShut
  {NoStop}%
\bibitem [{\citenamefont {Kowalewski}\ and\ \citenamefont
  {Mannel}(2014)}]{PDGVubVcb}%
  \BibitemOpen
  \bibfield  {author} {\bibinfo {author} {\bibfnamefont {R.}~\bibnamefont
  {Kowalewski}}\ and\ \bibinfo {author} {\bibfnamefont {T.}~\bibnamefont
  {Mannel}},\ }\href@noop {} {\enquote {\bibinfo {title} {{Semileptonic
  $B$-meson decays and the determination of $V_{cb}$ and $V_{ub}$}},}\
  }\bibinfo {howpublished}
  {\url{http://pdg.lbl.gov/2014/reviews/rpp2014-rev-vcb-vub.pdf}} (\bibinfo
  {year} {2014}),\ \bibinfo {note} {{review prepared for PDG 2014
  edition}}\BibitemShut {NoStop}%
\bibitem [{\citenamefont {Glattauer}\ \emph {et~al.}(2015)\citenamefont
  {Glattauer} \emph {et~al.}}]{Glattauer:2015teq}%
  \BibitemOpen
  \bibfield  {author} {\bibinfo {author} {\bibfnamefont {R.}~\bibnamefont
  {Glattauer}} \emph {et~al.} (\bibinfo {collaboration} {Belle}),\ }\href@noop
  {} {\  (\bibinfo {year} {2015})},\ \Eprint {http://arxiv.org/abs/1510.03657}
  {arXiv:1510.03657 [hep-ex]} \BibitemShut {NoStop}%
\end{thebibliography}

\end{document}